\documentclass[format=acmsmall, review=false, screen=true]{acmart}

\usepackage{booktabs} 
\usepackage{graphicx,epstopdf}

\usepackage{epsfig}
\usepackage{amsmath}
\usepackage{amssymb}
\usepackage{comment}
\usepackage{multirow}
\usepackage{wrapfig}
\usepackage{caption}
\usepackage{mathrsfs}
\usepackage{enumerate}
\usepackage{enumfrom}

\usepackage{array}
\newlength\Origarrayrulewidth
\newcommand{\ttline}[1]{%
  \noalign{\global\setlength\Origarrayrulewidth{\arrayrulewidth}}%
  \noalign{\global\setlength\arrayrulewidth{1.5pt}}\cline{#1}%
  \noalign{\global\setlength\arrayrulewidth{\Origarrayrulewidth}}%
}
\usepackage{hhline}

\newcommand\ltcell[1]{%
  \multicolumn{1}{!{\vrule width 1.5pt}c!{\vrule width 1pt}}{#1}%
}
\newcommand\rtcell[1]{%
  \multicolumn{1}{!{\vrule width 1pt}c!{\vrule width 1.5pt}}{#1}%
}

\usepackage{hyperref}
\hypersetup{
  colorlinks   = true, 
  urlcolor     = blue, 
  linkcolor    = blue, 
  citecolor   = blue 
}

\newcommand{\dproof}[2]{{\noindent\bf Proof of #1:~}#2}
\newcommand{\ignore}[1]{}
\newcommand{\ignoreT}[1]{}
\newcommand{\upext}[3]{\scalebox{1.5}{\ensuremath{\hat{\varepsilon}}}_{#1}^{#2}(#3)}
\newcommand{\downext}[3]{\scalebox{1.5}{\ensuremath{\check{\varepsilon}}}_{#1}^{#2}(#3)}
\newcommand{\bcq}{BCQ}
\newcommand{\cq}{CQ}
\newcommand{\ucq}{UCQ}

\newcommand{\boxtheorem}{\ensuremath{\hfill \Box}}
\newcommand{\mc}[1]{\mathcal{ #1}}
\newcommand{\mscr}[1]{\mathscr{ #1}}
\newcommand{\mf}[1]{\mathfrak{ #1}}
\newcommand{\nit}[1]{{\it #1}}

\newcommand{\bl}[1]{#1}
\newcommand{\rred}[1]{#1}

\newcommand{\nc}{{\em nc}}

\newcommand{\da}{Datalog}
\newcommand{\dpm}{{Datalog}$^\pm$}
\newcommand{\de}{{Datalog}$^\exists$}

\newcommand{\rules}{\Pi^{R}}
\newcommand{\constraints}{\Pi^{C}}
\newcommand{\prg}{\Pi}

\newcommand{\schema}{\mc{R}}
\newcommand{\rank}{\pi}
\newcommand{\finiteRank}{\rank_F}
\newcommand{\infiniteRank}{\rank_\infty}

\newcommand{\dplus}{{Datalog}$^+$}

\newcommand{\m}{\;\!\!}
\newcounter{rownum}

\newcommand{\SCh}{{\em S\m{}Ch}}

\newcommand{\WS}{{\em W\m{}S}}
\newcommand{\WA}{{\em W\m{}A}}

\newcommand{\CQQA}{{\sf QualityQA}}

\newcommand{\dg}{DG}
\newcommand{\qa}{QA}

\newcommand{\omd}{OMD}

\newcommand{\dl}{DL}

\newcommand{\owa}{OWA}
\newcommand{\cwa}{CWA}

\newcommand{\hm}{HM}

\newcommand{\md}{MD}
\newcommand{\fd}{FD}
\newcommand{\fds}{FDs}
\newcommand{\fo}{FO}
\newcommand{\ic}{IC}
\newcommand{\ics}{ICs}
\newcommand{\idep}{ID}
\newcommand{\ideps}{IDs}
\newcommand{\vectt}[1]{\bar{#1}}

\newcommand{\conp}{{\sc conp}}
\newcommand{\np}{{\sc np}}
\newcommand{\exptime}{{\sc exptime}}
\newcommand{\ptime}{{\sc ptime}}
\newcommand{\acz}{{\sc ac}$_0$}
\newcommand{\T}{\mc{T}}
\newcommand{\Tc}{\mc{T}^c}
\newcommand{\Tp}{\mc{T}'}

\newcommand{\egds}{{\em egds}}
\newcommand{\egd}{{\em egd}}
\newcommand{\tgds}{{\em tgds}}
\newcommand{\tgd}{{\em tgd}}
\newcommand{\ncs}{{\em ncs}}

\newcommand{\red}[1]{{#1}}
\newcommand{\blue}[1]{{#1}}

\newcommand{\comlb}[1]{{\vspace{2mm}\noindent \bf \blue{COMM(LEO):}}~ #1 \hfill {\bf
    END.}\\}
\newcommand{\commos}[1]{{\vspace{2mm}\noindent \bf \blue{COMM(MOSTAFA):}}~ #1 \hfill {\bf
    END.}\\}

\newcounter{mycounter}

\usepackage[ruled]{algorithm2e} 

\SetAlFnt{\small}
\SetAlCapFnt{\small}
\SetAlCapNameFnt{\small}
\SetAlCapHSkip{0pt}
\IncMargin{-\parindent}

\acmYear{2017}
\acmMonth{3}

\setcopyright{acmlicensed}

\acmDOI{0000001.0000001}

\received{March 2017}

\begin{document}
\title[Ontological \md \ Data Models and Contextual Data Quality]{Ontological Multidimensional Data Models and Contextual Data Quality}
\author{Leopoldo Bertossi}
\affiliation{%
\email{bertossi@scs.carleton.ca}
  \institution{Carleton University}
  \city{Ottawa}
  \country{Canada}}
\author{Mostafa Milani}
\affiliation{%
\email{mmilani@mcmaster.ca}
  \institution{McMaster University}
  \city{Hamilton}
  \country{Canada}}

\begin{abstract}
Data quality assessment and data cleaning are context-dependent activities. Motivated by this observation, we propose the {\em Ontological Multidimensional Data Model} (\omd \ model), which can be used to model and represent contexts as logic-based ontologies. The data under assessment  is mapped into the context, for additional analysis, processing, and quality data extraction. The resulting contexts allow for the representation of {\em dimensions}, and   multidimensional data quality assessment becomes possible. At the core of a multidimensional context we include a generalized multidimensional data model and a \dpm \ ontology with provably good properties in terms of {\em query answering}. These main components are  used to represent dimension hierarchies, dimensional constraints, dimensional rules, and define predicates for  quality data specification. Query answering relies upon and triggers  navigation through dimension hierarchies, and becomes the basic tool for the extraction of quality data. The \omd \ model is interesting {\em per se}, beyond applications to data quality. It allows for a logic-based, and computationally tractable representation of multidimensional data,  extending previous multidimensional data models with additional expressive power and functionalities.
\end{abstract}


\terms{Database Management, Multidimensional data, Contexts, Data quality, Data cleaning}
\thanks{This work is supported by NSERC Discovery Grant 2016-06148, and the NSERC Strategic Network on Business Intelligence (BIN)}
\maketitle
\renewcommand{\shortauthors}{L. Bertossi \& M. Milani}

\section{Introduction}\label{sec:intr}

Assessing the quality of data and performing data cleaning when the data are not up to the expected standards of quality have been and will
continue being common, difficult and costly problems in data management \cite{batini,eckerson,redman}. This is due, among other factors, to the fact that there is no uniform, general definition of quality data. Actually,
data quality  has several {\em dimensions}. Some of them are~\cite{batini}: (1) {\em Consistency}, which refers to the validity and integrity of data representing real-world entities, typically identified with satisfaction of integrity constraints. (2) {\em Currency} (or timeliness), which aims to identify the current values of entities represented by tuples in a (possibly stale) database, and to answer queries with the current values. (3) {\em Accuracy}, which refers to the closeness of values in a database to the true values for the entities that the data in the database represents; and (4) {\em Completeness}, which  is characterized in terms of the presence/absence of values. (5) {\em Redundancy}, e.g. multiple representations of external entities or of certain aspects thereof. Etc. (Cf. also \cite{lei,fan,geerts} for more on quality dimensions.)

In this work we consider data quality as referring  to the degree to which the data fits or fulfills a form of usage~\cite{batini}, relating our data quality concerns to the {\em production and the use} of data. We will elaborate more on this after the motivating example in this introduction.


Independently from the quality dimension we may consider, {\em data quality assessment and data cleaning are context-dependent activities}. This is our starting point, and the one leading our research. In more concrete terms, the quality of data has to be assessed with some form of  contextual knowledge; and whatever we do with the data in the direction of data cleaning also depends on contextual knowledge. For example, contextual knowledge can tell us if the data we have is incomplete or inconsistent. In the latter case, the context knowledge is provided by explicit semantic constraints.

\begin{figure}[h]
\begin{center}
\vspace{-2mm}
\includegraphics[width=7cm]{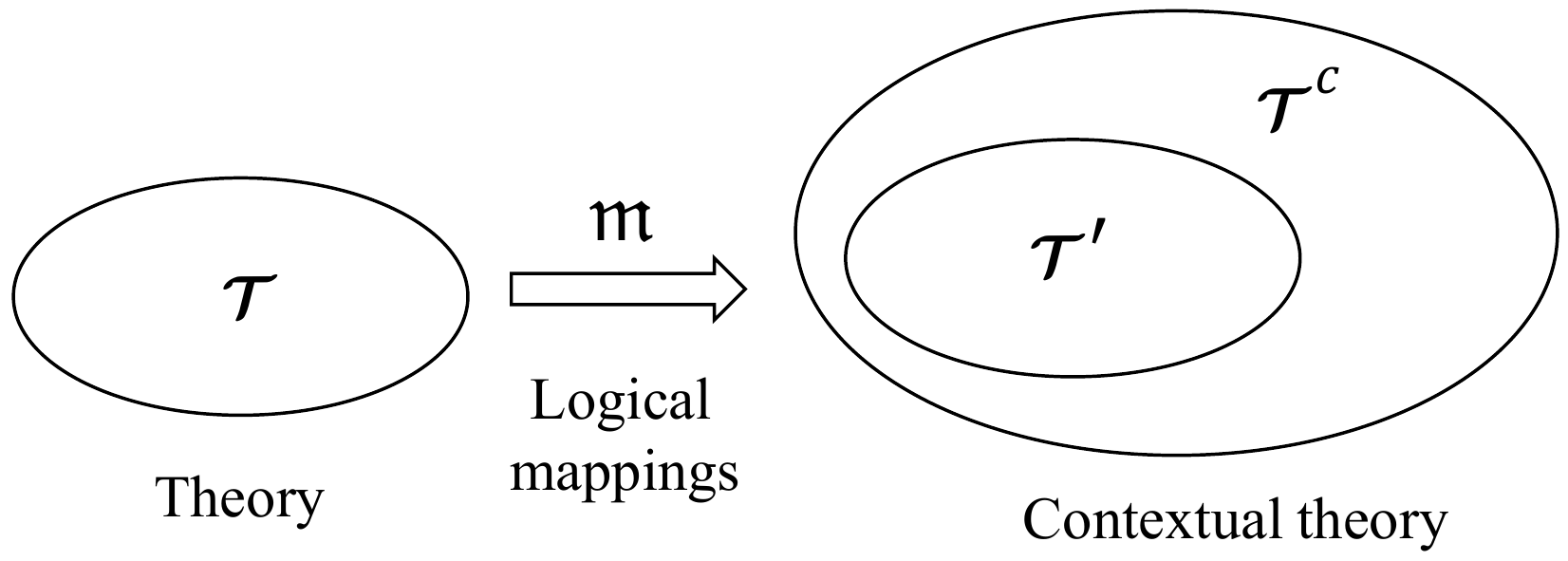}
\caption{Embedding into a contextual theory}\label{fig:emb}
\vspace{-6mm}
\end{center}
\end{figure}

In order to address contextual data quality issues, we need a formal model of context. In very general terms, the big picture is as follows. A database can be seen as a logical theory, $\T$, and a context for it, as another logical theory, $\Tc$, into which $\T$ is mapped by means of a set, $\mf{m}$, of {\em logical mappings}, \red{as shown in Figure \ref{fig:emb}. The image of $\T$ in $\Tc$ is $\Tp=\mf{m}(\T)$, which could be seen as an {\em interpretation} of $\T$ in $\Tc$.\footnote{\ Interpretations between logical theories have been investigated in mathematical logic \cite[sec. 2.7]{enderton} and used, e.g. to obtain (un)decidability results \cite{rabin}.} The contextual theory $\Tc$ provides extra knowledge about $\T$, as a logical  extension of its image \red{$\Tp$}}. For example, $\Tc$ may contain  additional semantic constraints on elements of $\T$ (or their images in $\Tc$) or extensions of  their definitions. In this way, $\Tc$ conveys more semantics or meaning about $\T$, contributing to {\em making more sense} of $\T$'s elements. $\Tc$ may also contain  data and logical rules that can be used for further processing or using knowledge in $\T$. The embedding of $\T$ into $\Tc$ can be achieved via predicates in common or, more complex logical formulas.

In this work, building upon and considerably extending the framework in \cite{bertossi-brite,bertossi16}, context-based data quality assessment, quality data extraction and data cleaning on a relational database $D$ \red{for a relational schema $\mc{R}$} are approached by creating a context model where $D$ is the theory $\T$ above (it could be expressed as a logical theory \cite{reiter}), the theory $\Tc$ is a (logical) ontology $\mc{O}^c$; and, considering that we are using theories around data, the mappings can be logical mappings as used in virtual data integration \cite{lenzerini} or data exchange \cite{barcelo}. In this work, the mappings turn out to be  quite simple: The ontology contains, among other predicates, {\em nicknames} for the predicates in \red{$\mc{R}$} (i.e. copies of them), so that each predicate $P$ in $\mc{R}$ is directly mapped to its copy $P'$ in $\mc{O}^c$.

 Once the data in $D$ is mapped into $\mc{O}^c$, i.e. {\em put in context}, the extra elements in it can be used to define alternative versions of $D$, in our case, {\em clean or quality versions}, $D^q$, of $D$ in terms of
 data quality. The data quality criteria are imposed within $\mc{O}^c$.  This may determine a class of possible quality versions of $D$, virtual or material. The existence of several quality versions reflects the uncertainty that emerges \red{from not having only quality data in $D$}.

\begin{figure}[h]
\begin{center}
\includegraphics[width=7cm]{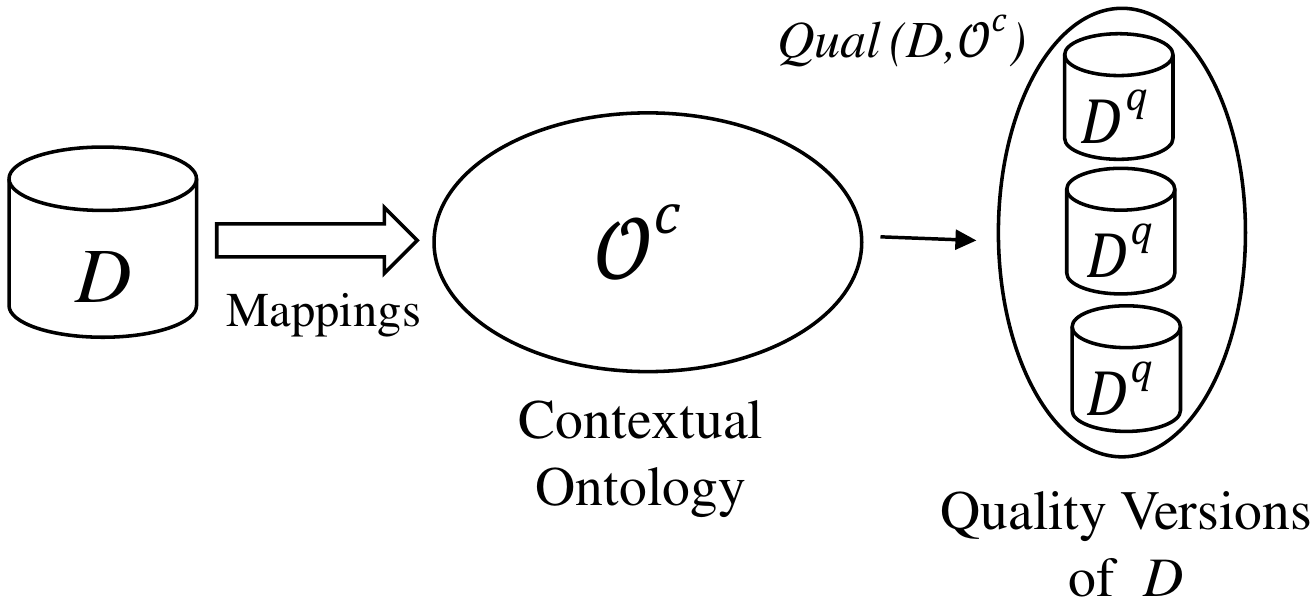}\vspace{-3mm}
\caption{Contextual ontology and quality versions}\label{fig:contOnto}
\vspace{-3mm}
\end{center}
\end{figure}

The whole class, $\nit{Qual}(D,\mc{O}^c)$, of quality versions of $D$ determines or characterizes the quality data in $D$, as the data that are {\em certain} with respect to $\nit{Qual}(D,\mc{O}^c)$. One way to go in this direction consists  in keeping only the data that are found in the intersection of all the instances in $\nit{Qual}(D,\mc{O}^c)$. A more relaxed alternative consists in  considering as quality data those that are obtained as {\em certain answers} to queries posed to $D$, but answered through $\nit{Qual}(D,\mc{O}^c)$: The query is posed to each of the instances in $\nit{Qual}(D,\mc{O}^c)$ (which essentially  have the same schema as $D$), but only those answers that are shared by those instances are considered to be certain \cite{imielinski}.\footnote{\ \label{ft:cqa}Those familiar with {\em database repairs and consistent query answering} \cite{bertossi11,bertossi06}, would notice that both can be formulated in this general stetting. Instance $D$ would be the inconsistent database, the ontology would provide the integrity constraints and the specification of repairs, say in answer set programming \cite{monica}, the class $\nit{Qual}(D,\mc{O}^c)$ would contain the repairs, and the general certain answers would become the consistent answers.} These answers become the {\em quality-answers} in our setting.

The main question is about the kind of contextual ontologies that are appropriate for our tasks. There are several basic conditions to satisfy.
First of all, $\mc{O}^c$ has to be written in a logical language. As a theory it has to be expressive enough, but not too much so that computational problems, such as (quality) data extraction via queries becomes intractable, if not impossible. It also has to combine well with relational data. And, as we emphasize and exploit in our work, it has to allow for the  representation and use of {\em dimensions of data}, \red{i.e. conceptual axes along which data are represented and analyzed. They are the basic elements in
multidimensional databases and data warehouses \cite{jensen}, where we usually find {\em time}, {\em location}, {\em product}, as three dimensions that give context to numerical data, e.g. {\em sales}.} Dimensions are almost essential elements of contexts, in general, and crucial if we want to analyze data from different perspectives or points of view. \red{We use dimensions as (possibly partially ordered) hierarchies of categories.}\footnote{\ Data dimensions were not considered in \cite{bertossi-brite,bertossi16}.} \red{For example, the {\em location} dimension could have categories, {\em city}, {\em province}, {\em country}, {\em continent}, in this hierarchical order of abstraction.}

The language of choice for the contextual ontologies will be Datalog$^\pm$ \cite{cali09}. As an extension of Datalog, a declarative query language for relational databases \cite{ceri}, it provides declarative
extensions of relational data by means of expressive rules and semantic constraints. Certain classes of \dpm \ programs have non-trivial expressive power and good computational properties at the same time.
One of those good classes is that of {\em weakly-sticky} \dpm~\cite{cali12}. Programs in that class allow us to represent a logic-based, relational reconstruction and extension of the Hurtado-Mendelzon  multidimensional data model \cite{hurtado-pods,hurtado-acm}, which allows us to bring data dimensions into contexts.

\begin{figure}[ht]
\vspace{-5mm}
\begin{center}
\includegraphics[width=9.5cm]{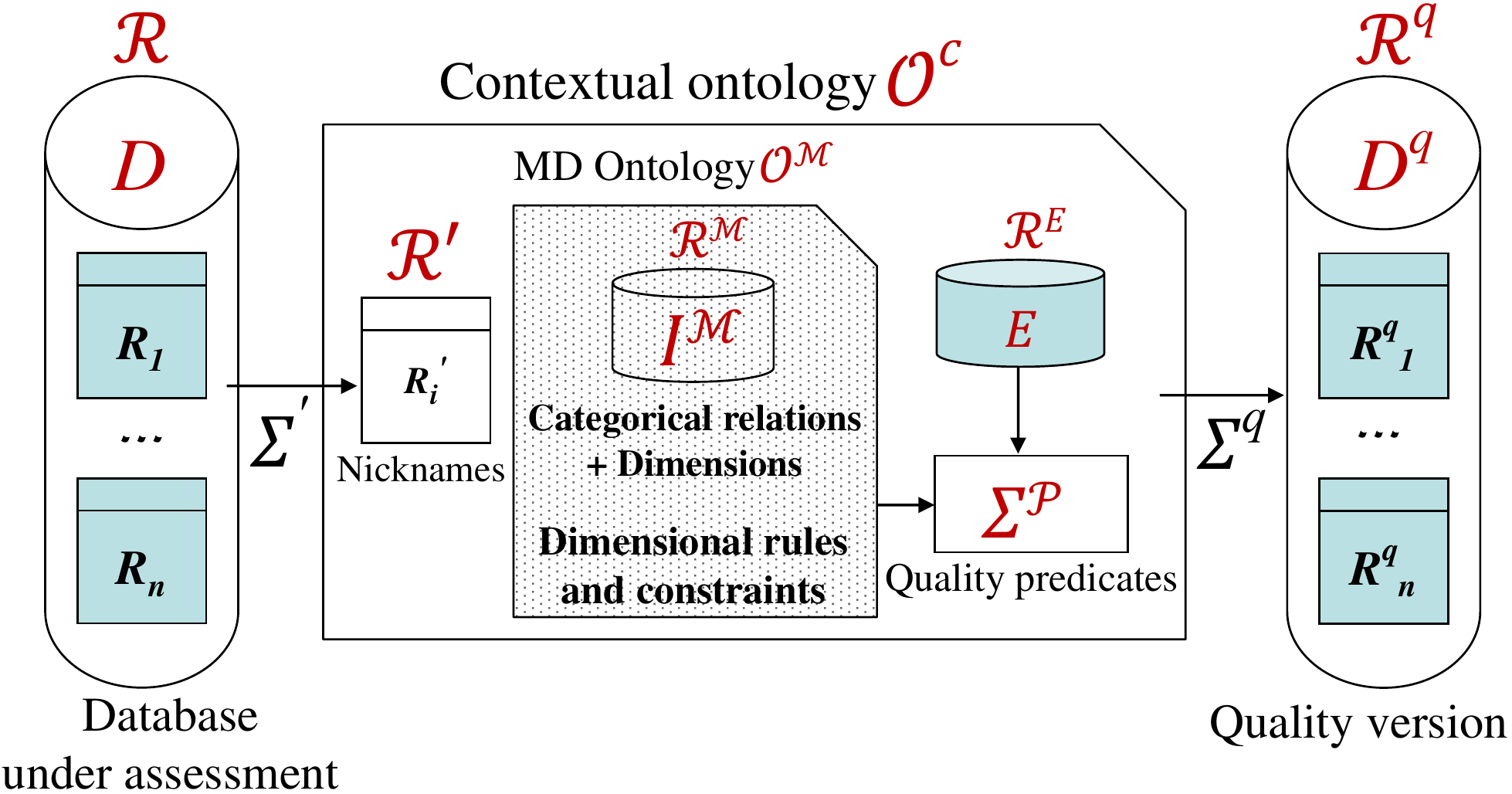}
\vspace{-5mm}
\caption{A  contextual ontology}
\label{fig:frm}
\end{center}
\vspace{-5mm}
\end{figure}

Every contextual ontology $\mc{O}^c$ contains its {\em multidimensional core ontology}, $\mc{O}^M$, which is written in \dpm \ and represents what we will call the {\em ontological multidimensional data model} (\omd \ model, in short), plus a quality-oriented sub-ontology, $\mc{O}^q$, containing extra relational data (shown as instance $E$ in Figure \ref{fig:frm}), Datalog  rules, and possibly additional constraints.  Both sub-ontologies are application dependent, but
$\mc{O}^M$ follows a relatively fixed format, and contains the dimensional structure and data that extend and supplement the data in the input instance $D$, without any explicit quality concerns in it. The \omd \ model is interesting {\em per se} in that it considerably extends the usual multidimensional data models (more on this later). Ontology $\mc{O}^q$ contains as main elements definitions of quality predicates that will be used to produce quality versions of the original tables, and to compute quality query answers.  Notice that the latter problem becomes a case  of
{\em ontology-based data access} (OBDA), i.e. about indirectly accessing underlying data through queries posed to the interface and elements of an ontology~\cite{poggi}.

\vspace{-2mm}

\begin{table}[ht]
\begin{minipage}[t]{0.45\textwidth}
\hspace{-1.2cm} \begin{center}
\includegraphics[width=5cm]{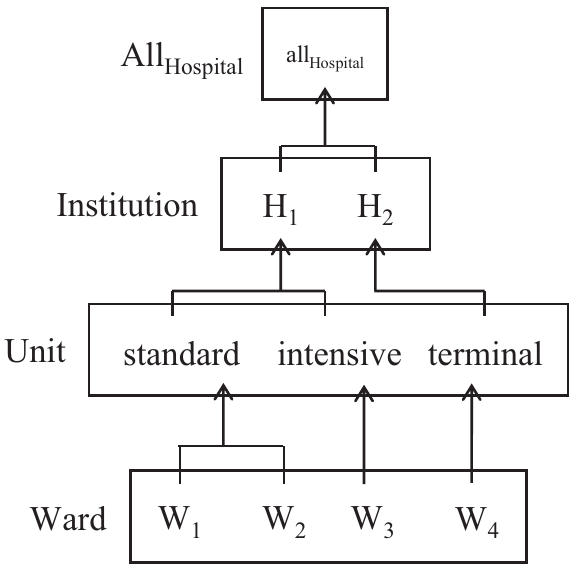}
 \captionof{figure}{The {\sf Hospital} dimension}\label{fig:dimension}
 \end{center}
\end{minipage}
\begin{minipage}[t]{0.45\textwidth}
\vspace{-2mm}
\begin{center}
\setlength{\tabcolsep}{0.2em}
\setcounter{rownum}{0}
\setlength{\arrayrulewidth}{0.75pt}
\renewcommand*\arraystretch{1.4}
\captionof{table}{$\nit{Temperatures}$}\label{tab:temperatures}
\vspace{-4mm}
{\scriptsize \begin{tabular}{c|c|c|c|c|}
\cline{2-5}
 & \textbf{Time} & \textbf{Patient} & \textbf{Value} & \textbf{Nurse}\\
\cline{2-5}
{\tiny \addtocounter{rownum}{1}\arabic{rownum}} & {\footnotesize 12:10-Sep/1/2016} & Tom Waits & 38.2  & Anna \\
\cline{2-5}
{\tiny \addtocounter{rownum}{1}\arabic{rownum}} & {\footnotesize 11:50-Sep/6/2016} & Tom Waits & 37.1  & Helen \\
\cline{2-5}
{\tiny \addtocounter{rownum}{1}\arabic{rownum}} & {\footnotesize 12:15-Nov/12/2016} & Tom Waits & 37.7  & Alan \\
\cline{2-5}
{\tiny \addtocounter{rownum}{1}\arabic{rownum}} & {\footnotesize 12:00-Aug/21/2016} & Tom Waits & 37.0  & Sara \\
\cline{2-5}
{\tiny \addtocounter{rownum}{1}\arabic{rownum}} & {\footnotesize 11:05-Sep/5/2016} & Lou Reed & 37.5  & Helen \\
\cline{2-5}
{\tiny \addtocounter{rownum}{1}\arabic{rownum}} & {\footnotesize 12:15-Aug/21/2016} & Lou Reed & 38.0  & Sara \\
\cline{2-5}
\end{tabular} }
\end{center}
\vspace{1mm}
\begin{center}
\setlength{\tabcolsep}{0.2em}
\setcounter{rownum}{0}
\setlength{\arrayrulewidth}{0.75pt}
\renewcommand*\arraystretch{1.2}
\captionof{table}{${\it Temperatures}^q$}\label{tab:temperaturesq}
\vspace{-4mm}
\begin{tabular}{c|c|c|c|c|}
\cline{2-5}
 & \textbf{Time} & \textbf{Patient} & \textbf{Value} & \textbf{Nurse}\\
 \cline{2-5}
{\tiny \addtocounter{rownum}{1}\arabic{rownum}} & {\footnotesize 12:15-Nov/12/2016} & Tom Waits & 37.7  & Alan \\
\cline{2-5}
{\tiny \addtocounter{rownum}{1}\arabic{rownum}} & {\footnotesize 12:00-Aug/21/2016} & Tom Waits & 37.0  & Sara \\
\cline{2-5}
{\tiny \addtocounter{rownum}{1}\arabic{rownum}} & {\footnotesize 12:15-Aug/21/2016} & Lou Reed & 38.0  & Sara \\
\cline{2-5}
\end{tabular}
\end{center}
\end{minipage}
\end{table}

\vspace{-4mm}

\begin{example}\label{ex:intr} The relational table {\it Temperatures} (Table~\ref{tab:temperatures}) shows body temperatures of patients in an institution. A doctor wants to know {\em ``The body temperatures of Tom Waits for August 21 taken around noon with a thermometer of brand $B_1$"}  (as he expected). Possibly a nurse, unaware of this requirement, used a thermometer of brand $B_2$, storing the data in {\it Temperatures}. In this case, not all the temperature measurements in the table are up to the expected quality. However, table  {\it Temperatures} alone does not discriminate between intended values (those taken with brand $B_1$) and the others.

For assessing the quality of the data or for extracting quality data in/from the table {\it Temperatures} according to the doctor's quality requirement, extra contextual information about the thermometers in use may help. In this case, we may have contextual information in the form of a {\em guideline} prescribing that: {\em ``Nurses in intensive care unit use thermometers of Brand $B_1$"\!.} We still cannot combine this guideline with the data in the table. However, if we know that nurses work in {\em wards}, and those wards are associated to {units}, then we may be in position to combine the table with the given contextual information. \red{Actually, as shown   in Figure \ref{fig:omdm0}, the context contains {\em dimensional data}, in categorical relations linked to dimensions.}

\begin{figure}[ht]
\begin{center}
\includegraphics[width=10cm]{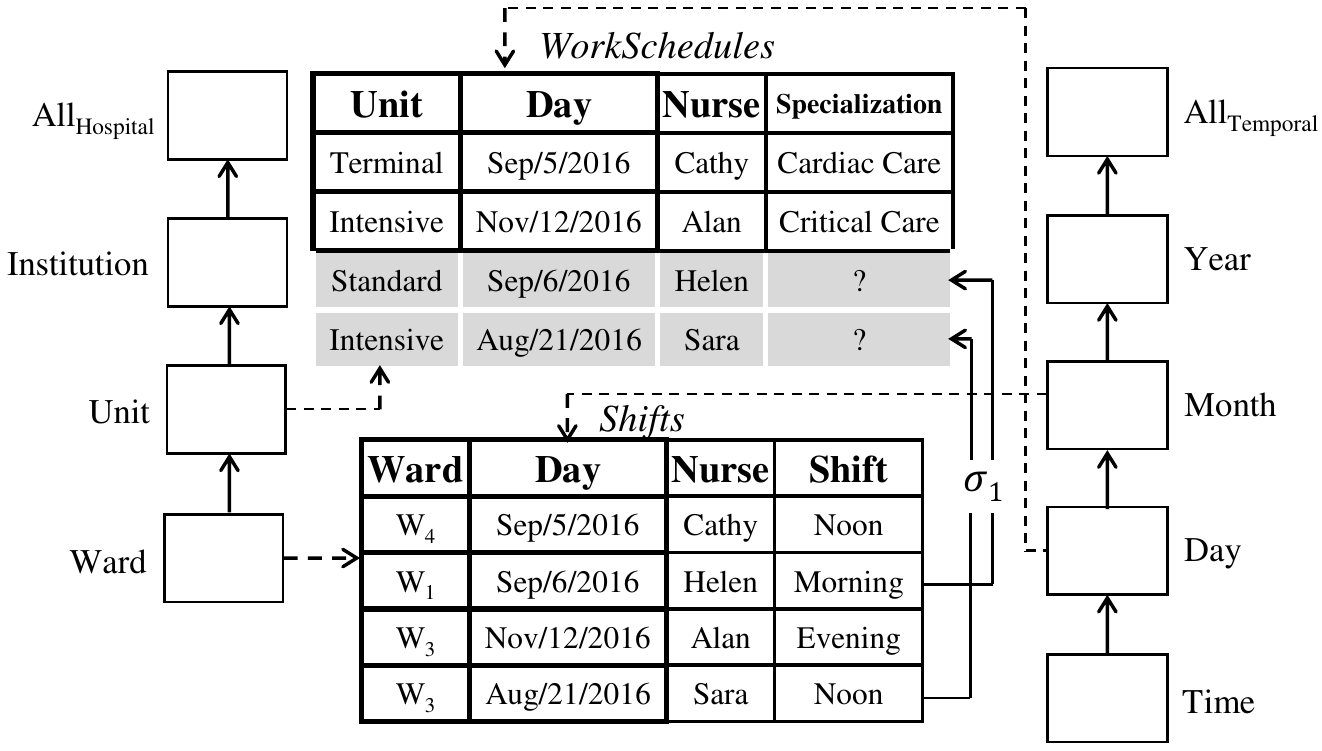}
 \caption{Dimensional data with categorical relations}\label{fig:omdm0} \vspace{-4mm}
\end{center}
\end{figure}

In it we find two dimensions, $\sf{Hospital}$, on the left-hand side, and $\sf{Temporal}$, on the right-hand side. For example, the $\sf{Hospital}$ dimension's instance is found in Figure \ref{fig:dimension}. In the middle of Figure \ref{fig:omdm0} we find {\em categorical relations} (shown as solid tables and initially excluding the two rows shaded
 in gray at the bottom of the top table). They  are associated to categories in the dimensions.

Now we have all the necessary information to discriminate between quality and non-quality entries in Table \ref{tab:temperatures}: Nurses appearing in it are associated to wards, as shown in table
{\em Shifts}; and the wards are associated to units, as shown in Figure \ref{fig:dimension}. Table {\em WorkSchedules} may be incomplete, and new -possibly virtual- entries can be produced for it, showing \nit{Helen} and \nit{Sara} working for the \nit{Standard} and \nit{Intensive} units, resp. (These correspond to the two (potential) extra, shaded  tuples in Figure~\ref{fig:omdm0}.) This is done by {\em upward navigation and data propagation} through the dimension hierarchy. At this point we are in position to take advantage of the guideline, inferring that {\it Alan} and {\it Sara} used thermometers of brand $B_1$, as expected by the physician.

\ignore{
\comlb{In Figures it would be better to use lower case for category members, as we do outside the figures.}
\commos{I applied it to the rest of the paper. I keep the letters, e.g. $H_1$ and $W_1$, as before. Let me know if you also wanted to make them small letters.}
}

As expected, in order to do upward navigation  and use the guideline, they have to be represented in our multidimensional contextual ontology. Accordingly, the latter contains,  in addition to the data in
Figure \ref{fig:omdm0}, the two rules, for upward data propagation and the guideline, resp.:
 \begin{eqnarray}
\hspace{-5mm}\sigma_1\!: \ \  \nit{Shifts}(w,d;n,s),\nit{WardUnit}(w,u) &\rightarrow& \exists t\;\nit{WorkSchedules}(u,d;n,t).\label{eq:s1intr}\\
\hspace{-3mm}\nit{WorkTimes}({\sf intensive},t;n,y) &\rightarrow& \nit{TakenWithTherm}(t,n,{\sf b1}).\label{frm:qp-intr}
\end{eqnarray}

Here, $\nit{WorkTimes}$ is a categorical relation linked to the Time category in the {\sf Temporal} dimension. It contains the schedules as in relation \nit{WorkSchedules}, but at the time of the day level, say ``14:30 on Feb/08, 2017", rather than the day level.

\ignore{
\comlb{This time vs. day issue looks a bit strange, at least here. Is it solved? Also, not a good idea to have "Time" to denote the dimension and one of its categories.}
\commos{I changed the relation name from WorkTimes to WorkTimes, and the attribute from Time to TimeOfDay to prevent confusion with the Time dimension name. If you agree, I will have to change the Time level in the figures as well.}
}

Rule (\ref{eq:s1intr}) tells that: \ {\em ``If a nurse has shifts in a ward on a specific day, he/she has a work schedule in the unit of that ward on the same day"}. Notice that the use of
(\ref{eq:s1intr}) introduces unknown, actually {\em null}, values in attribute \nit{Specialization}, which is due to the existential variable ranging over the attribute domain. Existential rules of this kind already make us depart from classic Datalog, taking us to \dpm.

 Also notice that in (\ref{eq:s1intr}) we are making use of the binary {\em dimensional predicate} \nit{WardUnit} that represents in the ontology the child-parent relation between members of the \nit{Ward} and \nit{Unit} categories.\footnote{\ In the
$\nit{WorkSchedules(Unit,Day;Nurse,Speciality)}$ predicate, attributes \nit{Unit} and \nit{Day} are called {\em categorical attributes}, because they take values from categories in dimension.
They are separated by a semi-colon ($;$) from the non-categorical \nit{Nurse} and \nit{Speciality}.}

Rule (\ref{eq:s1intr}) properly belong to a contextual, multidimensional, core ontology $\mc{O}^M$ in the sense that it describes properly dimensional information. Now, rule (\ref{frm:qp-intr}), the guideline, could also belong to $\mc{O}^M$, but it is less clear that it convey strictly dimensional information. Actually, in our case we intend to use it for data quality purposes (cf. Example
\ref{ex:quality} below), and as such we will place it in the quality-oriented ontology $\mc{O}^q$. In any case, the separation is always application dependent. However, under certain conditions on the contents of $\mc{O}^M$, we will be able to guarantee (in Section \ref{sec:complexity}) that the latter has good computational properties. \boxtheorem
\end{example}

The contextual ontology $\mc{O}^c$ can be used to support the specification and extraction of quality data,  as shown in Figure~\ref{fig:frm}.  A database instance $D$ for a relational schema $\schema=\{R_1,...,R_n\}$ is mapped into $\mc{O}^c$ for quality data specification and extraction. The ontology contains a copy,  $\schema'=\{R'_1,...,R'_n\}$, of schema $\schema$ with predicates that are nicknames  for those in $\schema$. The nickname predicates are directly populated with the data in the corresponding relations (tables) in $D$.

In addition to the multidimensional (MD) ontology, $\mc{O}^M$, the contextual ontology contains, in ontology $\mc{O}^q$, definitions of application-dependent {\em quality predicates} $\mc{P}$, those in $\Sigma^\mc{P}$. Together with application-dependent, not directly dimensional rules, e.g. capturing  guidelines as in Example \ref{ex:intr}, they  capture data quality concerns. \red{Figure~\ref{fig:frm} also shows $E$ as a possible {\em extra contextual database}, with schema $\mc{R}^E$, whose data could be used at the contextual level in combination with the data strictly associated to the multidimensional ontology (cf. Section~\ref{sec:fw} for more details).}

Data originally
obtained from $D$ is processed through the contextual ontology, producing, possibly virtual, extensions for copies, $R^q$, of the original predicates $R$ in $\schema$. Predicates $R^q \in \schema^q$ are the ``quality versions" of predicates $R \in \schema$.  The following example shows the gist. 

\begin{example} \label{ex:quality} (ex. \ref{ex:intr} cont.) \ $\nit{Temperatures}'$, the nickname  for predicate $\nit{Temperatures}$ in the original instance, is defined by the rule: \begin{equation}\nit{Temperatures}(t,p,v,n) \rightarrow \nit{Temperatures}'(t,p,v,n).\label{eq:nick}\end{equation}
Furthermore, $\mc{O}^q$ contains rule (\ref{frm:qp-intr}) as a definition of quality predicate $\nit{TakenWithTherm}$.
Now, $\nit{Temperatures}^q$, the quality-version of predicate $\nit{Temperatures}$, is defined by means of:

\vspace{-4mm}
\begin{equation}
\hspace{-4mm}\nit{Temperatures}'(t,p,v,n),\nit{TakenWithTherm}(t,n,{\sf b1}) \rightarrow \nit{Temperatures}^q(t,p,v,n). \label{frm:qv-intr}
\end{equation}

The extension of $\nit{Temperatures}^q$ can be computed, and is shown in \ Table~\ref{tab:temperaturesq}. It contains ``quality data"  from the original relation $\nit{Temperatures}$.
The second and the third tuples in $\nit{Temperatures}^q$ are obtained through the fact that {\it Sara} was in the intensive care unit on {\it Aug/21}, as reported by the last in-gray shaded tuple in {\it WorkSchedules} in Figure \ref{fig:omdm0}, which was created by upward data propagation with the dimensional ontology.

It is not mandatory to materialize relation $\nit{Temperatures}^q$. Actually, the doctor's query:
{\small \begin{equation}\!\!\!\mc{Q}(v)\!:\exists n\;\exists t\;(\nit{Temperatures}(t,{\sf tom\;waits},v,n) \ \wedge \ {\sf 11\!:\!45\mbox{-}aug\mbox{/}21\mbox{/}2016} \le t \le {\sf 12\!:\!15\mbox{-}aug\mbox{/}21\mbox{/}2016})\label{frm:q}\vspace{-4mm}\end{equation}}
\phantom{oo}

\noindent
can be answered by, (a) replacing \nit{Temperatures} by $\nit{Temperatures}^q$, (b) unfolding the definition of $\nit{Temperatures}^q$ in (\ref{frm:qv-intr}), obtaining a query in terms of  $\nit{TakenWithTherm}$ and
$\nit{Temperatures}'$; and (c) using (\ref{frm:qp-intr}) and (\ref{eq:nick}) to compute the answers through $\mc{O}^M$ and $D$. The quality answer is
the second tuple in Table \ref{tab:temperaturesq}. \red{(This procedure is described in general in Section \ref{sec:extrac}).}
\boxtheorem
\end{example}

Due to the simple ontological rules and the use of them in the example above, we obtain a single quality instance. In other cases, we may obtain several of them, and quality query answering amounts to doing certain query answering (\qa) on \dpm \ ontologies, in particular on the the MD ontologies. Query answering on \dpm \ ontologies has been investigated in the literature for different classes of \dpm \ programs. For some of them,
query answering is tractable and there are efficient algorithms. For others, the problem is know to be tractable, but still practical algorithms are needed. For some classes, the problem is known to be intractable. For this reason, it becomes important to characterize the kind of \dpm \ ontologies used for the \omd \ model.

The promising application of the \omd \ model that we investigate in this work is related to data quality concerns as pertaining to {\em the use and production of data} \cite{batini}. By this we mean that the available data are not bad or good a priori or {\em prima facie}, but their quality depends on how they were created or how they  will be used, and this information is obtained from a contextual ontology. This form of data quality has been mentioned in the literature. For example, in~\cite{wang} {\em contextual data quality dimensions} are described  as those quality dimensions that are relevant to the context of data usage. In \cite{herzog} and \cite{juran},  quality is characterized as ``fitness for use".

Our motivating example already shows the gist of our approach to this form of data quality: Nothing looks wrong with the data in Table \ref{tab:temperatures} (the data source), but in order to assess the quality of the source's data or to extract quality data from it, we need to provide additional data and knowledge that do not exist at the source; and they are both provided by the context. From this point of view, we are implicitly addressing a problem of {\em incomplete data}, one of the common data quality dimensions \cite{batini}.
However, this is not the form of explicit incompleteness that we face with null or missing values in a table \cite{libkin}. (Cf. Section \ref{sec:repairs} for an additional discussion on the data quality dimensions addressed by the \omd \ model.)

As we pointed out before (cf. Footnote \ref{ft:cqa}), our contextual approach can be used, depending on the elements we introduce in a contextual ontology, to address other data quality concerns, such as inconsistency, redundancy,\footnote{\ In the case of duplicate records in a data source, the context could contain an {\em answer set program}  or a Datalog program to enforce matching dependencies for entity resolution \cite{kr12}.} and
the more typical and direct form  of incompleteness, say obtaining from the context data values for  null or missing values in tables.

In this work we concentrate mostly on the \omd \ model by itself, but also on its combination and use with quality-oriented ontologies for quality QA. We do not go into data quality {\em assessment}, which is also an interesting subject.\footnote{\ The quality of $D$ can be measured in terms of how much $D$ departs from (its quality versions in) $\mc{D}^q$: \ $\nit{dist}(D,\mc{D}^q)$. Of course, different distance measures may be used for this purpose \cite{bertossi-brite,bertossi16}.} Next, we summarize the main contributions of this work.

\vspace{2mm}

\noindent (A) \ We propose and formalize the {\em Ontological Multidimensional Data Model} (\omd \ model), which is based on a relational extension via \dpm \ of the HM model for multidimensional data. \
The \omd \ allows for: (a) Categorical relations linked to dimension categories (at any level), which go beyond the bottom-level, numerical fact tables found in data warehouses. (b) Incomplete data (and complete data as usual). (c) A logical integration and simultaneous representation of dimensional data and metadata, the latter by means of semantic dimensional constrains and dimensional rules. (d) Dimensional navigation and data generation, both upwards and downwards (the examples above show only the upward case).

\vspace{1mm}
\noindent (B) \ We establish that, under natural assumptions that  \md \ ontologies belong to the class of {\em weakly-sticky}  (WS) \dpm \ programs~\cite{cali12}, for which conjunctive \qa \ is tractable (in data).
\
The class of \WS \ programs is an extension of {\em sticky} \dpm \cite{cali12} \ and {\em weakly-acyclic} programs \cite{fagin}. Actually, \WS \  \dpm \ is defined through restrictions on join variables occurring in {\em infinite-rank} positions, as introduced in \cite{fagin}.

In this work, we do not provide algorithms for (tractable) \qa \ on {\em weakly-sticky} \dpm \ programs. However, in \cite{milani16rr} a practical algorithm was proposed, together with a methodology for {\em magic-set- based} query optimization.

\vspace{1mm}
\noindent (C) \ We analyze the interaction between dimensional constraints and the dimensional rules, and their effect on \qa. Most importantly, the combination of constraints that are {\em equality-generating dependencies} (\egds) and the rules, which are {\em tuple-generating dependencies} (\tgds) \cite{cali03}, may lead to undecidability of QA. \ {\em Separability}~\cite{cali12} is a semantic condition on \egds \ and \tgds \ that guarantees the interaction between them does not harm the tractability of QA. \
    Separability is an application-dependent issue. However, we show that, under reasonable syntactic conditions on \egds \ in MD ontologies,  separability holds.

 \vspace{1mm}
\noindent (D) \ We propose a general ontology-based approach to contextual quality data specification and extraction. The methodology takes advantage of  a MD ontology and a process of  dimensional navigation and data generation that is triggered by queries about quality data. \ We show that under natural conditions the elements of the quality-oriented ontology $\mc{O}^q$, in form of additional \dpm rules and constraints, do not affect the good computational
    properties of the core MD ontology $\mc{O}^M$.


\vspace{2mm}
The closest work related to our \omd \ model can be found in the {\em dimensional relational algebra} proposed in \cite{\ignore{martinenghi-qa,martinenghi-er,}martinenghi-vldb}, which is subsumed by the \omd \ model \cite[chap. 4]{milaniThesis}. The contextual and dimensional data representation framework in \cite{bolchini-is} is also close to our \omd \ model in that it uses dimensions for modeling context. However, in their work dimensions are different from the dimensions in the \hm \ data model. Actually, they use the notion of {\em context dimension trees} (CDTs) for modeling multidimensional contexts. \red{Section~\ref{sec:related} includes more details on related work.}

This paper is structured as follows. Section~\ref{sec:background} contains a review of databases,  \dpm, and the \hm \ data model. Section~\ref{sec:omd}  formalizes the \omd \ data model. Section \ref{sec:complexity} analyzes the computational properties of the \omd \ model. Section \ref{sec:fw} extends the \omd \ model with additional contextual elements for specifying and extracting quality data, and show how to use the extension for this task. Section \ref{sec:cfw} discusses additional related work, draws some final conclusions, and includes a discussion of  possible extensions of the \omd \ model. \
This paper considerably extends results previously reported in \cite{milani15ruleml}.

\section{Background} \label{sec:background}

In this section, we briefly review relational databases and the multidimensional data model.

\subsection{Relational Databases}\label{sec:relational}

We always start with a  relational schema $\schema$ with two disjoint domains:  $\mc{C}$, with possibly infinitely many {\em constants}, and $\mc{N}$, of infinitely many {\em labeled nulls}. $\schema$ also contains predicates of fixed finite arities. If $P$ is an $n$-ary predicate (i.e. with $n$ arguments) and $1\leq i \leq n$, $P[i]$ denotes its $i$-th position. $\schema$ gives rise to a language $\mf{L}(\schema)$ of first-order (FO)  predicate logic with equality ($=$).  Variables are usually denoted with $x, y, z, ...$, and sequences thereof by $\bar{x}, ...$. Constants are usually denoted with $a, b, c, ...$; and nulls are denoted with $\zeta, \zeta_1, ...$. An {\em atom} is of the form $P(t_1, \ldots, t_n)$, with $P$ an $n$-ary predicate  and $t_1, \ldots, t_n$ {\em terms}, i.e. constants, nulls, or variables. The atom is {\em ground} (aka. a tuple) if it contains no variables. An {\em instance} $I$ for schema $\mc{R}$ is a possibly infinite set of ground atoms; \red{this set $I$ is also called {\em an extension for the schema}. In particular, the extension of a predicate $P$ in an instance $I$, denoted by $P(I)$, is the set of atoms in $I$ whose predicate is $P$.} A {\em database instance} is a {\em finite} instance that contains no  nulls. The {\em active domain} of an  instance $I$, denoted ${\it Adom}(I)$, is the set of \red{constants or nulls} that appear in atoms of $I$. Instances can be used as interpretation structures for language $\mf{L}(\schema)$.

\red{An instance $I$ may be {\em closed} or {\em incomplete} (a.k.a. {\em open} or {\em partial}). In the former case, one makes the meta-level assumption, the so-called {\em closed-world-assumption} (CWA) \cite{reiter,abiteboul}, that the only positive ground atoms that are true w.r.t. $I$ are those explicitly given as members of $I$. In the latter case, those explicit atoms may form only a proper subset of those positive atoms that could be true w.r.t. $I$.}\footnote{\ \red{In the most common scenario one starts with a (finite) open database instance $D$ that is combined with an ontology whose \tgds \ are used to create new tuples. This process may lead to an infinite instance $I$. Hence the distinction between database instances and instances.}}

 A {\em homomorphism} is a structure-preserving mapping, $h\!\!:\mc{C}\cup\mc{N} \!\rightarrow \!\mc{C}\cup\mc{N}$,  between two instances $I$ and $I'$ for  schema $\mc{R}$ such that: (a) $t \in \mc{C}$ implies $h(t)=t$, and (b) for every ground atom $P(\vectt{t})$: if $P(\vectt{t}) \in I$, then $P(h(\vectt{t})) \in I'$. 

A {\em conjunctive query} (\cq) is an \fo \ formula,  $\mc{Q}(\vectt{x})$, of the form:

\vspace{-3mm}
\begin{align}\label{frm:cq}\exists  \vectt{y}\;(P_1(\vectt{x}_1)\wedge \dots \wedge P_n(\vectt{x}_n)),\end{align}
\vspace{-3mm}

\noindent with \red{$P_i \in \mc{R}$,} and (distinct) free variables $\vectt{x} := \bigcup \vectt{x}_i \smallsetminus \vectt{y}$. If $\mc{Q}$ has $m$ (free) variables, for an instance $I$, $\vectt{t} \in (\mc{C} \cup \mc{N})^m$ \ is an {\em answer} to $\mc{Q}$ if $I \models \mc{Q}[\vectt{t}]$, meaning that  $Q[\vectt{t}]$ becomes true in $I$  when the variables in $\vectt{x}$ are componentwise replaced by the values in $\vectt{t}$. $\mc{Q}(I)$ denotes the set of answers to $\mc{Q}$ in $I$. $\mc{Q}$ is a {\em boolean conjunctive query} (\bcq) when $\vectt{x}$ is empty, and if it is {\em true} in $I$, in which case $\mc{Q}(I) := \{\nit{true}\}$. Otherwise, $\mc{Q}(I) = \emptyset$, and we say it is
{\em false}.

A {\em tuple-generating dependency} (\tgd), also called  a {\em rule}, is an implicitly universally quantified  sentence of $\mf{L}(\schema)$ of the form:

\vspace{-4mm}
\begin{align}\sigma\!: \ \ P_1(\vectt{x}_1), \ldots, P_n(\vectt{x}_n) \ \rightarrow \ \exists \vectt{y} \ P(\vectt{x},\vectt{y}),\label{frm:tgd}\end{align}
\vspace{-4mm}

\noindent  with \red{$P_i \in \mc{R}$, and} $\bar{x} \subseteq \bigcup_i \bar{x}_i$, and the dots in the antecedent standing for conjunctions. The variables in~$\vectt{y}$ (that could be empty) are the {\em existential variables}. We assume $\vectt{y}\cap \cup \vectt{x}_i=\emptyset$. With ${\it head}(\sigma)$ and ${\it body}(\sigma)$ we denote the atom in the consequent and the set of atoms in the antecedent of $\sigma$, respectively.

A {\em constraint} is an {\em equality-generating dependency} (\egd\;\m) or a {\em negative constraint} ({\em nc}), which are also sentences of $\mf{L}(\schema)$, respectively, of the forms:

\vspace{-3mm}
\begin{align}
P_1(\vectt{x}_1), \ldots, P_n(\vectt{x}_n) \ \rightarrow \  x=x',\label{frm:egd}\\
P_1(\vectt{x}_1), \ldots, P_n(\vectt{x}_n) \ \rightarrow \ \bot,\label{frm:nc}
\end{align}
\vspace{-3mm}

\noindent  with \red{$P_i \in \mc{R}$, and} $x,x' \in \bigcup_i \bar{x}_i$, and $\bot$ is a symbol that denotes the Boolean constant (propositional variable) that is always false. Satisfaction of constraints by an instance is  as in \fo \ logic. \red{In Section~\ref{sec:omd} we will use \ncs \ with negated body atoms (i.e. negative literals), in a limited manner. Their semantics is also as in \fo \ logic, i.e. the body cannot be made true in a consistent instance, for any data values for the variables in it.}

{\em Tgds}, \egds, and {\em ncs} are particular kinds of relational {\em integrity constraints} (ICs)~\cite{abiteboul}. In particular,  \egds \ include {\em key constraints} and {\em functional dependencies} (\fds).
ICs also include {\em inclusion dependencies} (\ideps): For an $n$-ary predicate $P$ and an $m$-ary predicate $S$, the \idep \ $P[j] \subseteq S[k]$, with $j \leq n, \ k \leq m$, means that -in \red{the extensions of $P$ and $S$ in an instance}- the values appearing in the
$j$th position (attribute) of $P$ must also appear in the $k$th position of $S$.

\red{Relational databases work under the CWA, i.e. ground atoms not belonging to a database instance are assumed to be false. As a consequence, an IC is true or false when checked for satisfaction on a (closed) database instance, never undetermined. However, as we will see below, if instances are allowed to be incomplete, i.e. with undetermined or missing ground atoms, ICs may not be false, but only undetermined in relation to their truth status. Actually, they can be used, by enforcing them, to generate new tuples for the (open) instance.}

\da \ is a declarative query language for relational databases that is based on the logic programming paradigm, and allows to define recursive views~\cite{abiteboul,ceri}. A \da \ program $\prg$ for schema $\schema$ is a finite set of non-existential rules, i.e. as in (\ref{frm:tgd}) but without $\exists$-variables. Some of the predicates in $\prg$ are {\em extensional}, i.e. they do not appear in rule heads, and their \red{{\em complete extensions}} are given by a \red{database instance} $D$ (for a subschema of $\schema$), that is  \red{called the program's {\em extensional database}}. \ignore{\red{A predicate $P$ has the complete extension in $D$ if for every model $I$ of $\Pi$, $P(I)=P(D)$.}} The program's {\em intentional} predicates \red{are those that are defined by the program by appearing in \tgds' heads. The program's extensional database $D$ may give to them only partial extensions (additional tuples for them may be computed by the application of the program's \tgds). However, without loss of generality, it is common with Datalog to make the assumption that intensional predicates do not have an explicit extension, i.e. explicit ground atoms in $D$. }

The {\em minimal-model semantics} of a \da \ program w.r.t. an extensional database instance $D$ is given by a fix-point semantics\red{~\cite{abiteboul}}: the extensions of the intentional predicates are obtained by, starting from $D$, iteratively enforcing the rules and creating tuples for the intentional predicates, i.e. whenever a ground (or instantiated) rule body becomes true in the extension obtained so far, but not the head, the corresponding ground head atom is added to the extension under computation. If the set of initial ground atoms is finite, the process reaches a fix-point after a finite number of steps. \red{The database instance obtained in this way turns out to be the unique minimal model of the Datalog program: it extends the extensional database $D$, makes all the rules true, and no proper subset  has the two previous properties. Notice that the constants in a minimal model of a Datalog program are those already appearing in the active domain of $D$ or in the program rules; no new data values of any kind are introduced.}

\red{One can pose a CQ to a Datalog program by evaluating it on the minimal model of the program, seen as a database instance. However, it is common to add the query to the program, and the  minimal model of the combined program gives us the set of answers to the query. In order to do this,} a  \cq \ as in~(\ref{frm:cq}) is expressed as a \da \ rule of the form\ignore{ \red{(this is an alternative for defining \cq s over a program)}}:

\vspace{-4mm}
\begin{align}P_1(\vectt{x}_1),...,P_n(\vectt{x}_n)\rightarrow \nit{ans}_\mc{Q}(\vectt{x}),\end{align}
\vspace{-4mm}

\noindent where $\nit{ans}_\mc{Q}(\cdot)$ \ignore{ \notin \mc{R}$} is an auxiliary, answer-collecting predicate.  The answers to query $\mc{Q}$ form the extension of  predicate $\nit{ans}_\mc{Q}(\cdot)$ in the minimal model \red{of the original program extended with the query rule.} When $\mc{Q}$  is a \bcq,  $\nit{ans}_{\mc{Q}}$ is a propositional atom; and $\mc{Q}$ is true in the undelying instance exactly when the atom $\nit{ans}_\mc{Q}$ belongs to the minimal model of the program.

\begin{example} \label{ex:datalog} A \da \ program $\prg$ containing the rules \ $P(x,y)~\rightarrow~R(x,y)$, \ and \
$P(x,y),R(y,z)$ \ $~\rightarrow~R(x,z)$ \
\ignore{
\vspace{-3mm}
\begin{align*}
P(x,y)~\rightarrow~R(x,y),\\
P(x,y),R(y,z)~\rightarrow~R(x,z),
\end{align*}
\vspace{-3mm} }
recursively defines, \red{on top of an extension for predicate $P$,} the intentional predicate $R$ as the transitive closure of  $P$. With $D=\{P(a,b),P(b,d)\}$ as the extensional database, the extension of $R$ can be computed by iteratively adding tuples enforcing the program rules, which results in the \red{ instance $I=\{P(a,b),P(b,d),R(a,b),R(b,d),R(a,d)\}$, which is the minimal model of the program}.

The CQ $\mc{Q}(x)\!: \ R(x,b) \wedge R(x,d)$ can be expressed by the rule \ $R(x,b), R(x,d) \ \rightarrow \ \nit{ans}_{\mc{Q}}(x)$. \red{The set of answers is the computed extension for $\nit{ans}_{\mc{Q}}(x)$ \red{on instance $D$}, namely $\{a\}$. Equivalently, the query rule can be added to the program, and the minimal model of the resulting program will contain the extension for the auxiliary predicate $\nit{ans}_{\mc{Q}}$: $I'=\{P(a,b),P(b,d),R(a,b),R(b,d),R(a,d),\nit{ans}_{\mc{Q}}(a)\}$.}
\boxtheorem
\end{example}


\subsection{Datalog$^\pm$}\label{sec:dpm}


\dpm \ is an extension of Datalog. The ``$+$" stands for the extension, and the ``$-$", for some syntactic restrictions on the program that guarantee some good computational properties. We will refer to some of those restrictions in Section \ref{sec:complexity}. Accordingly, until then we will consider \dplus \ programs.

A \dplus \ program may contain, in addition to (non-existential) Datalog rules, also existential rules rules of the form (\ref{frm:tgd}), and constraints of the forms (\ref{frm:egd}) and (\ref{frm:nc}). A \dplus \ program has an extensional database $D$. \red{In a \dplus \ program $\Pi$, unlike plain Datalog, predicates are not necessarily partitioned into extensional and intentional ones: any predicate may appear in the head of a rule. As a consequence, some predicates may have {\em partial extensions} in the extensional database $D$, and their extensions will  be completed via rule enforcements. \ignore{A predicate $P$ has a partial extension in $D$ if there is a model $I$ of $\Pi$ in which $P(I)\neq P(D)$.}}

The semantics of a \dplus \ program $\Pi$ with \red{an extensional database instance $D$} is model-theoretic, and given by the class $\nit{Mod}(\Pi,\red{D})$ of all, \red{possibly infinite}, instances $\red{I}$ for the program's schema (in particular, with domain contained in $\mc{C} \cup \mc{N}$) that extend $D$ and make $\Pi$ true. \red{Notice that, and in contrast to Datalog,  the combination of the ``open-world assumption" and the use of $\exists$-variables in rule heads makes us consider possibly infinite models for a \dplus program, actually  with domains that go beyond the active domain of the extensional database.}

\blue{If a Datalog$^\pm$  program $\Pi$ has an extensional database instance $D$, a set $\Pi^R$ of \tgds, and a set $\Pi^C$ of constraints  of the forms (\ref{frm:egd}) or (\ref{frm:nc}), then $\Pi$ is {\em consistent} if
 $\nit{Mod}(\Pi,\red{D})$ is non-empty, i.e. the program has at least one model.}


\red{Given a \dplus program $\Pi$ with database instance $D$ and  an $n$-ary CQ $\mc{Q}(\bar{x})$, $\bar{t} \in (\mc{C} \cup \mc{N})^n$ is an answer \red{w.r.t. $\Pi$} iff $\red{I} \models \mc{Q}[\bar{t}]$ for every $\red{I} \in \nit{Mod}(\Pi,\red{D})$, which is equivalent to $\Pi \cup D \models \mc{Q}[\bar{t}]$.} Accordingly, this is {\em certain answer} semantics. \blue{In particular, a BCQ $\mc{Q}$ is true w.r.t. $\Pi$ if it is true in every  $\red{I} \in \nit{Mod}(\Pi,\red{D})$.} \blue{In the rest of this paper, unless otherwise stated, CQs are BCQs, and  CQA is the problem of deciding if a BCQ is true w.r.t. a given program.\footnote{\ \red{For \dplus \ programs, \cq \ answering, i.e. checking if a tuple is an answer to a CQ query,  can be reduced to \bcq \ answering as shown in~\cite{cali13}, and they have the same data complexity.}}}

Without any syntactic restrictions on the program, and even for programs without constraints, {\em conjunctive query answering} (CQA) may be undecidable \cite{beeri-icalp}.
CQA appeals to all possible models of the program. However, the {\em chase procedure} \cite{maier} can be used to generate a single, \red{possibly infinite}, instance that represents the class $\nit{Mod}(\Pi,\red{D})$ for this purpose. We show it by means of an example.

\begin{example} \label{ex:chaseNEW} Consider a program $\Pi$ with the set of rules \ $\sigma\!:  R(x,y) \ \rightarrow \ \exists z \ R(y,z)$, \ and \
$\sigma'\!:  R(x,y),R(y,z) \ \rightarrow \ S(x,y,z)$,
\ignore{\vspace{-4mm}
\begin{align*}
\sigma&:&\hspace{-2cm}R(x,y) \! ~&\rightarrow~ \! \exists z \ R(y,z),\\
\sigma'&:&\hspace{-2cm}R(x,y),R(y,z) \! ~&\rightarrow~ \! S(x,y,z),
\end{align*}  }
and an extensional database instance $\red{D}=\{R(a,b)\}$, providing an incomplete extension for the program's schema.
With the instance $I_0:=\red{D}$, the pair $(\sigma, \theta_1)$, with (value) {\em assignment} (for variables) $\theta_1\!: \ x\mapsto a, y\mapsto b$, is {\em applicable}: $\theta_1(\nit{body}(\sigma))=\{R(a,b)\} \subseteq I_0$. The chase {\em enforces} $\sigma$ by inserting a new tuple $R(b,\zeta_1)$ into $I_0$ ($\zeta_1$ is a fresh null, i.e. not in $I_0$), resulting in instance $I_1$. 

Now, $(\sigma',\theta_2)$, with $\theta_2\!: \ x\mapsto a, y\mapsto b, z\mapsto \zeta_1$, is applicable, because $\theta_2(\nit{body}(\sigma'))$ $=$ $\{R(a,b),R(b,$ \ $\zeta_1)\} \subseteq I_1$. The chase adds $S(a,b,\zeta_1)$ into $I_1$, resulting in $I_2$.
The chase continues, without stopping, creating an infinite instance, usually called {\em the chase} (instance): ${\it chase}(\Pi,\red{D}) = \{R(a,b), R(b,\zeta_1), S(a,b,\zeta_1),$ \ $ R(\zeta_1,\zeta_2), R(\zeta_2,\zeta_3), S(b,\zeta_1,\zeta_2), \ldots\}$.
\boxtheorem
\end{example}
For some programs an instance obtained through the chase may be finite. Different orders of chase steps may result in different sequences and instances. However, it is possible to define a {\em canonical chase procedure} that determines  a  canonical sequence of chase steps, and consequently, a canonical chase instance~\cite{cali13}. 



Given a program $\Pi$ and extensional database $\red{D}$, its chase (instance) is a \emph{universal model}~\cite{fagin}: For every  $\red{I} \in \nit{Mod}(\Pi,\red{D})$, there is a homomorphism from the chase into $\red{I}$. For this reason, the (certain) answers to a \cq \ $\mc{Q}$ under $\Pi$ and $\red{D}$ can be computed by evaluating $\mc{Q}$ over the chase instance (and discarding the answers containing nulls)~\cite{fagin}. \red{Universal models
 of Datalog programs are finite and coincide with the minimal models. However, the universal model of a \dplus \ program may be infinite, this is when the chase procedure does not stop, as  shown in Example~\ref{ex:chaseNEW}. This is a consequence  of the \owa \ underlying \dplus \ programs and the presence of existential variables.}

\ignore{
++++

\blue{For a program $\Pi$ with a set $\Pi^C$ of \ncs, the latter must hold in all the models in $\nit{Mod}(\Pi,D)$ (if models exist). According to~\cite[Theorem~11]{cali12jws}, \cq \ answering under $\Pi$ can be reduced to CQA under $\Pi \smallsetminus \Pi^C$. This is done by, }

\begin{itemize}
  \item [(a)] \red{Checking if $\Pi \smallsetminus \Pi^C$ entails the \ncs, which done by posing to $\Pi \smallsetminus \Pi^C$ the \bcq s obtained from the bodies of the constraints in $\Pi^C$. They should all be (certainly) {\em false}; otherwise  If $\Pi'$ does not satisfy the constraints, $\Pi$ is inconsistent, and thus \qa \ is trivial since every query is entailed.}
  \item [(b)] \red{If the \ncs \ are satisfied by $\Pi'$, for every \bcq \ $\mc{Q}$, $\Pi \models \mc{Q}$ if and only if $\Pi' \models \mc{Q}$, i.e. we can answer queries over $\Pi'$, ignoring the \ncs.}
\end{itemize}

+++  }

\blue{If a program $\Pi$ consists of a set of \tgds \ $\Pi^R$ and a set of \ncs \ $\Pi^C$, then CQA amounts to deciding if $D \cup \Pi^R \cup \Pi^C \models \mc{Q}$. However, this is equivalent  to deciding if:
(a) $D \cup \Pi^R \models \mc{Q}$, \ or \ (b) for some $\eta \in \Pi^C$, $D \cup \Pi^R \models \mc{Q}_\eta$, where $\mc{Q}_\eta$ is the BCQ obtained as the existential closure of the body of $\eta$ \cite[theo. 6.1]{cali12}.\ignore{\footnote{\ We haven't found an explicit proof of this claim in the literature. So, we do it here. Assume (b) does not hold, then $\nit{Mod}(\Pi^R \cup \Pi^C,D) \neq \emptyset$. We have to show that $D \cup \Pi^R \cup \Pi^C \models \mc{Q}$ iff $D \cup \Pi^R \models \mc{Q}$. From right to left is obvious. Now, from left to right, assume $D \cup \Pi^R \cup \Pi^C \models \mc{Q}$, $I \in \nit{Mod}(\Pi^R,D)$. We have to show that $I \models \mc{Q}$. Let $I^\nit{ch}$ be $\nit{chase}(\Pi^R,D)$, for which $I^\nit{ch} \models \Pi^C$ holds (otherwise, due to the universality of the chase and preservation of CQA under homomorphisms,  $\Pi^R \cup \Pi^C \cup D$ would be inconsistent). Then, $I^\nit{ch} \in \nit{Mod}(\Pi^R \cup \Pi^C,D)$, and then, by hypothesis, $I^\nit{ch} \models \mc{Q}$. As a consequence, $I \models \mc{Q}$.}}
\ignore{\ \footnote{That is,  the BCQ $\overline{\exists}(P_1(\vectt{x}_1) \wedge \cdots \wedge P_n(\vectt{x}_n))$, the existential closure of the formula obtained from  the body of (\ref{frm:nc}).}} In the latter case, $D \cup \Pi$ is inconsistent, and $\mc{Q}$ becomes trivially true. This shows that CQA evaluation under \ncs \ can be reduced to the same problem without \ncs, and the data complexity of CQA does not change. Furthermore, \ncs \ may have an effect on CQA only if they are mutually inconsistent with the rest of the program, in which case every BCQ becomes trivially true. }


\ignore{they are expected to be satisfied by the chase obtained applying its \tgds, i.e. each BCQ $\exists\vectt{x}_1 \cdots \vectt{x}_n(P_1(\vectt{x}_1)\wedge \cdots \wedge P_n(\vectt{x}_n))$ obtained from the body of (\ref{frm:nc}) has to be false in the chase, which is equivalent to the BCQ being false in all the models of the program without constraints. This happens exactly when the program with constraints is consistent.

\blue{As expected from the fact that \ncs \ do not participate in the chase, they can have an effect on CQ answers only if they are mutually inconsistent with the \tgds \ (in which case, every query becomes trivially true).
If the \ncs \ are consistent with the rest of the program, they cannot change the status of a possible query answer w.r.t. the program without them.} }



\blue{If $\Pi$ has \egds, they are  expected to be satisfied by a modified (canonical) chase \cite{cali13} that also enforces the \egds. This enforcement may become impossible at some point, in which case we say the {\em chase fails} (cf. Example \ref{ex:non-separableNEW}).} \
\red{Notice that consistency of a \dplus \ program is defined independently from the chase procedure, but can be characterized in terms of the chase. Furthermore, if the canonical chase procedure terminates (finitely or by failure) the result can be used to decide if the program is consistent.} \blue{The next example shows that \egds \ may have an effect on CQA even with consistent programs.}

\begin{example} \label{ex:non-separableNEW} Consider a program $\prg$ with $\red{D}=\{R(a,b)\}$ with two rules and an \egd:

\vspace{-4mm}
\begin{align}
R(x,y) ~&\rightarrow~ \exists z\; \exists w\;S(y,z,w).\label{frm:sep1NEW}\\
S(x,y,y) ~&\rightarrow~ P(x,y).\label{frm:sep2NEW}\\
S(x,y,z) ~&\rightarrow~ y=z.\label{frm:sep3NEW}
\end{align}
The chase of $\prg$ first applies (\ref{frm:sep1NEW}) and results in $I_1=\{R(a,b),S(b,\zeta_1,\zeta_2)\}$. There are no more tgd/assignment applicable pairs. But, if we enforce the \egd~(\ref{frm:sep3NEW}), equating $\zeta_1$ and $\zeta_2$, we obtain $I_2=\{R(a,b),S(b,\zeta_1,\zeta_1)\}$.  Now, (\ref{frm:sep2NEW}) and $\theta': x\mapsto b, y\mapsto \zeta_1$ are applicable, so we add $P(b,\zeta_1)$ to $I_2$, generating $I_3=\{R(a,b),S(b,\zeta_1,\zeta_1),P(b,\zeta_1)\}$.
The chase terminates (no applicable \tgds \ or \egds), obtaining $\nit{chase}(\prg,\red{D}) = I_3$.

Notice that the program consisting only of (\ref{frm:sep1NEW}) and (\ref{frm:sep2NEW}) produces $I_1$ as the chase, which makes the BCQ $\exists x \exists y\;P(x,y)$ evaluate to {\em false}. With the program also including the \egd \ (\ref{frm:sep3NEW}) the answer is now {\em true}, \blue{which shows  that consistent \egds \ may affect \cq \ answers. This is in line with the use of a modified chase procedure that applies them along with the \tgds.}

Now consider  program $\prg'$ that is $\prg$ with the extra rule \ $R(x,y) ~\rightarrow~ \exists z\;S(z,x,y)$, which enforced on $I_3$ results in $I_4=\{R(a,b),S(b,\zeta_1,\zeta_1),P(b,\zeta_1),S(\zeta_3,a,b)\}$.
Now (\ref{frm:sep3NEW}) is applied, which creates a chase failure as it tries to equate constants $a$ and $b$. This is case where the set of \tgds \ and the \egd \ are mutually inconsistent. \boxtheorem\end{example}

\ignore{\vspace{-4mm}
\bl{\begin{align}
R(x,y) ~&\rightarrow~ \exists z\;S(z,x,y).\label{frm:sep4NEW}
\end{align}
The chase of $\prg'$ additionally applies (\ref{frm:sep4NEW}) and}

\comlb{Can the example be modified to produce also a  failing chase?}
\commos{The additional part highlighted in blue makes for chase failure.}
}

\subsection{The Hurtado-Mendelzon Multidimensional Data Model} \label{sec:hm}

According to the {\em Hurtado-Mendelzon  multidimensional data model} (in short, the \hm model)~\cite{hurtado-pods}, a \nit{dimension schema}, $\mscr{H}=\langle \mscr{K},\nearrow \rangle$, consists of a finite set $\mscr{K}$ of {\em categories}, and an irreflexive, binary relation $\nearrow$, called the {\em child-parent relation}, between categories (the first category is a child and the second category is a parent).  The transitive and reflexive closure of $\nearrow$ is denoted by $\nearrow^{*}$, and is a partial order (a lattice) with a {\em top category}, {\em All}, which is reachable from every other category: $K\!\nearrow^*\!\nit{All}$, for every category $K \in \mscr{K}$. There is a unique {\em base category}, $K^b$, that has no children. There are no {\em ``shortcuts"}, i.e. if $K \nearrow K'$, there is no category $K''$, different from $K$ and $K'$, with $K \nearrow^* K''$, $K'' \nearrow^* K'$.

A {\em dimension instance} for schema $\mscr{H}$ is a structure $\mscr{L}=\langle\;\!\mc{U},<,m\;\!\rangle$, where $\mc{U}$ is a non-empty, finite set of data values called {\em members}, $<$ is an irreflexive binary relation between members, also called a {\em child-parent relation} (the first member is a child and the second member is a parent),\footnote{\ There are two child-parent relations in a dimension: $\nearrow$, between categories; and $<$, between category members.} and $m\!: \mc{U} \rightarrow \mscr{K}$ is the total {\em membership function}. Relation $<$ parallels (is consistent with) relation $\nearrow$: $e < e'$ implies $m(e) \nearrow m(e')$. The statement $m(e)=K$ is also expressed as $e \in K$. $<^{*}$ is the transitive and reflexive closure of $<$, and is a partial order over the members. There is a unique member \nit{all}, the only member of \nit{All}, which is reachable via $<^*$ from any other member: $e <^{*} \nit{all}$, for every member $e$. A child member in $<$ has only one parent member in the same category: for members $e$, $e_1$, and $e_2$, if $e < e_1$, $e < e_2$ and $e_1,e_2$ are in the same category (i.e. $m(e_1)=m(e_2)$), then $e_1=e_2$. $<^{*}$ is used to define the {\em roll-up} {\em relations} for any pair of distinct categories  $K \nearrow^* K'$: \ $L_{K}^{K'}(\mscr{L})=\{(e,e')~|~e \in K, \ e' \in K' \mbox{ and } e <^* e'\}$.

\begin{figure}[ht]
\vspace{-4mm}
\begin{center}
\includegraphics[width=6cm]{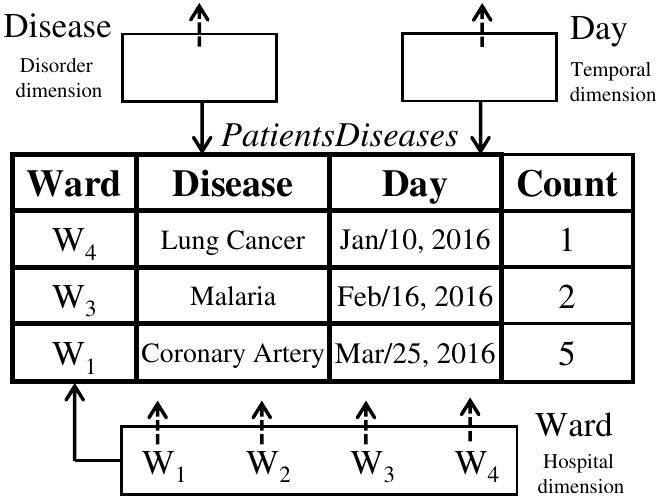}
 \caption{An \hm \ model}\label{fig:hm-dimension}
 \end{center}
 \vspace{-5mm}
\end{figure}




\begin{example}\label{ex:dim} \red{The \hm \ model in Figure~\ref{fig:hm-dimension} includes three dimension instances: {\sf Temporal} and {\sf Disorder} (at the top) and {\sf Hospital} (at the bottom). They are not shown in full detail, but only their base categories \nit{Day}, \nit{Disease}, and \nit{Ward}, resp. We will use four different dimensions in our running example, the three just mentioned and also  {\sf Instrument} (cf. Example \ref{ex:ont}).} For the {\sf Hospital} dimension,  shown in detail in Figure~\ref{fig:dimension}, $\mscr{K}=\{\nit{Ward}, \nit{Unit}, \nit{Institution}, \nit{All}_\nit{Hospital}\}$, with base category \nit{Ward} and top category $\nit{All}_\nit{Hospital}$. The child-parent relation $\nearrow$ contains $(\nit{Institution}$ \ $,\nit{All}_\nit{Hospital})$, $(\nit{Unit},\nit{Institution})$, and $(\nit{Ward},\nit{Unit})$. The category of each member is specified by $m$, e.g. $m({\sf H}_1)=\nit{Institution}$. The child-parent relation $<$ between  members contains $({\sf W}_1, {\sf standard})$, $({\sf W}_2, {\sf standard})$, $({\sf W}_3, {\sf intensive})$, $({\sf W}_4, {\sf terminal})$, $({\sf standard}, {\sf H}_1)$, $({\sf intensive}, {\sf H}_1)$, $({\sf terminal}, {\sf H}_2)$, $({\sf H}_1,$ \ ${\sf all}_\nit{Hospital})$, and $({\sf H}_2,$ \ ${\sf all}_\nit{Hospital})$. Finally, $L_\nit{Ward}^\nit{Institution}$ is one of the roll-up relations and contains $({\sf W}_1,{\sf H}_1)$, $({\sf W}_2,{\sf H}_1)$, $({\sf W}_3,{\sf H}_1)$, and $({\sf W}_4,{\sf H}_2)$.\boxtheorem
\end{example}

In the rest of this section we show how to represent an HM model in relational terms. This representation will be used in the rest of this paper, in particular to extend the \hm \ model. We introduce a relational dimension schema $\mc{H}=\mc{K}\cup \mc{L}$, where $\mc{K}$ is a set of unary {\em category predicates}, and $\mc{L}$ (for ``lattice") is a set of binary {\em child-parent predicates}, with the first attribute as the child and the second as the  parent. The data domain of the schema is $\mc{U}$ (the set of category members).
Accordingly, a dimension instance is a database instance $D^\mc{H}$ for $\mc{H}$ that gives extensions to predicates in $\mc{H}$. The extensions of the category predicates form a partition of $\mc{U}$.

 In particular, for each category $K \in \mscr{K}$ there is a category predicate $K(\cdot) \in \mc{K}$, and the extension of the predicate contains the members of the category. Also, for every pair of categories $K$, $K'$ with $K \nearrow K'$, there is a corresponding child-parent predicate, say $K\!K'(\cdot,\cdot)$, in $\mc{L}$, whose extension contains the child-parent, $<$-relationships between members of $K$ and $K'$. In other words, each child-parent predicate in $\mc{L}$ stands for a roll-up relation between two  categories in child-parent relationship.

\vspace{-2mm}
\begin{example} \label{ex:hm} (ex~\ref{ex:dim} cont.) In the relational representation of the {\sf Hospital} dimension \red{(cf. Figure~\ref{fig:dimension})}, schema $\mc{K}$ contains unary predicates $\nit{Ward}(\cdot)$, $\nit{Unit}(\cdot)$, $\nit{Institution}(\cdot)$  and $\nit{All}_\nit{Hospital}(\cdot)$. The instance $D^\mc{H}$ gives them the  extensions:  $\nit{Ward} = \{{\sf W}_1,{\sf W}_2,{\sf W}_3,{\sf W}_4\}$, $\nit{Unit}$ $=$ $\{{\sf standard},{\sf intensive},{\sf terminal}\}$, $\nit{Institution}$ $=$ $\{{\sf H}_1,{\sf H}_2\}$  and $\nit{All}_\nit{Hospital}$ $=$ $\{{\sf all}_\nit{Hospital}\}$. $\mc{L}$ contains binary predicates: $\nit{WardUnit}(\cdot,\cdot)$, $\nit{UnitInstitution}(\cdot,\cdot)$, and $\nit{Institution}\nit{All}_\nit{Hospital}(\cdot,\cdot)$, with the following extensions: \   $\nit{WardUnit}=\{({\sf W}_1,$ \ $ {\sf standard}),$ $({\sf W}_2, {\sf standard}),$ $({\sf W}_3, {\sf intensive}),$
  $({\sf W}_4, {\sf terminal})\}$,   $\nit{UnitInstitution} = \{({\sf standard},{\sf H}_1),$ \\ $({\sf intensive}, {\sf H}_1),$ $({\sf terminal}, {\sf H}_2)\}$, and $\nit{InstitutionAll}_\nit{Hospital}$ $=$ $\{({\sf H}_1,{\sf all}_\nit{Hospital}),$ $ ({\sf H}_2,{\sf all}_\nit{Hospital})\}$.
\boxtheorem\end{example}
\vspace{-2mm}

In order to recover the hierarchy of a dimension in its relational representation, we have to impose some ICs. First, {\em inclusion dependencies} (\ideps) associate the child-parent predicates to the category predicates. For example, the following \ideps:  $\nit{WardUnit}[1] \subseteq \nit{Ward}[1]$, and
$\nit{WardUnit}[2] \subseteq \nit{Unit}[1]$. We need key constraints for the child-parent predicates, with the first attribute (child) as the {\em key}. For example, $\nit{WardUnit}[1]$ is the key attribute for $\nit{WardUnit}(\cdot,\cdot)$, which can be represented as the \egd: \
$\nit{WardUnit}(x,y), \nit{WardUnit}(x,z) \rightarrow y = z$.

\ignore{We can have multiple dimensions reflected with disjoint relational dimensional schemas, one for each dimension. They can be put together into a single multidimensional schema that is the union of the individual ones. In particular, there are now top and base categories predicates in $\mc{K}$, for each dimension. }

Assume $\mc{H}$ is the relational schema with multiple dimensions. A {\em fact-table schema} over $\mc{H}$ is a predicate $T(C_1,...,C_n,M)$, where $C_1,...,C_n$ are attributes with domains $\mc{U}_i$ for subdimensions $\mc{H}_i$, and $M$ is the {\em measure} attribute with a numerical domain. Attribute $C_i$ is associated with base-category predicate $K^b_i(\cdot) \in \mc{K}_i$ through the \idep: \ $T[i] \subseteq K^b_i[1]$. Additionally, $\{C_1,...,C_n\}$ is a key for $T$, i.e. each point in the base multidimensional space is mapped to at most one measurement. A {\em fact-table} provides an extension (or instance) for $T$. For example, \red{in the center of Figure~\ref{fig:hm-dimension}, the fact table $\nit{PatientsDiseases}$ is linked to the base categories of the three participating dimensions through its attributes \nit{Ward}, \nit{Disease}, and \nit{ Day}, upon which its measure attribute \nit{Count} functionally depends.}

This multidimensional representation enables aggregation of numerical data at different levels of granularity, i.e. at different levels of the hierarchies of categories. The roll-up relations can be used for aggregation.

\section{The Ontological Multidimensional  Data Model}\label{sec:omd}

In this section, we present the \omd \ model as an ontological, \dplus-based extension of the \hm \ model.  In this section we will be referring to the working example from Section~\ref{sec:intr}, extending it along the way when necessary to illustrate elements of the \omd \ model.




An \omd \ model has a  {\em database schema}  $\schema^\mc{M}=\mc{H} \cup \mc{R}^c$, where $\mc{H}$ is a relational schema with multiple dimensions, with sets $\mc{K}$ of unary category predicates,  and sets $\mc{L}$ of binary, child-parent predicates (cf. Section~\ref{sec:hm}); and $\mc{R}^c$ is a set of {\em categorical predicates}, whose categorical relations can be seen as extensions of  the fact-tables in the \hm \ model.

Attributes of categorical predicates are either {\em categorical}, whose values are members of dimension categories, or {\em non-categorical}, taking values from arbitrary domains. Categorical predicate are represented in the form  $R(C_1,\ldots,C_m;N_1, \ldots, N_n)$, with categorical attributes (the $C_i$) all before the semi-colon (``;''), and non-categorical attributes (the $N_i$) all after it.

The extensional data, i.e the instance for the schema $\schema^\mc{M}$, is $I^\mc{M}=D^\mc{H}\cup I^c$, where $D^\mc{H}$ is a complete database instance for subschema $\mc{H}$ containing the dimensional predicates (i.e. category and child-parent predicates); and  sub-instance $I^c$ contains possibly partial, incomplete extensions for the categorical predicates, i.e. those in $\mc{R}^c$.

Every schema $\schema^\mc{M}=\mc{H} \cup \mc{R}^c$ for an \omd \ model comes with some basic, application-independent semantic constraints. We list them next, represented as ICs.

\vspace{2mm}
\noindent {\bf 1.} \  Dimensional child-parent predicates must take their values from categories. Accordingly, if child-parent predicate $P \in \mc{L}$ is associated to category predicates $K,K' \in \mc{K}$, in this order, we
introduce \ideps \ $P[1] \subseteq K[1]$ and $P[2] \subseteq K'[1]$), as \ncs:

\vspace{-4mm}
\begin{align}
P(x,x'), \lnot K(x) \rightarrow \bot, \ \ \ \ \mbox{ and } \ \ \ \  
P(x,x'), \lnot K'(x') \rightarrow \bot.\label{frm:refh2}
\end{align}
We do not represent them as the \tgds \ $P(x,x') \rightarrow K(x)$, etc., because we reserve the use of \tgds \ for predicates (in their right-hand sides) that may be incomplete. This is not the case for $K$ or $K'$, which have complete extensions in every instance. \red{For this same reason, as mentioned right after introducing \ncs \ in (\ref{frm:egd}), we use here \ncs \ with negative literals: they are harmless in the sense that they are checked against complete extensions for predicates that do not appear in rule heads. Then, this form of negation is the simplest case of {\em stratified} negation \cite{abiteboul}.\footnote{\ \blue{\dplus with stratified negation, i.e. that is not intertwined with recursion, is considered in \cite{cali13}.}}}
Checking any of these constraints amounts to posing a non-conjunctive query to the instance at hand (we retake this issue in Section \ref{sec:withCons}).

\vspace{2mm}
\noindent {\bf 2.} \ Key constraints on dimensional child-parent predicates $P \in \mc{K}$, as \egds:
\begin{align}
P(x,x_1),P(x,x_2) ~\rightarrow~ x_1=x_2.\label{frm:key}
\end{align}

\noindent {\bf 3.} \
The connections between categorical attributes and the category predicates are specified by means of  \ideps \ represented as \ncs. More precisely, for the $i$th categorical position of predicate $R$ taking values in category $K$, the \idep \ $R[i] \subseteq K[1]$ is represented by:
\begin{align}
R(\bar{x};\bar{y}), \lnot K(x) ~\rightarrow~ \bot,\label{frm:referential}
\end{align}
where $x$ is the $i$th variable in the list $\bar{x}$.

\begin{figure}[ht]
\begin{center}
\vspace{-1mm}
 \includegraphics[width=11cm]{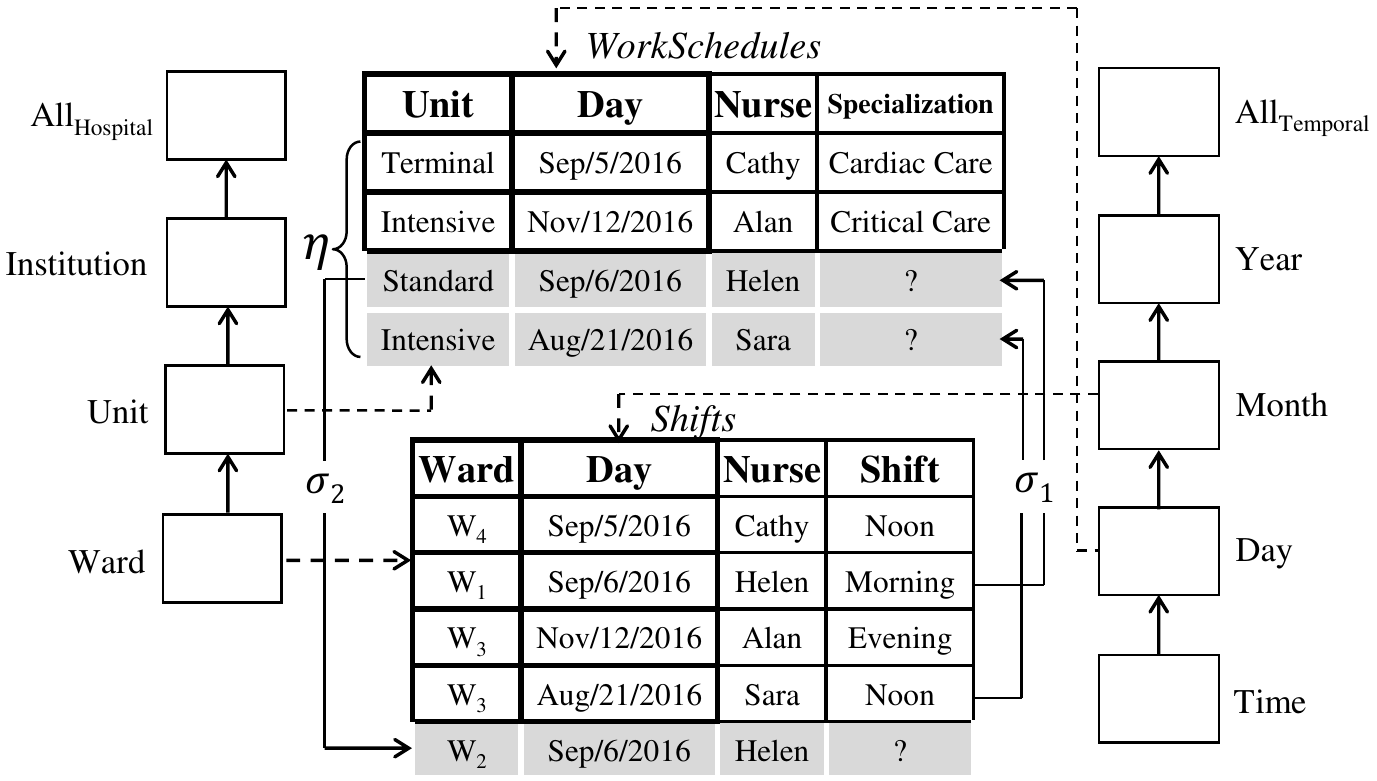}
 \caption{An \omd \ model with categorical relations, dimensional rules, and constraints}\label{fig:omdm} \vspace{-5mm}
\end{center}
\end{figure}

\begin{example}\label{ex:md-relational} \bl{(ex.~\ref{ex:intr} cont.)} The categorical attributes {\it Unit} and {\it Day} of categorical predicate $\nit{WorkSchedules}(\!\nit{Unit},\!\nit{Day};\!\nit{Nurse}\!,\nit{Speciality})$ in $\mc{R}^c$ \ignore{and $\nit{Shifts}(\!\nit{Ward},\!\nit{Day};\!\nit{Nurse},\!\nit{Shift})$ in Figure~\ref{fig:omdm} are categorical relations. In \nit{WorkSchedules}, {\it Unit} and {\it Day} are categorical attributes,} are connected to the {\sf Hospital} and {\sf Temporal} dimensions, resp., which is captured by the  \ideps \ $\nit{WorkSchedules}[1]\subseteq \nit{Unit}[1]$, and $\nit{WorkSchedules}[2]\subseteq \nit{Day}[1]$. The former is written in \dplus \ as in (\ref{frm:referential}):

\vspace{-4mm}
\begin{align}
{\it WorkSchedules(u,d;n,t)},\lnot {\it Unit}(u)  ~\rightarrow&~ \bot. \label{frm:ref}
\end{align}



For the {\sf Hospital} dimension,  one of the two \ideps \ for the child-parent predicate $\nit{WardUnit}$ is $\nit{WardUnit}[2] \subseteq \nit{Unit}[1]$, which  is expressed by an \nc \ of the form (\ref{frm:refh2}):

\vspace{-4mm}
\begin{align*}
{\it WardUnit(w,u)},\lnot {\it Unit}(u)  ~\rightarrow&~ \bot.
\end{align*}
\vspace{-4mm}

\noindent The key constraint of \nit{WardUnit} is captured by an \egd \ of the form (\ref{frm:key}):

\vspace{-4mm}
\begin{align*}
\hspace*{3.3cm}{\it WardUnit(w,u)},{\it WardUnit(w,u')}  ~\rightarrow&~ \ u=u'. \hspace{3.3cm} \boxtheorem
\end{align*}

\end{example}

The \omd \ model allows us to build  {\em multidimensional ontologies}, $\mc{O}^\mc{M}$. Each of them, in addition to an instance  $I^\mc{M}$ for a schema $\mc{R}^\mc{M}$, includes a set $\Omega^\mc{M}$ of {\em basic constraints} as in {\bf 1.}-{\bf 3.} above, a  set $\Sigma^\mc{M}$ of {\em dimensional rules}, and a set $\kappa^\mc{M}$ of {\em dimensional constraints}. All these rules and constraints are expressed in the \dplus \ language associated to schema $\schema^\mc{M}$. Below we introduce the general forms for dimensional rules in $\Sigma^\mc{M}$ (those in {\bf 4.}) and the dimensional constraints in $\kappa^\mc{M}$ (in {\bf 5.}), which are all application-dependent.

\vspace{2mm}
\noindent {\bf 4.} \ {\em Dimensional rules} as \dplus \ \tgds:

\vspace{-4mm}
\begin{align}
R_1(\bar{x}_1;\bar{y}_1),...,R_n(\bar{x}_n;\bar{y}_n),P_1(x_1,x'_1),...,P_m(x_m,x'_m) \ \rightarrow \ \exists \bar{y}' \ \red{R'}(\red{\bar{x}'};\bar{y}).\label{frm:dimensional-rule}
\end{align}
Here, $R_i(\bar{x}_i;\bar{y}_i)$ and $\red{R'}(\red{\bar{x}'};\bar{y})$ are categorical predicates, the $P_i$ are child-parent predicates, $\bar{y}' \subseteq \bar{y}$, \ $\red{\bar{x}'} \subseteq \bar{x}_1 \cup ... \cup \bar{x}_n \cup \{x_1,...,x_m, x'_1,...,x'_m\}$, \ $\bar{y} \! \smallsetminus \! \bar{y}' \subseteq \bar{y}_1 \cup ... \cup \bar{y}_n$; repeated variables in bodies (join variables) appear only in categorical positions in the categorical relations and attributes in child-parent predicates.\footnote{\ This is a natural restriction to capture dimensional navigation as captured by the joins (cf. Example \ref{ex:ont}).}

Notice that existential variables appear only in non-categorical attributes. \ \red{The main reason for this condition is that in some applications we may have an existing, fixed and closed-world multidimensional database providing the multidimensional structure and data. In particular, we may not want to create new category elements via value invention, but only values for non-categorical attributes, which do not belong to categories. We will discuss this condition in more detail and its possible relaxation in Section~\ref{sec:clo}.}

\vspace{2mm}
\noindent {\bf 5.} \
{\em Dimensional constraints}, as \egds \ or \ncs, of the forms:
\begin{align}
R_1(\bar{x}_1;\bar{y}_1),...,R_n(\bar{x}_n;\bar{y}_n),P_1(x_1,x'_1),...,P_m(x_m,x'_m) ~\rightarrow~& z=z'.\label{frm:dimensional-egd} \\
R_1(\bar{x}_1;\bar{y}_1),...,R_n(\bar{x}_n;\bar{y}_n), P_1(x_1,x'_1),...,P_m(x_m,x'_m) ~\rightarrow~& \bot. \label{frm:dimensional-nc}
\end{align}
Here, $R_i \in \mc{R}^c$, $P_j \in \mc{L}$, and $z,z' \in \bigcup \bar{x}_i \cup \bigcup \bar{y}_j$. 

\vspace{2mm}

Some of the lists in the bodies of (\ref{frm:dimensional-rule})-(\ref{frm:dimensional-egd}) may be empty, i.e. $n=0$ or $m=0$. This allows us to represent, in addition to properly ``dimensional" constraints, also classical constraints on categorical relations, e.g. keys or \fds.


\blue{A general \tgd \ of the form (\ref{frm:dimensional-rule}) can be used for  {\em upward-} or {\em downward-navigation} (or, more precisely, upward or downward data generation) depending on the joins in the body. The direction is determined by both the difference of category levels in a dimension of the categorical variables that appear in the body joins, and the value propagation to the rule head. To be more precise, consider the simplest case where (\ref{frm:dimensional-rule}) is of the form
\begin{align*}
R(\bar{x};\bar{y}_1), P(x_1,x'_1) \ \rightarrow \ \exists \bar{y}' \ \red{R'}(\red{\bar{x}'};\bar{y}),
\end{align*}
with a join between $R(\bar{x};\bar{y}_1)$ and $P(x_1,x'_1)$ (via a categorical variable in $\bar{x}$). When $x_1 \in \bar{x}$ and $x'_1 \in \bar{x}'$, {\em one-step upward-navigation} is enabled, from (the level of) $x_1$ to (the level of) $x'_1$. An example is $\sigma_1$ in (\ref{eq:s1intr}). Now, when $x'_1 \in \bar{x}$ and $x_1 \in \bar{x}'$, {\em one-step downward-navigation} is enabled. An example is $\sigma_2$ in (\ref{eq:sigma2}). \
More generally, {\em multi-step navigation}, between a category and an ancestor or descendant category, can be captured through a  chain of joins with adjacent  child-parent dimensional predicates in the body of a \tgd \ (an example is (\ref{eq:multi}) below). However, a general dimensional rule of the  form (\ref{frm:dimensional-rule}) may contain joins in mixed directions, even on the same dimension.}

\ignore{(\ref{if there is a single child-parent predicate $P_j \in \mc{L}$ in the body of (\ref{frm:dimensional-rule}), and the join is between $R_i(\bar{x}_i;\bar{y}_i)$ and $P_j(x_j,x'_j)$ (via a categorical variable in $\bar{x}_i$), then {\em one-step upward-navigation} is enabled, from $x'_j$ to $x_j$, when $x'_j \in \bar{x}_i$ (i.e. $x'_j$ appears in $R_i(\bar{x}_i;\bar{y}_i)$) and $x_j \in \bar{x}'$, i.e in the head). An example is $\sigma_1$ in (\ref{eq:s1intr}). \  {\em One-step downward-navigation} is enabled,  from $x_j$ to $x'_j$, when $x_j$ occurs in $R_i$ and $x'_j$ occurs in $R_k$.} }

\begin{example} \bl{(ex.~\ref{ex:md-relational} cont.)} \ \label{ex:ont} The left-hand-side of Figure~\ref{fig:omdm} shows a  {\em dimensional constraint} $\eta$
categorical relation  {\it WorkSchedules}, which is  linked to  the {\sf Temporal} dimension via the {\it Day} category. It tells us (possibly because the {\it Intensive} care unit was closed during January) that: \ {\em ``No personnel was working in the Intensive care unit in January"}. It is a constraint of the form (\ref{frm:dimensional-nc}):

\vspace{-4mm}
\begin{align}
\eta\!: \ \nit{WorkSchedules}({\sf intensive},d;n,s),\nit{DayMonth}(d,{\sf jan})  ~\rightarrow~ \bot.\label{frm:expnc}
\end{align}
\vspace{-4mm}

The dimensional rule  $\sigma_1$ in Figure~\ref{fig:omdm} and given in (\ref{eq:s1intr}) (as a \tgd \ of the general form (\ref{frm:dimensional-rule})) can be used to generate new tuples for relation {\it WorkSchedules}. Then, constraint $\eta$ is expected to be satisfied both by the initial extensional tuples for {\it WorkSchedules} and its tuples generated through $\sigma_1$, i.e. by its non-shaded tuples and shaded tuples in Figure~\ref{fig:omdm}, resp.
In this example, $\eta$ is satisfied.

Notice that {\it WorkSchedules} refers to the $\nit{Day}$ attribute of the {\sf Temporal} dimensions, whereas $\eta$ involves the \nit{Month} attribute. Then, checking $\eta$ requires upward navigation through the {\sf Temporal} dimension.  Also the {\sf Hospital} dimension  is involved in the satisfaction of $\eta$: \ The \tgd \ $\sigma_1$ in  may generate new tuples for {\it WorkSchedules}, by upward navigation from {\it Ward} to {\it Unit}.

Furthermore, we have an additional \tgd:
\begin{align}
\sigma_2&:&\!\!\nit{WorkSchedules}(u,d;n,t),\nit{WardUnit}(w,u)  ~\rightarrow~ \exists s\;\nit{Shifts}(w,d;n,s) \label{eq:sigma2}
\end{align}
that can be used with  \nit{WorkSchedules}  to generate data for categorical relation \nit{Shifts}. The shaded tuple in it is one of those.
 This \tgd \ reflects the institutional guideline stating that {\em ``If a nurse works in a unit on a specific day, he/she has shifts in every ward of that unit on the same day"}. \ Accordingly, $\sigma_2$ relies on downward navigation for tuple generation, from the {\it Unit} category level down to the {\it Ward} category level.

\blue{Here, $\sigma_1$ and $\sigma_2$ in (\ref{eq:s1intr}) and (\ref{eq:sigma2}) are examples of \tgds \ enabling upward and downward, one-step dimension navigation, resp.  The following dimensional rule enables multi-step navigation,  propagating doctors at the unit level all the way up to the hospital level: \
\begin{align}\nit{WardDoc}(\nit{ward};\nit{na},\nit{sp}),\nit{WardUnit}(\nit{ward,unit}),\nit{UnitInst}(\nit{unit,ins})  \rightarrow \nit{HospDoc}(\nit{ins};\nit{na},\nit{sp}). \label{eq:multi}
\end{align}}

\vspace{-4mm}
Assuming the ontology also has  a categorical relation, ${\it Therm(Ward,Thertype};\nit{Nurse})$, with \nit{Ward} and \nit{Thertype} categorical attributes, the latter for an ${\sf Instrument}$  dimension,
the following should be an \egd \ of the form (\ref{frm:dimensional-egd}) saying that {\em ``All thermometers in a unit are of the same type"}:

\vspace{-4mm}
\begin{align} \label{frm:expegd}
{\it Therm(w,t;n)},{\it Therm(w',t';n')},\!{\it WardUnit(w,u)},\!{\it WardUnit(w',u)} \rightarrow t=t'.
\end{align}
\vspace{-4mm}

\noindent Notice that our ontological language allows us to impose a condition at the {\it Unit} level without having it as an attribute in the categorical relation.\footnote{\ If we have that relation, then (\ref{frm:expegd}) could be replaced by a ``static", non-dimensional \fd.}

Notice that existential variables in dimensional rules, such as  $t$ and $s$ as in (\ref{eq:s1intr}) and (\ref{eq:sigma2}), resp., make up for the missing, non-categorical attributes {\it Speciality} and {\it Shift} in {\it WorkSchedules} and {\it Shifts}, resp.\boxtheorem \end{example}

\begin{example} \label{ex:dn} (ex. \ref{ex:ont} cont.) \ Rule $\sigma_2$ supports downward tuple-generation. When enforcing it on a tuple ${\it WorkSchedules}(u,d;n,t)$, via category member $u$ (for Unit), a tuple for {\it Shifts} is generated {\em for each} child $w$ of $u$ in the \nit{Ward} category for which the body of $\sigma_2$ is true.
For example, chasing $\sigma_2$ with the third tuple in {\it WorkSchedules} generates two new tuples in \nit{Shifts}: \ ${\it Shifts}({\sf W}_2, {\sf sep/6/2016},{\sf helen},\zeta)$ and ${\it Shifts}({\sf W}_1, {\sf sep/6/2016},{\sf helen},\zeta')$, with fresh nulls, $\zeta$ and $\zeta'$. The latter tuple is not shown in Figure~\ref{fig:omdm} \red{since} it is dominated by the third tuple, ${\it Shifts}({\sf W}_1, {\sf sep/6/2016},{\sf helen},{\sf morning})$, in \nit{Shifts} \red{(i.e. the existing tuple is more general or informative than the one that would be introduced with a null value, and also it already serves as a witness for the existential statement).\footnote{\ \red{Eliminating those dominated tuples does not have any impact on certain query answering.}}} With the old and new tuples we can obtain the answers to the query about the wards of {\it Helen}  on \nit{Sep/6/2016}: \ $\mc{Q}'(w)\!: \ \exists s\;{\it Shifts}(w, {\sf sep/6/2016}, {\sf helen}, s)$.
 They are $W_1$ and $W_2$.



In contrast, the join between {\it Shifts} and {\it WardUnit} in $\sigma_1$ enables upward-dimensional navigation; and the generation of only one tuple for {\it WorkSchedules} from each tuple in {\it Shifts}, because each {\it Ward} member has at most one {\it Unit} parent.
\boxtheorem\end{example}

\ignore{
\commos{We assumed strictness in adjacent child-parent categories (in (\ref{frm:key})), but not in general, i.e. between ancestors and descendants. So I think if we do not consider strictness, even in upward navigation we might generate multiple tuples.

But the issue I tired to explain here also relates to the fact that we employed oblivious chase that does not check satisfaction and blindly adds atoms (e.g. ${\it Shifts}({\sf W}_1, {\sf sep/6/2016},{\sf helen},\zeta')$). However, the new atom do not change \cq \ answers because of the already existing atom (e.g. ${\it Shifts}({\sf W}_1, {\sf sep/6/2016},{\sf helen},{\sf morning})$).}  }

\section{Computational Properties of the OMD Model} \label{sec:complexity}

As mentioned before, without any restrictions \dplus \ programs {\em conjunctive query answering} (CQA) may be undecidable, even without constraints \cite{cali03}. Accordingly, it is important to identify classes of programs for which CQA is decidable, and hopefully in polynomial time in the size of the underlying database, i.e. in data complexity. Some classes of this kind have been identified. In the rest of this section we introduce some of them that are particularly relevant for our research. We show that under natural assumptions or \omd \ ontologies belong to those classes. In general, those program classes  do not consider constraints. At the end of the section we consider the presence of them in terms of their effect on QA.

\subsection{Weakly-Acyclic, Sticky and Weakly-Sticky Programs} \label{sec:wa}

{\em Weakly-acyclic}  \dpm \ programs (without constraints) form a syntactic class of \dplus \ programs that is defined appealing to the notion of dependency graph \cite{fagin}.
The {\em dependency graph} (DG) of a \dplus \ program $\prg$ is a directed graph whose vertices are the {\em positions} of the program's schema. Edges are defined as follows. For every $\sigma \in \prg$ and universally quantified variable ($\forall$-variable) $x$ in ${\it head}(\sigma)$ and  position $p$ in ${\it body}(\sigma)$ where $x$ appears: \ (a)  Create an edge from $p$ to position $p'$ in ${\it head}(\sigma)$ where $x$ appears \red{(representing the propagation of a value from a position in the body of a rule to a position in its head).} \ (b) Create a {\it special edge} from $p$ to position $p''$ in ${\it head}(\sigma)$ where an $\exists$-variable $z$ appears \red{(representing a value invention in the position of an  existential variable in the rule head)}.

The {\it rank of a position} $p$, $\nit{rank}(p)$, is the maximum number of special edges on (finite or infinite) paths ending at $p$. $\finiteRank(\prg)$ denotes the set of finite-rank positions in $\prg$. A program is {\em Weakly-Acyclic}~(WA) if all of the positions have finite-rank.

\begin{example} \label{ex:dgNEW} Program $\Pi$ below has the DG in Figure ~\ref{fig:dgNEW}, with dashed special edges.

\vspace{2mm}
\begin{minipage}[t]{0.45\textwidth}
\resizebox{!}{2.0cm}{\begin{picture}(0,0)%
\includegraphics{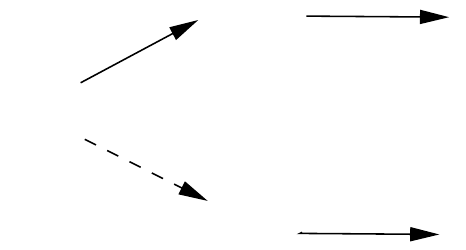}%
\end{picture}%
%
%
\setlength{\unitlength}{3947sp}%
\begingroup\makeatletter\ifx\SetFigFont\undefined%
\gdef\SetFigFont#1#2#3#4#5{%
  \reset@font\fontsize{#1}{#2pt}%
  \fontfamily{#3}\fontseries{#4}\fontshape{#5}%
  \selectfont}%
\fi\endgroup%
\begin{picture}(2220,1194)(631,-1597)
\put(2836,-578){\makebox(0,0)[lb]{\smash{{\SetFigFont{12}{14.4}{\rmdefault}{\mddefault}{\updefault}{\color[rgb]{0,0,0}$P[1]$}%
}}}}
\put(2836,-1523){\makebox(0,0)[lb]{\smash{{\SetFigFont{12}{14.4}{\rmdefault}{\mddefault}{\updefault}{\color[rgb]{0,0,0}$P[2]$}%
}}}}
\put(1655,-578){\makebox(0,0)[lb]{\smash{{\SetFigFont{12}{14.4}{\rmdefault}{\mddefault}{\updefault}{\color[rgb]{0,0,0}$R[1]$}%
}}}}
\put(1655,-1523){\makebox(0,0)[lb]{\smash{{\SetFigFont{12}{14.4}{\rmdefault}{\mddefault}{\updefault}{\color[rgb]{0,0,0}$R[2]$}%
}}}}
\put(646,-1021){\makebox(0,0)[lb]{\smash{{\SetFigFont{12}{14.4}{\rmdefault}{\mddefault}{\updefault}{\color[rgb]{0,0,0}$U[1]$}%
}}}}
\end{picture}%
}
\captionof{figure}{Dependency graph}\label{fig:dgNEW}
\vspace{3mm}
\end{minipage}\hfill
\begin{minipage}[t]{0.5\textwidth}
{\centering
\vspace{-2.3cm}
\begin{align*}
  U(x) &~\rightarrow~ \exists y\;R(x,y),\\
  R(x,y) &~\rightarrow~ P(x,y).
\end{align*}
}

\vspace{-6mm}\phantom{oo}

\noindent $U[1]$, $R[1]$ and $P[1]$ have rank $0$. \ $R[2]$ and $P[2]$ have rank  $1$. Then, $\finiteRank(\prg) = \{U[1], R[1], P[1], R[2], P[2]\}$, and   $\Pi$  is \WA.\boxtheorem
\end{minipage}
\end{example}

\vspace{-3mm}The chase for these programs stops in polynomial time in the size of the extensional data, making  CQA  \ptime-complete in data complexity~\cite{fagin}, but  2\exptime-complete in combined complexity, i.e. in the combined size of the program, query  and data~\cite{kolaitis}.




Sticky \dplus \ programs (without constraints) are characterized through a  {\em marking procedure} on body variables program rules. For a program $\prg$, the procedure has two steps:
\begin{enumerate}[(a)]
  \item {\em Preliminary step}: For every $\sigma \in \prg$ and variable $x$ in ${\it body}(\sigma)$, if there is an atom in ${\it head}(\sigma)$ where $x$ does not appear, mark  every occurrence $x$ in  ${\it body}(\sigma)$.
  \item {\em Propagation step}: For every $\sigma \in \prg$, if a marked variable in ${\it body}(\sigma)$ appears in position $p$, then for every $\sigma' \in \prg$, mark every occurrence of a variable in ${\it body}(\sigma')$ that also appears in ${\it head}(\sigma')$ in  position $p$.
\end{enumerate}

\begin{example} \label{example:stickyNEW}
 Consider program $\Pi$ on the left-hand side below., with its second rule already showing marked variables (with a hat) after the preliminary step. The set of rules on the right-hand side show the result of whole marking procedure.
\[ \arraycolsep=0pt
\begin{array}{rl c rl}
R(x,y),P(x,z) ~\rightarrow&~ S(x,y,z). &\hspace*{1.0cm}& R(\hat{x},y),P(\hat{x},\hat{z}) ~\rightarrow&~ S(x,y,z).\\
S(\hat{x},y,\hat{z}) ~\rightarrow&~ U(y).&\hspace*{1.0cm}&S(\hat{x},y,\hat{z}) ~\rightarrow&~ U(y).\\
U(x) ~\rightarrow&~ \exists y\;R(y,x).&\hspace*{1.0cm}&U(x) ~\rightarrow&~ \exists y\;R(y,x).
\end{array}
\]

 For example, $x$ is marked in $S[1]$ in the body of the second rule (after the preliminary step). For the propagation step, we find $S[1]$ in the head of the first rule, containing $x$ (it could have been a different variable). Then the occurrences of $x$ in the body of the first rule have to be marked too, in positions $R[1]$ and $P[1]$.
\ignore{
 and $z$ in the first rule-body end up marked  after the propagation step: they appear in the same rule's head, in positions where marked variables appear in the second rule ($S[1]$ and $S[3]$).}
 \boxtheorem
\end{example}

A \dplus \ program $\prg$ is {\em sticky} when, after applying the marking procedure, there is no rule with a marked variable appearing more than once in its body. (Notice that a variable never appears both marked and unmarked in a  same body.) Accordingly, the program in Example \ref{example:stickyNEW} is {\em not} sticky: marked variable $x$ in the first rule's body appears in a join (in $R[1]$ and $P[1]$).

The stickiness property for a program guarantees that, given a \cq, a finite initial fragment of the possibly infinite chase can be used for answering the query; \ignore{\footnote{\ \red{This is the result of the fact that, with the chase of a \tgd, if a value replaces a join variable in the body, the value is propagated through all the possible subsequent steps and will appear in the query atoms that make it true. Therefore the number of nulls is limited by the number of $\exists$-variables in the query. Cf. \cite{milani16rr} for a more detailed discussion.}}} actually, a fragment of polynomial size in that of the extensional data \red{(cf. \cite{cali12} and \cite{milani16rr} for a more detailed discussion)}. As a consequence,
CQA  on sticky programs is in \ptime \ in data. (It is \exptime-complete in combined complexity~\cite{cali12}.) Even more, CQA  over sticky programs enjoys {\em first-order rewritable}~\cite{gottlob11}, that is,
  a CQ posed to the program can be rewritten into an \fo \ query that can be evaluated directly on the extensional data. \ignore{This makes CQA belong to the tractable class \acz \ (in data complexity)~\cite{vardi}.}
\ignore{  Sticky programs can have an infinite chase, but given a CQ only a finite, initial, query-dependent  fragment of it suffices to answer the query.}

None of the well-behaved classes of weakly-acyclic and sticky programs contain the other, but they can be combined into a new syntactic class of {\em weakly-sticky} (WS) programs that extends both original classes.  Again, its characterization does not depend on the extensional data, and uses the already introduced notions of finite-rank and marked variable: A program $\prg$ (without constraints) is {\em weakly-sticky} if every repeated variable in a
rule body is either non-marked  or appears in some position in $\finiteRank(\prg)$ (in that body).

\begin{example} Consider program $\prg$ already showing the marked variables:

\vspace{-4mm}
\begin{align*}
R(\hat{x},\hat{y}) ~\rightarrow&~ \exists z\; R(y,z).\\
R(\hat{x},\hat{y}),U(\hat{y}),R(\hat{y},\hat{z}) ~\rightarrow&~ R(x,z).
\end{align*}
Here, $\finiteRank(\prg)=\{U[1]\}$. The only join variable is $y$ in the second rule, which appears in $U[1]$. Since $U[1] \in \finiteRank(\prg)$,  $\prg$ is WS.
\ Now, let $\prg'$ be obtained from $\Pi$ by  replacing the second rule by (the already marked) rule: \
\ignore{\vspace{-4mm}
\begin{align*}}
$R(\hat{x},\hat{y}),R(\hat{y},\hat{z}) ~\rightarrow~ R(x,z)$. \
Now, $\finiteRank(\prg')=\emptyset$, and the marked join variable $y$ in the second rule appears in $R[1]$ and $R[2]$, both non-finite (i.e. infinite) positions. Then, $\prg'$ is {\em not} WS.\boxtheorem\end{example}

The \WS \ conditions basically prevent marked join variables from appearing only in infinite (i.e. infinite-rank) positions. With \WS \ programs the chase may not terminate, due to an infinite generation and propagation of null values,
but in finite positions only finitely many nulls may appear, which restricts the values that the possibly problematic variables, i.e. those marked in joins, may take (cf. \cite{milani16rr} for a discussion).
For \WS \ programs CQA is tractable. Actually, CQA  can be done on initial,  query-dependent fragments of the chase  of polynomial size in data.  CQA is tractable, but \ptime-complete in data, and 2\exptime-complete in combined complexity~\cite{cali12}.

\ignore{
\vspace{1cm}
Table~\ref{tab:complexity} is a summary of complexity of \bcq \ answering under programs that have been reviewed in this section.

\begin{table}[h]
  \centering
\setlength{\tabcolsep}{0.3em}
\setlength{\arrayrulewidth}{0.75pt}
\renewcommand*\arraystretch{1.35}
\begin{tabular}{p{1.5cm} p{3cm} p{4cm}}
\hline
& {\bf Data complexity} & {\bf Combined complexity}\\
\hline
\WA    & \ptime-complete \hfill & 2\exptime-complete \hfill \\
{\em sticky}    & in \acz & \exptime-complete \\
\WS    & \ptime-complete & 2\exptime-complete\\
\hline
\end{tabular}
\caption{Complexity of \bcq \ answering under programs in Section~\ref{sec:pclasses}}\label{tab:complexity}
\end{table}
}

{\em In the following and as usual with a \dpm \ program $\Pi$, we say $\Pi$ is weakly-acyclic, sticky or weakly-sticky, etc.,  if its set of \tgds \ has those properties.}

\subsection{\omd \ Ontologies as Weakly-Sticky \dpm \ Programs}

In this section we investigate  the  ontologies $\mc{O}^\mc{M}$ used by the \omd \ model as \dpm \ programs. We start by considering only their subontologies $\Sigma^\mc{M}$ formed by their \tgds.
The impact of the set $\kappa^\mc{M}$ of constraints in $\mc{O}^\mc{M}$ is analyzed in Section \ref{sec:withCons}.

It turns out that the MD ontologies are weakly-sticky. Intuitively, the main reason is that the join variables in the dimensional \tgds \ are in the categorical positions, where finitely many members of dimensions can appear during the chase, because  no existential variable ($\exists$-variable) occurs in  a categorical position; so, no new values are invented in them positions during the chase.

\begin{proposition}\label{prop:weaklysticky}\em \md \ ontologies are weakly-sticky  \dpm \ programs.
\boxtheorem\end{proposition}

\dproof{Proposition~\ref{prop:weaklysticky}}{\red{The \tgds \ are of the form (\ref{frm:dimensional-rule}):}
\red{\begin{align*}
R_1(\bar{x}_1;\bar{y}_1),...,R_n(\bar{x}_n;\bar{y}_n),P_1(x_1,x'_1),...,P_m(x_m,x'_m) \ \rightarrow \ \exists \bar{y}' \ R'(\bar{x}';\bar{y}),
 \end{align*}}where: (a) $\bar{y}' \subseteq \bar{y}$, (b) $\bar{x}' \subseteq \bar{x}_1 \cup ... \cup \bar{x}_n \cup \{x_1,...,x_m, x'_1,...,x'_m\}$, (c) $\bar{y} \! \smallsetminus \! \bar{y}' \subseteq \bar{y}_1 \cup ... \cup \bar{y}_n$, and (d) repeated (i.e. join) variables in bodies are only in positions of categorical attributes.

We have to show that every join variable in such a \tgd \ either appears at least once in a finite-rank position or it is not marked. Actually,  the former always holds, because, by condition (d), join variables appear only in categorical positions; and categorical positions,  as we will show next, have finite, actually $0$, rank (so, no need to investigate marked positions).\footnote{\ Actually, we could extend the \md \ ontologies by relaxing the condition on join variables in the dimensional rules, i.e. condition (d), while still preserving the weakly-sticky condition. It is by allowing non-marked joins variables in non-categorical positions.}



In fact, condition (a) guarantees that there is no special edge in the dependency graph of a set of dimensional rules $\Sigma^\mc{M}$ that ends at a categorical position. Also, (b) ensures that there is no path from a non-categorical position to a categorical position, i.e. categorical positions are connected only to categorical positions. Consequently, every categorical position has a finite-rank, namely $0$. \boxtheorem}\\


The proof establishes that every position involved in join in the body of a \tgd \ has finite rank. However, non-join body variables in a \tgd \ might still have infinite rank.

\begin{example}\label{ex:ranks} (ex. \ref{ex:ont} cont.) For  the \md \ ontology  with $\sigma_1$ and $\sigma_2$, $\nit{WorkSchedules}[4]$ and $\nit{Shifts}[4]$ have infinite rank; and all the other positions have finite rank.\boxtheorem
\end{example}

\ignore{+++{A \md \ ontology includes dimensional rules of the form (\ref{frm:dimensional-rule}): $R_1(\bar{e}_1;\bar{a}_1),...,R_n(\bar{e}_n;\bar{a}_n),P_1(e_1,e'_1),...,P_m(e_m,e'_m)  \rightarrow \exists \bar{a}_z \ R_k(\bar{e}_k;\bar{a}_k),$ in which (a) $\bar{a}_z \subseteq \bar{a}_k$, (b) $\bar{e}_k \subseteq \bar{e}_1 \cup ... \cup \bar{e}_n \cup \{e_1,...,e_m, e'_1,...,e'_m\}$, (c) $\bar{a}_k \! \smallsetminus \! \bar{a}_z \subseteq \bar{a}_1 \cup ... \cup \bar{a}_n$, and (d) repeated variables in bodies are only in positions of categorical attributes.

In particular, (a) guarantees that no null values are invented in the categorical positions during the chase of $\mc{O}^\mc{M}$. Also, (b) ensures that the variables in the non-categorical positions in the bodies do not appear in the heads in the categorical positions. As a result, no null value can replace a variable in a categorical position during the chase. This, in addition with (d), proves that no null value can replace a repeated body variable, which proves a set of dimensional rules is \WS.\boxtheorem}\\ +++}

\begin{corollary} \label{cor:ws} \em Conjunctive query answering on MD ontologies (without constraints) can be done in polynomial-time in data complexity. \boxtheorem 
\end{corollary}

The tractability (in data) of CQA under \WS \ programs was established in \cite{cali12} on theoretical grounds, without providing a practical algorithm. An implementable, polynomial-time algorithm
for CQA under \WS \ programs is presented in~\cite{milani16rr}. Given a \cq \ posed to the program, they apply a query-driven chase of the program, generating a finite initial portion of the chase instance that
 suffices to answer the query at hand. Actually, the algorithm can be applied to a class that not only extends WS, but is also closed under {\em magic-set} rewriting of \dplus \ programs \cite{alviano12-datalog}, which allows for query-dependent optimizations of the program
 \cite{milani16rr}.

Unlike sticky programs, for complexity-theoretic reasons, \WS \ programs do not allow \fo \ rewritability for CQA. However, a hybrid algorithm is proposed in~\cite{milani16rr-cali}. It is based on partial grounding of the \tgds \ using the extensional data, obtaining a sticky program, and a subsequent rewriting of the query. These algorithms can be used for CQA under our MD ontologies. However, presenting the details  of these algorithms is beyond the scope of this paper.



\subsection{\omd \ Ontologies with Constraints}\label{sec:withCons}

\red{In order to analyze  the impact of the constraints  on MD ontologies, i.e. those in {\bf 1.}-{\bf 3.}, {\bf 5.} in Section \ref{sec:omd}, on CQA, we have to make and summarize some general considerations on constraints in \dplus \ programs. First, the whole discussion on constraints of Section \ref{sec:dpm} apply here. In particular,
the presence of constraints may make the ontology inconsistent, in which case CQA becomes trivial. Furthermore, in comparison to a program without constraints, the addition of the latter to the same program may change query answers, because some models of the ontology may be discarded. Furthermore, CQA under \ncs \ can be reduced to CQA without them.  }

First, those in (\ref{frm:refh2}) and (\ref{frm:referential}) are \ncs \ with negative literals in their bodies. The former capture the structure of the underlying multidimensional data model (as opposed to the ontological one). They can be checked against the extensional database $D$. If they are satisfied, they will stay as such, because the dimensional \tgds \ in (\ref{frm:dimensional-rule}) do not invent category members. If the underlying multidimensional database has been properly created, those constraints will be satisfied and preserved as such.  The same applies to the negative constraints in (\ref{frm:referential}): the dimensional \tgds \ may invent only non-categorical values in categorical relations. (cf. Section \ref{sec:clo} for a discussion.)

As discussed in Section \ref{sec:omd}, \egds \ may be more problematic since there may be \ignore{
Including \egds \ on \dplus \ programs is different from imposing \ncs, due to the possible} interactions between  \egds \ and \tgds \ during the chase procedure: the enforcement of a \tgd \ may activate an \egd, which in turn may make some \tgds \ applicable, etc. (cf. Section \ref{sec:dpm}).
\ignore{+++
\begin{example} \label{ex:non-separable} Consider a program $\prg$ with $D=\{R(a,b)\}$ and the following rules:
\vspace{-4mm}
\begin{align}
R(x,y) ~&\rightarrow~ \exists z\; \exists w\;S(y,z,w).\label{frm:sep1}\\
S(x,y,y) ~&\rightarrow~ P(x,y).\label{frm:sep2}
\end{align}
\noindent $\nit{chase}(\prg)=\{R(a,b),S(b,\zeta_1,\zeta_2)\}$, with $\zeta_1$ and $\zeta_2$ fresh nulls. Rule~(\ref{frm:sep2}) is not applied since $\zeta_1$ and $\zeta_2$ are not equal, as required by the body. The answer to a \bcq \ $\mc{Q}:\exists x\;\exists y\;P(x,y)$ is {\em false} under $\prg$ as $\nit{chase}(\prg) \not \models \mc{Q}$. Now consider $\prg'$ that is obtained by adding the following \egd \ to $\prg$:
\vspace{-0.7cm}
\begin{align}
S(x,y,z) ~&\rightarrow~ y=z.\label{frm:sep3}
\end{align}
\noindent The chase of $\prg'$ first applies rule~(\ref{frm:sep1}) and results in $I_1=\{R(a,b),S(b,\zeta_1,\zeta_2)\}$. Now, there is no more tgd/assignment applicable pair. But, if we apply the \egd~(\ref{frm:sep3}), it equates $\zeta_1$ and $\zeta_2$, and results in $I_2=\{R(a,b),S(b,\zeta_1,\zeta_1)\}$. (This kind of egd-chase step explained in Section~\ref{sec:kappas}.) Now, rule~(\ref{frm:sep2}) and $\theta': x\mapsto b, y\mapsto \zeta_1$ are applicable and they add $P(b,\zeta_1)$ to $I_2$, generating $I_3=\{R(a,b),S(b,\zeta_1,\zeta_1),P(b,\zeta_1)\}$.\\
The procedure terminates since no more \tgds \ or \egds \ can be applied. The chase result, $\nit{chase}(\prg')$, is $I_3$. $\mc{Q}$ holds under $\prg'$: $\nit{chase}(\prg') \models \mc{Q}$.\boxtheorem\end{example}
+++}
 Actually, these interactions between \tgds \ and  \egds, make it in general impossible to postpone \egd \ checking or enforcement until all \tgds \ have been applied: {\em tgd-chase steps} and  {\em egd-chase steps} may have to be interleaved. When the (combined) chase does not fail, the result is a possibly infinite universal model that satisfies both the \tgds \ and \egds~\cite{cali13}.

The interaction of \tgds \ and \egds \  may lead to undecidability of CQA~\cite{cali03,chandra,johnson,mitchell}. \ignore{In fact, this is true even in simple cases, such as combinations of {\em functional dependencies} (\fds) and {\em inclusion dependencies} (\ideps)~\cite{chandra}, or {\em key constraints} and \ideps~\cite{cali03}.} However, a {\em separability property} of the combination of \egds \ and \tgds \ guarantees a harmless interaction that makes  CQA  decidable and preserves CQA~\cite{cali12}: \ For a program $\prg$ with extensional database $D$, a set of \tgds \ $\rules$, and a set of \egds \ $\constraints$,  $\rules$ and $\constraints$ are {\it separable} if either (a) the chase with $\prg$ fails, or (b) for any \bcq \ $\mc{Q}$, $\prg \models \mc{Q}$ if and only if $\rules \cup D \models \mc{Q}$.

In Example~\ref{ex:non-separableNEW}, the \tgds \ and the \egd \ are not separable as the chase does not fail, and the \egd \ changes \cq \ answers (in that case, $\prg \not\models \mc{Q}$ and $\prg' \models \mc{Q}$).

Separability is a semantic property, relative to the chase, and depends on a program's extensional data. If separability holds, combined chase failure can be detected by posing
\bcq s (with $\neq$, and obtained from the \egds' \ bodies)  to the program without the \egds~\cite[theo. 1]{cali12amw}. However, separability is undecidable \cite{cali12amw}. Hence the need for an alternative, syntactic,  decidable, sufficient condition for separability. Such a condition has been identified for \egds \ that are key constraints~\cite{cali13}; it is that of {\em non-conflicting} interaction.\footnote{\ The notion has been extended to FDs in ~\cite{cali12}: A set of \tgds \ $\rules$ and a set $\constraints$ of \fds \ are {\it non-conflicting} if, for every  \tgd \ $\sigma$, with set $U_\sigma$ of
 non-existential(lly quantified variables for) attributes in $\nit{head}(\sigma)$, and \fd \ $\epsilon$ of the form $R:\vectt{A} \rightarrow \vectt{B}$, at least one of the following holds: (a) $\nit{head}(\sigma)$ is not an $R$-atom, (b) $U_\sigma \not\supseteq \vectt{A}$, or (c) $U_\sigma = \vectt{A}$ and each $\exists$-variable in $\sigma$ occurs just once in the head of $\sigma$. }  Intuitively,  the condition guarantees that the \tgds \ can only generate tuples with new key values, so they cannot violate the key dependencies.

Back to our \omd \ ontologies, it is easy to check that the \egds \ of the form~(\ref{frm:key}) in {\bf 2.}, actually key constraints, are non-conflicting,  because they satisfy the first of the conditions for non-conflicting interaction. Then, they are  separable
from the dimensional constraints as \egds.

More interesting and crucial are the dimensional constraints under {\bf 5.}. They are application-dependent \ncs \ or \egds. Accordingly, the discussion in Section \ref{sec:withCons} applies to them, and not much can be said in general. However,  for the combination of dimensional \tgds \ and dimensional \egds \ in \omd \ ontologies, separability holds when the \egds \ satisfy a simple condition.

\ignore{\comlb{Below is stated in terms of egds. Check.}
\commos{I checked and it is fine.}
\comlb{\red{NEW}: You mean you get separability without going through non-conflictingness (which we do not have for general egds)? If right, the last line in red of the previous paragraph is wrong: it is not done via non-interaction.}  }

\begin{proposition}\label{prop:sep}\em For an \md \ ontology $\mc{O}^\mc{M}$ with a set $\Sigma^\mc{M}$ of \tgds \ as in (\ref{frm:dimensional-rule}) and set $\kappa^\mc{M}$ of \egds \ as in (\ref{frm:dimensional-egd}), if for every \egd \ in $\kappa^\mc{M}$ the variables in the  head occur in categorical positions in the body, then separability holds. \boxtheorem\end{proposition}

\dproof{Proposition~\ref{prop:sep}}{Let $D^\mc{M}$ be the ontology's extensional data. We have to show that if ${\it chase}(\mc{O}^\mc{M})$ does not fail, then for every BCQ $\mc{Q}$, ${\it chase}(\mc{O}^\mc{M}) \models \mc{Q}$ if and only if ${\it chase}(\Sigma^\mc{M},D^\mc{M}) \models \mc{Q}$.

Let's assume that the chase with $\mc{O}^\mc{M}$ does not fail.  As we argued before Proposition~\ref{prop:weaklysticky}, no null value replaces a variable in a categorical position during the chase with $\mc{O}^\mc{M}$. For this reason, the variables in the heads of \egds \ are never  replaced by nulls. As a result, the \egds \ can only equate constants, leading to chase failure if they are different. Since we assumed the chase with $\mc{O}^\mc{M}$
does not fail, the \egds \ are never applicable during the chase or they do not produce anything new (when the two constants are indeed the same), so they can be ignored, and the same result for the chase with or without \egds.\boxtheorem}\\


An example of dimensional \egd \ as in Proposition ~\ref{prop:sep} is (\ref{frm:expegd}). Also the key constraints in (\ref{frm:key}) satisfy the syntactic condition.
In combination with Proposition~\ref{prop:weaklysticky}, we obtain:

\begin{corollary} \em \label{cor:ptime}
Under the hypothesis of Proposition~\ref{prop:sep}, CQA from an \md \ ontology can be done in polynomial-time in data. \boxtheorem
\end{corollary}
\dproof{Corollary~\ref{cor:ptime}}{From the proof of Proposition \ref{prop:sep}, we have that the ${\it chase}(\mc{O}^\mc{M})$ never fails, and the \egds \ can be eliminated. Then CQA can be correctly done with
the extensional database and the \tgds, which can de done in polynomial time in data. \boxtheorem}

\section{Contextual Data Quality Specification and Extraction} \label{sec:fw}

The use of the \omd \ model for quality data specification and extraction generalizes a previous approach to- and work on context-based data quality assessment and extraction \cite{bertossi-brite,bertossi16}, which was briefly described in Section \ref{sec:intr}. The important new element in comparison to previous work is the presence in an ontological context $\mc{O}^c$ as in Figure \ref{fig:frm}  of the {\em core} multi-dimensional (MD) ontology $\mc{O}^M$ represented by an \omd \ model as introduced in Section \ref{sec:omd}.

In the rest of this section we show in detail the components and use of an \md \ context in quality data specification and extraction, for which we refer to Figure \ref{fig:frm}. For motivation and illustration we use a running example that extends those  in Sections~\ref{sec:intr} and \ref{sec:omd}.

 On the LHS of Figure~\ref{fig:frm}, we find a database instance, $D$, for a relational schema $\schema=\{R_1,...,R_n\}$. The goal is to specify and extract quality data from $D$. For this we use
the contextual ontology $\mc{O}^c$ shown in the middle, which contains the following elements and components:

\begin{enumerate}[(a)]
  \item Nickname predicates $R'$ in a nickname schema $\schema'$ for predicates $R$ in $\schema$. These are copies of the predicates for $D$ and are populated exactly as in $D$, by means of the simple mappings (rules) forming a set  $\Sigma'$ of \tgds, of the form:

\vspace{-4mm}
\begin{align}
R(\vectt{x})~\rightarrow~R'(\vectt{x}).\label{eq:exp}
\end{align}
\vspace{-4mm}

\noindent whose enforcement producing a material or virtual instance $D'$ within $\mc{O}^c$.

  \item The core MD ontology, $\mc{O}^\mc{M}$, as in Section \ref{sec:omd},  with an instance $I^\mc{M}=D^\mc{H}\cup I^c$, a set $\Sigma^\mc{M}$ of dimensional \tgds, and a set $\kappa^\mc{M}$ of dimensional constraints, among them \egds \ and \ncs.

      \item  There can be, for data quality use, {\em extra contextual data} forming an instance $E$, with schema $\mc{R}^E$,  that is not necessarily part of (or related to)    the \omd \ ontology $\mc{O}^\mc{M}$. It is shown in Figure \ref{fig:frm} on the RHS of the middle box.



\item A set of {\em quality predicates}, $\mc{P}$, with their definitions as Datalog rules forming a set $\Sigma^\mc{P}$ of \tgds. They may be defined in terms of  predicates in $\mc{R}^E$, built-ins, and  dimensional predicates in $\mc{R}^\mc{M}$. \red{We will assume that quality predicates in $\mc{P}$ do not appear in the core dimensional ontology $\mc{O}^M$ that defines the dimensional predicates in $\mc{R}^\mc{M}$. As a consequence, the program defining quality predicates can be seen as a ``top layer", or top sub-program, that can be computed after the core (or base) ontological program has been computed.}\footnote{\  \red{This assumption does not guarantee that the resulting, combined ontology has the same syntactic properties of the core \md \ ontology, e.g. being \WS \ (cf. Example~\ref{ex:nws}), but the analysis of the combined ontology becomes easier, and in some cases it allows us to establish that the combination inherits the good computational properties from the MD ontology. We could allow definitions of quality predicates in Datalog with stratified negation ($\nit{not}$) or even in \dplus. In the former case, the complexity of CQA would not increase, but in the latter we cannot say anything general about the complexity of CQA.}} For a quality predicate $P \in \mc{P}$, its definition of the form:

\vspace{-4mm}
\begin{align}
\varphi_P^E(\vectt{x}),\varphi^\mc{M}_P(\vectt{x})~\rightarrow~P(\vectt{x}).\label{frm:qpd}
\end{align}
\vspace{-4mm}

\noindent Here, $\varphi_P^E(\vectt{x})$ is a conjunction of atoms with predicates in \bl{$\mc{R}^E$} or plus built-ins, and $\varphi^\mc{M}_P(\vectt{x})$ is a conjunction of atoms with predicates in $\mc{R}^\mc{M}$.\footnote{\ We could also have predicates from $\mc{P}$ in the body if we allow mutual or even recursive dependencies between quality predicates.}

Due to their definitions, quality predicates in the context can be syntactically told apart from dimensional predicates. Quality predicate reflect application dependent, specific quality concerns.
\end{enumerate}

\ignore{
 \comlb{You had an example in the middle of the def. of quality predicates saying ``... in terms of predicates  (\red{e.g.} \nit{WorkingScheduels} and \nit{Personnel} in Example~\ref{ex:case})". Put all that in the example.}
\comlb{This is good place to start with the running example, illustrating what we have so far.}
\commos{Here is the example about the elements 1-4 in the context.}
}

\begin{example}\label{ex:frw1} (ex.~\ref{ex:intr} and \ref{ex:md-relational} cont.) Predicate for $\nit{Temperatures} \in \schema$, the initial schema, has $\nit{Temperatures}' \in \schema'$ as a nickname, and defined by
$\nit{Temperatures}(\bar{x}) \rightarrow \nit{Temperatures}'(\bar{x})$. The former has Table \ref{tab:temperatures} as extension in instance $D$, which is under quality assessment, and with this rule, the data are copied into the context.

The core \md \ ontology $\mc{O}^\mc{M}$  has \nit{WorkSchedules} and \nit{Shifts} as categorical relations, linked to the {\sf Hospital} and {\sf Temporal} dimensions (cf. Figure \ref{fig:omdm0}). \ignore{\nit{WardUnit}, \nit{TimeDay} are child-parent relations in the , resp.} $\mc{O}^\mc{M}$ has a set of dimensional \tgds, $\Sigma^\mc{M}$, that includes $\sigma_1$ and $\sigma_2$, and also a
dimensional rule defining a categorical relation \nit{WorkTimes}, as a view in terms of {\it WorkSchedules} the  \nit{TimeDay} child-parent dimensional relation, to create data from the day level down to the time (of the day) level:

\vspace{-4mm}
\begin{align}
{\it WorkSchedules}(u,d;n,s),{\it TimeDay}(t,d) &\rightarrow \nit{WorkTimes}(u,t;n,s).\label{frm:dr1}
\end{align}
\vspace{-3mm}

 $\mc{O}^\mc{M}$ also has a  set $\kappa^\mc{M}$ of {\em dimensional constraints}, including the dimensional \nc \ and \egd, (\ref{frm:expnc}) and (\ref{frm:expegd}), resp.

Now, in order to address data quality concerns, e.g.  about certified nurses or thermometers, we introduce quality predicates, e.g. ${\it TakenWithTherm}$, about times at which nurses use certain thermometers, with a definition of the form (\ref{frm:qpd}):

\vspace{-4mm}
\begin{align}
\hspace{-1cm}\nit{WorkTimes}(\bl{{\sf intensive}},t;n,y) &\rightarrow {\it TakenWithTherm}(t,n,{\sf b1}),\hspace{-7mm}\label{eq:qua1}
\end{align}
\vspace{-5mm}

 \noindent which captures the guideline about thermometers used in \bl{intensive} care units;  and becomes a member of  $\Sigma^\mc{P}$\! (cf. Figure~\ref{fig:frm}).

In this case, we are not using  any contextual database $E$ outside the MD ontology, but we could have an extension for a predicate $\nit{Supply(Ins,Th)} \in \mc{R}^E$, showing thermometer brands ($\nit{Th}$) supplied to hospital institutions ($\nit{Ins}$), in the {\sf Hospital} dimension.\footnote{\ $E$ could represent data brought from external sources, possible at query answering time \cite{bertossi-brite,bertossi16}. In this example, it governmental data about hospital supplies.} It could be used to define (or supplement the previous definition of) {\it TakenWithTherm(t,n,th)}:

\vspace{-4mm}
\begin{align}
\nit{Supply(ins,th),\!UnitInstitution(u,ins)},\!\nit{WorkTimes}(u,t;n,y) \rightarrow {\it TakenWithTherm(t,n,th)}.\label{eq:second}
\end{align}
\boxtheorem\end{example}

Now the main idea consists in using the data brought into the context via the nickname predicates and all the contextual elements to specify quality data for the original schema $\mc{R}$, as a quality alternative to instance $D$.

\begin{itemize}
\item[(e)]
We introduce a ``quality schema", $\mc{R}^q$, a copy of schema $\mc{R}$, with a predicate $R^q$ for each predicate $R \in \mc{R}$. These are
{\em quality versions} of the original predicates. They are defined, and populated if needed, through {\em quality data extraction rules} that form a set, $\Sigma^q$ (cf. Figure ~\ref{fig:frm}), of \da \ rules of the form:

\vspace{-4mm}
\begin{align}
R'(\vectt{x}),\psi^\mc{P}_{R'}(\vectt{x})~\rightarrow~R^q(\vectt{x}).\label{frm:qvd}
\end{align}
\vspace{-4mm}

\noindent Here, $\psi^\mc{P}_{R'}(\vectt{x})$ is an {\em ad hoc} for predicate $R$  conjunction of  quality predicates (in $\mc{P}$) and built-ins. The connection with the data in the corresponding original predicate is captured with the join with its nickname predicate $R'$.\footnote{\ As in the previous item, these definitions could be made more general, but we keep them like this to fix ideas. In particular,  $R^q$ could be defined not only in terms of $R$ (or its nickname $R'$), but also from other predicates in the original (or, better, nickname) schema.}
\end{itemize}

Definitions of the initial predicates' quality versions impose conditions corresponding to user's data quality profiles, and their extensions form the quality data (instance).

\begin{example} \label{ex:framework} (ex.~\ref{ex:frw1} cont.) The quality version of the original predicate $\nit{Temperatures}$ is ${\it Tempera}$- \ ${\it tures}^q \in \schema^q$,  defined by:
\begin{align}
{\it Temperatures}'(t,p,v,n),{\it TakenWithTherm}(t,n,{\sf b1}) ~\rightarrow~ {\it Temperatures}^q(t,p,v,n),\label{frm:expm3}
\end{align}
imposing extra quality conditions on the former. This is a definition of the form (\ref{frm:qvd}) in $\Sigma^q$ (cf. also Example \ref{ex:quality}).
\boxtheorem\end{example}

\ignore{
Notice that the connection between the quality versions in $\mc{R}^q$, categorical relations in $\mc{O}^\mc{M}$, and contextual relations in \bl{$E$} is through quality predicates $\mc{P}$\red{. Since} the latter are defined by general and flexible rules, through \red{them we can also} access the ontology $\mc{O}^\mc{M}$ and the contextual instance \bl{$E$}.

The external sources $\mc{E}=\{E_1,...,E_j\}$ are of different types and contribute with data to the contextual schema. These data can be materialized and stored at the context level by the contextual instance \bl{$E$}, or left at the sources and accessed through mappings. }

\subsection{Computational Properties of the Contextual Ontology}\label{sec:ext}

\ignore{\comlb{PLEASE REVISE THIS SECTION. MAYBE SHORTEN TOO.}

\comlb{We have to say something short about the preservation of good computational properties of the \omd context when extra things are added (we had something in the thesis). Actually, it is better that we summarize here the assumptions about the extra rules, to make the algorithm in the next section more clear.}

\comlb{I brought here what was at the end of the next section. Better dispatch this subject here.}

\commos{I modified the part in blue to explain the assumptions and the effect of them.}   }

In Section~\ref{sec:complexity}, we studied the computational properties of \md \ ontologies without considering additional rules defining quality predicates and quality versions of  tables in $D$. In this regard,
 it may happen that the combination of \dpm \ ontologies that enjoy good computational properties may be an ontology without such properties~\cite{baget11ai,baget15}. Actually, in our case,
 the contextual ontology may not preserve the syntactic properties of the core \md \ ontology.

\begin{example} \label{ex:nws} (ex.~ \ref{ex:ranks} and \ref{ex:framework} cont.) To the \WS \ MD ontology containing the dimensional rules $\sigma_1$, $\sigma_2$. we can add a non-recursive \da \ rule defining  a quality predicate ${\it SameShift}({\it Ward},{\it Day};{\it Nurse}_1,$ \ ${\it Nurse}_2)$ saying that ${\it Nurse}_1$ and ${\it Nurse}_2$ have the same shifts at the same ward and on the same day:

\vspace{-4mm}
\begin{align*}
\sigma_3\!: \ {\it Shifts}(w,d;n,s), {\it Shifts}(w,d;n',s) \ \rightarrow \ {\it SameShift}(w,d;n,n').
\end{align*}
\vspace{-4mm}

\noindent Now, $\Sigma=\{\sigma_1,\sigma_2,\sigma_3\}$ is not \WS \ since variable $s$ in the body of $\sigma_3$ is a repeated marked body variable only appearing in infinite-rank position ${\it Shifts}[4]$. This shows that the even the definition of a quality predicates in plain \da \ may break the WS \ property. \boxtheorem \end{example}

 Under our layered (or modular) approach \red{(cf. item (d) at the beginning of this section)}, according to which  definitions in $\Sigma^\mc{P}$ and $\Sigma^q$ belong to Datalog programs that call predicates defined in the MD ontology $\mc{O}^M$ as  extensional predicates, we can guarantee that the good computational properties
 of the core MD ontology still hold for the contextual ontology $\mc{O}^c$. \red{In fact, the top Datalog program can be computed in terms of CQs and iteration starting from extensions for the dimensional predicates. In the end, all this can be done in polynomial time in the size of the initial extensional database.} The data in the non-dimensional, contextual, relational instance $E$ are also called as extensional data by  $\Sigma^\mc{P}$ and $\Sigma^q$. Consequently, this is not a source of additional complexity. Thus, even when weak-stickiness does not hold for the combined contextual ontology, CQA is still tractable.

\subsection{Query-Based Extraction of Quality Data}\label{sec:extrac}

In this section we present a methodology to obtain quality data through the context on the basis of data that has origin in the initial instance $D$. The approach is query based, i.e. queries are posed to the contextual
ontology $\mc{O}^c$, and in its language. In principle, any query can be posed to this ontology, assuming one knows its elements. However, most typically a user will know about $D$'s schema $\mc{R}$ only, and the (conjunctive) query, $\mc{Q}$, will be expressed in  language $\mf{L}(\mc{R})$, but (s)he will still expect quality answers. For this reason, $\mc{Q}$ is rewritten into a query $\mc{Q}^q$, the {\em quality version} of $\mc{Q}$,  that is obtained by replacing
every predicate $R \in \mc{R}$ in it by its quality version $R^q$ (notice that $\mc{Q}^q$ is also conjunctive). This idea leads as to the following notion of {\em quality answer} to a query.

\begin{definition}\label{rm:qqa} Given instance $D$ of schema $\mc{R}$ and a conjunctive query $\mc{Q} \in \mf{L}(\mc{R})$, a sequence of constants $\bar{c}$ is a {\em quality answer} to $\mc{Q}$
from $D$ via $\mc{O}^c$ iff \ $\mc{O}^c \models \mc{Q}^q[\bar{c}]$, where $\mc{Q}^q$ is the quality version of $\mc{Q}$, and $\mc{O}^c$ is the contextual ontology containing the MD ontology $\mc{O}^M$, and into which $D$ is mapped via rules (\ref{eq:exp}). \ $\nit{Q\!Ans}(\mc{Q},D,\mc{O}^c)$ denotes the set of quality answers to $\mc{Q}$ from $D$ via $\mc{O}^c$. \boxtheorem
\end{definition}

\ignore{
\vspace{-4mm}
\begin{align*}\red{
{\it QAns}^{\mf{C}}_D(\mc{Q}) \ =\ \{\bar{c}\;|\; D \cup \Sigma' \cup \mc{M} \cup \bl{E} \cup \Sigma^\mc{P} \cup \Sigma^q  \
\models \ \mc{Q}^q[\bar{c}]\}.}
\end{align*}
\vspace{-4mm}

This formulation of clean answers corresponds to a {\em model-theoretic \red{definition} of clean answers}. Note that in this formulation, $\mc{O}^\mc{M}$ can have multiple models that can only be represented by a canonical model \red{(the chase)} for the purpose of computing certain answers to \cq s.\boxtheorem\end{remark}
}

A particular case of this definition occurs when the query is an open atomic query, say $\mc{Q}\!: \ R(\bar{x})$, with $R \in \mc{R}$.  We could define the {\em core quality version of} $D$, denoted by $\nit{Core}^q(D)$, as the database instance for schema $\mc{R}$ obtained by collecting  the quality answers for these queries:
\begin{equation}
\nit{Core}^q(D) := \{R(\bar{c}) ~|~ \mc{O}^c \models R^q[\bar{c}] \mbox{ and } R \in \mc{R}\}. \label{eq:core}
\end{equation}

 We just gave a model-theoretic definition of quality answer. Actually, a clean  answer to a query holds in every quality instance in the class  $\nit{Qual}(D,\mc{O}^c)$ (cf. Figure \ref{fig:contOnto}).
  This semantic definition has a computational counterpart: quality answers can be obtained by conjunctive query answering from ontology $\mc{O}^c$\!, a process that in general will inherit the  good computational properties of the MD ontology $\mc{O}^M$, as discussed earlier in this section.

In the rest of this section, we describe the \CQQA \ algorithm (cf. Algorithm~\ref{ag:qualityalg}), given a CQ $\mc{Q} \in \mf{L}(\mc{R})$ and a contextual ontology $\mc{O}^c$ that imports data from instance $D$,  computes $\nit{Q\!Ans}(\mc{Q},D,\mc{O}^c)$. The assumption is that we have an algorithm for CQA from the MD ontology $\mc{O}^M$. If it is a weakly-sticky \dpm \ ontology, we can use the chase-based algorithm introduced in \cite{milani16rr}.\footnote{Actually the algorithm applies to a larger class of \dpm \ ontologies, that of {\em join-weakly-sticky} programs that is closed under magic-sets optimizations \cite{milani16rr}.} We also assume that a separability check takes place before calling the algorithm
(cf. Sections \ref{sec:withCons} and \ref{sec:inco}).

 For the unfolding-based steps~2 and 3, we are assuming the predicate definitions in $\Sigma^\mc{P}$ and $\Sigma^q$ are given in non-recursive Datalog.\footnote{\ If they are more general, but under the modularity assumption of Section \ref{sec:ext}, we do not unfold, but do first CQA on the Datalog programs defining the top, non-dimensional predicates, and next, when the ``extensional" dimensional predicates have to be evaluated, we call the algorithm for CQA for the MD ontology.}
 Starting from the CQ $\mc{Q}^q$, unfolding results into a union of conjunctive queries (\ucq s) (a union in the case predicates are defined by more than one Datalog rule).
  Next, according to Step~4,  each (conjunctive) disjunct of $\mc{Q}^\mc{M}$ can be answered by the given algorithm for CQA from $\mc{O}^\mc{M}$ with extensional data in $E$ and $D'$ (the latter obtained by importing $D$ into context $\mc{O}^c$). The algorithm can be applied in particular to compute the core clean version, $\nit{Core}^q(D)$, of  $D$.

  \ignore{More complex rules \red{than non-recursive \da \ } in $\Sigma^\mc{P}$ and $\Sigma^q$ require different \qa \ approaches, that might affect other steps. In particular, $\mc{Q}^\mc{M}$ in Step~4 might be a more complex query \red{than} a \ucq, for which \qa \ \red{may} not be done by \red{just} answering \cq s \red{over} $\mc{O}^\mc{M}$, in addition to $D'$ and \bl{$E$}.}

\ignore{Regarding \qa \ under $\mc{O}^\mc{M}$ in Step~4, we \red{can} use the algorithm in~\cite{milani16rr-cali,milani16rr}. This algorithm is a chase-based \qa \ algorithm that imposes \cq s on a canonical model that represents multiple models of the ontology. This means, in general, there are multiple clean instances, $\mc{D}^q$, for which we implicitly use certain answers for quality query answering, by utilizing the \qa \ algorithm under the \md \ ontology in \CQQA.}


\vspace{3mm}
\begin{algorithm}[t]
{\bf Step~1:} Replace each predicate $R$ in $\mc{Q}$ with its corresponding quality version $R^q$, obtaining a \cq \ $\mc{Q}^q$ over schema $\mc{R}^q$.

\vspace{2mm}
{\bf Step~2:} Unfold in $\mc{Q}^q$ the definitions of quality-version predicates $R^q$ given by the rules (\ref{frm:qvd}) in $\Sigma^q$.  Obtain a  \ucq \ $\mc{Q}^c$ in terms of predicates in $\mc{R}'\cup\mc{P}$ and built-ins.

\vspace{2mm}
{\bf Step~3:} Unfold in $\mc{Q}^c$ the definitions of quality predicates given by the rules  (\ref{frm:qpd}) in $\Sigma^\mc{P}$. Obtain a \ucq \ $\mc{Q}^\mc{M}$ in terms of predicates in $\mc{R}'\cup\bl{\mc{R}^E}\cup\mc{R}^\mc{M}$, and built-ins.

\vspace{2mm}
{\bf Step~4:} Answer $\mc{Q}^\mc{M}$ by CQA (for each of $\mc{Q}^\mc{M}$'s disjuncts) over the extensional database $E \cup D'$ and  the \md \ ontology $\mc{O}^\mc{M}$.
\caption{The \CQQA \ algorithm }
\label{ag:qualityalg}
\end{algorithm}


\begin{example} \label{ex:framework-qa} (ex.~\ref{ex:framework} cont.)  The initial query in (\ref{frm:q}), asking for (quality) values for Tom Waits' temperature, is, according to Step~1 of \CQQA,  first rewritten into:
{\small $$\mc{Q}^q(v)\!: \exists n\;\exists t\;({\it Temperatures^q(t,{\sf tom\;waits},v,n)} \; \wedge \; {\sf 11\!:\!45\mbox{-}aug\mbox{/}21\mbox{/}2016} \le t \le {\sf 12\!:\!15\mbox{-}aug\mbox{/}21\mbox{/}2016}),$$}

\vspace{-4mm} \noindent which can be answered using (\ref{frm:expm3}) to unfold according to Step~2 of \CQQA, obtaining:

\vspace{-4mm}
\begin{align*}
\mc{Q}^c(v)\!:\;\exists n\;\exists t\;(&\nit{Temperatures}'(t,{\sf tom\;waits},v,n) \wedge \nit{TakenWithTherm}(t,n,{\sf b1}) \ \wedge\\
                     &~~~~~~~~~~~~~~~~~~~~~~~~~~~~~~~~~~~{\sf 11\!:\!45\mbox{-}aug\mbox{/}21\mbox{/}2016} \le t \le {\sf 12\!:\!15\mbox{-}aug\mbox{/}21\mbox{/}2016}).
\end{align*}
\vspace{-4mm}

Step~3 of \CQQA \ uses the quality predicate definition (\ref{eq:qua1}) for unfolding, obtaining the query:

\vspace{-4mm}
{\small \begin{align*}
\mc{Q}^{\mc{M}}(v)\!:\exists n\;\exists t\;\exists y\;(&\nit{Temperatures}'(t,{\sf tom\;waits},v,n)  \wedge  \nit{WorkTimes}({\sf intensive},t;n,y) \ \wedge \\
&~~~~~~~~~~~~~~~~~~~~~~~~~~~~~~~~~~~~~{\sf 11\!:\!45\mbox{-}aug\mbox{/}21\mbox{/}2016} \le t \le {\sf 12\!:\!15\mbox{-}aug\mbox{/}21\mbox{/}2016}),
\end{align*}}

\vspace{-4mm}
\noindent expressed in terms of $\nit{Temperatures}'$, predicates in $\mc{R}^\mc{M}$, and built-ins.

Finally, at Step~4 of \CQQA, $\mc{Q}^\mc{M}$ is answered as a CQ over $\mc{O}^\mc{M}$ and database $D'$,\footnote{\ The predicates in the nickname schema $\mc{R}'$ act as extensional predicates at this point, without creating any computational problems.} using, for example, the \qa \ algorithms in~\cite{milani16rr-cali,milani16rr}.

\red{Predicate unfolding may produce a UCQ rather than a \cq. For example, if we unfold predicate {\it TakenWithTherm} according to both definitions (\ref{eq:qua1}) and (\ref{eq:second}), we obtain the following UCQ:}

\red{\vspace{-4mm}
{\small \begin{align*}
\mc{Q}^{\mc{M}}(v)\!:\ \ \ \exists n\;\exists t\;\exists y\;(&\nit{Temperatures}'(t,{\sf tom\;waits},v,n)  \wedge  \nit{WorkTimes}({\sf intensive},t;n,y) \ \wedge \\
&~~~~~~~~~~~~~~~~~~~~~~~~~~~~~~~~~~~~~{\sf 11\!:\!45\mbox{-}aug\mbox{/}21\mbox{/}2016} \le t \le {\sf 12\!:\!15\mbox{-}aug\mbox{/}21\mbox{/}2016}) \ \ \ \ \ \vee \\
\hspace*{5mm}\exists i\;\exists n\;\exists t\;\exists u\;\exists y\;(&\nit{Temperatures}'(t,{\sf tom\;waits},v,n)  \wedge  \nit{Supply}(i,{\sf b1}) \wedge \nit{UnitInstitution}(u,i) \ \wedge\\
& \nit{WorkTimes}(u,t;n,y) \wedge {\sf 11\!:\!45\mbox{-}aug\mbox{/}21\mbox{/}2016} \le t \le {\sf 12\!:\!15\mbox{-}aug\mbox{/}21\mbox{/}2016}). 
\end{align*}}}


\vspace*{-1cm}\boxtheorem\end{example}

\ignore{
\red{Under} our approach, data cleaning \red{(or extraction of quality data from an initial table)} amounts to obtaining a clean instance $D^q$ from the dirty target instance $D$, and it is done by collecting clean extensions $R_1^q,...,R_n^q$ \red{of $R_1,...,R_n \in \mc{R}$}. The clean extension $R^q$ of possibly dirty relation $R$ in $D$ is obtained by answering the atomic query $\mc{Q}(\vectt{x})\!:R(\vectt{x})$ using \CQQA. In particular, the algorithm uses a rule of the general form~(\ref{frm:qvd}) to collect the clean data of $R^q$ by applying conditions in $\psi^\mc{P}_{R'}$ on $R'$.
}

\section{Discussion and Conclusions} \label{sec:cfw}

In this paper, we started from the idea that data quality is context-dependent. As a consequence, we needed a formal model of context for context-based data quality assessment and quality data extraction. For that, we followed and extended the approach in~\cite{bertossi-brite,bertossi16}, by proposing ontological contexts, and embedding multidimensional (MD) data models in them.
For the latter, we took advantage of our relational reconstruction of the \hm \ data model~\cite{hurtado-pods,hurtado-acm}.

 The MD data model was extended with categorical relations, which are linked to categories at different levels of dimension hierarchies, and also with dimensional constraints rules. The latter add the capability of navigating multiple dimensions in both upward and downward directions. Although not shown here (but cf. \cite{milaniThesis}), it is possible to include in the ontological contexts
semantic constraints usually present in  the \hm \ model, such as {\em strictness} and {\em homogeneity},\footnote{\ A dimension is {\em strict} when every category member has at most one parent in each higher category. It is {\em homogeneous} (a.k.a. covering) when every category member has at least one parent in each parent category.} which guarantee {\em summarizability} (or aggregation) for the correct computation of cube views~\cite{hurtado-pods}.

We represented \md \ ontologies using the \dpm \ ontological language, and we showed that they fall in the syntactic class of \WS \ \dpm \ programs, for which CQA is tractable. We also unveiled conditions under which separability of \egds \ and rules holds.

We used and extended  the \md \ ontologies with rules for data quality specification and extraction, and proposed a general methodology for quality data extraction via query answering.
Our underlying approach to data quality is that the given database instance does not have all the necessary elements to assess the quality of data or to extract quality data. The context is used for that purpose, and provides additional information about the origin and intended use of data. Notice that from this point of view, our contexts  can also be seen as enabling {\em tailoring and focusing} of given data for a particular application. This is a idea that deserves additional investigation.

\red{Our approach to quality data specification and extraction is declarative~\cite{geerts,bertossi13}. It uses logic-based languages, namely relational calculus, \da \ and \dpm to specify quality data. These languages have a precise and clear semantics and their scope of applicability can be easily analyzed. It is also independent of any procedural mechanism for quality data extraction and data cleaning, but computational methods can be extracted from (or be based on) the specifications}

\red{The implementation of the \CQQA \ algorithm and experiments to evaluate its performance correspond to ongoing work. The algorithm and its optimization is based on our work on \qa \ under \WS \ programs~\cite{milani16rr,milani16rr-cali}.}

Some important possible extensions of- and issues about our \omd \ data model that deserve further investigation, have to do with: (a) Having categorical attributes in categorical relations forming a key. \ (b) Adopting and using a {\em repair semantics} when the MD ontology becomes inconsistent. \ (c) Analyzing and implementing data quality extraction as a data cleaning or repair problem. \ (d) Allowing some predicates to be closed and the related problem of non-deterministic or uncertain value invention, mainly for downwards navigation. We briefly elaborate on each of them in Sections
\ref{sec:catKeys}, \ref{sec:inco}, \ref{sec:repairs} and \ref{sec:clo}, respectively. \red{They correspond all to open areas of research. Hence the speculative style of the discussion.}

\subsection{Categorical Keys}\label{sec:catKeys}

In our running example, the categorical relation \nit{WorkSchedules(Unit,Day;Nurse,Speciality)}, does not have $\{\nit{Unit},\nit{Day}\}$ as a key: multiple nurses might have work schedules in the same unit and on the same day.
However, in many applications it may make sense to have the categorical attributes forming a key for a categorical relation. (For example, in the
\hm \ model, the non-measure attributes in a fact-table form a key.) This is not required by the basic \omd \ model, and such a key constraint has to be added.

\ignore{
There are still cases when categorical attributes define key attributes in categorical relations. For example in \nit{InstitutionBoard(Institution;Chair,President,CEO)}, if there is only one board directory (i.e. chair, ceo, and president) for an institution, then attribute \nit{Institution} is a key attribute for the relation.

Here, we discuss the effect of considering {\em categorical keys}, i.e. making categorical attributes form a key for their categorical relations. More precisely, we}

 If we assume that in a categorical relation $R(C_1,...,C_n;A_1,...,A_m)$, \ $\{C_1,...,C_n\}$ is a key for $R$, we have to include  \egds \ in the \md \ ontology, one for each pair $y_i \in \vectt{y}$, $y'_i \in \vectt{y}'$:

\vspace{-4mm}
\begin{align}
R(\vectt{x};\vectt{y}),R(\vectt{x};\vectt{y}')~\rightarrow~ y'_i=y_i.\label{frm:key2}
\end{align}
\vspace{-4mm}

We can use our running example to show that dimensional rules and categorical keys of the form (\ref{frm:key2}) may  not be separable (cf. Section~\ref{sec:withCons}).

\begin{example} \label{ex:non-sep-ex}Consider the categorical relation \nit{InstitutionBoard(Institution;Chair,} \nit{President,CEO)} with \nit{Institution} as a key. In particular, we have the \egd:

\vspace{-4mm}
\begin{align*}
\nit{InstitutionBoard}(i;c,p,e),\nit{InstitutionBoard}(i;c',p',e')~\rightarrow~ c'=c.
\end{align*}
\vspace{-4mm}

We also have the dimensional rules (they differ on the underlined $\exists$-variables on the RHS):

\vspace{-4mm}
\begin{align}
\nit{PatientUnit}(u,d;p),\nit{UnitInstiution}(u,i) \rightarrow \exists c\;\exists n\;\nit{InstitutionBoard}(i;c,\underline{c},n).\label{frm:key1e}\\
\nit{PatientUnit}(u,d;p),\nit{UnitInstiution}(u,i) \rightarrow \exists c\;\exists n\;\nit{InstitutionBoard}(i;c,\underline{n},n).\label{frm:key2e}
\end{align}
\vspace{-4mm}

Let $({\sf standard},{\sf sep/5};{\sf tom\; waits})$ be the only tuple in the extension of \nit{PatientUnit}. The \egds \ defining \nit{Institution} as a key  are not separable from the dimensional \tgds \ (\ref{frm:key1e}) and (\ref{frm:key2e}) (cf. Section \ref{sec:withCons}), because: (a) the chase does not fail since the \egds \ only equate nulls invented by (\ref{frm:key1e}) and (\ref{frm:key2e}), and (b) the \bcq \ $\mc{Q}: \exists i\;\exists c\;\nit{InstitutionBoard}(i,c,c,c)$ has a negative answer without the categorical key, but a positive answer with the categorical key. Actually, the combination of \tgds \ and \egds \ here is {\em conflicting} (cf. Section \ref{sec:withCons}), because no $\exists$-variable in (\ref{frm:key1e}) or (\ref{frm:key2e}) appears in a key position, and then the \tgds \ may generate different tuples with the same key values. \boxtheorem \end{example}

Despite the possible non-separability introduced by categorical keys, CQA is still in \ptime \ in data complexity, because no null value appears in categorical positions. As a consequence, there are polynomially many (in the size of data) key categorical  values. There are also polynomially many \tgd-chase steps, including those that are applicable after the \egd-chase steps or due to non-separability. This shows that the chase procedure, with \tgd- and \egd-chase steps (as we explained in Sections~\ref{sec:dpm}) runs in polynomial time for an \md \ ontology under categorical keys, and CQA \ can be done on the resulting chase instance.

\ignore{
 \comlb{NEW: Better give a bit more details for what follows. What kind of chase? Why does it stop? In failure or not? What's the size of the chase? So, the algorithm is basically querying the chase when there is no failure?}
 \comlb{So, how would be the algorithm in this case? Can we say more? In particular, anything that can be reused rom our RR paper (not with Cali)? }
 \commos{Here, we lose separability, but it does not matter because there at finitely many keys (made of categorical values) and the chase always stops. \qa \ is simple as we have terminating chase. Basically, separability only matters when there is infinite chase and we would like to guarantee the egds do not interact with this infinite chase and destroy the good computational properties of the infinite chase.}  }

\begin{proposition}\em \label{prop:key} The data complexity of CQA on \md \ ontologies with categorical keys is in \ptime.\boxtheorem\end{proposition}

\subsection{Inconsistent MD Ontologies}\label{sec:inco}

 We discussed in Section~\ref{sec:withCons}  the presence of dimensional \ncs \ and \egds \ may lead to an inconsistent MD ontology (cf. Section \ref{sec:dpm}). In this case, the ontology can be {\em repaired}  according to an {\em inconsistency-tolerant semantics}, so that it still gives semantically meaningful and non-trivial answers to queries under inconsistency. A common approach to DL or \dpm \ ontology repair has been based on repairing the extensional database in the case of \dpm \ \cite{lukasiewicz12}, and the A-Box in the case of DL~\cite{lembo10,rosati,bienvenu14,lembo15,bienvenu16} ontologies.

 According to this semantics, a repair of an inconsistent ontology $\mc{O}$ including an extensional instance $I$, is a consistent ontology with the same rules and constraints as $\mc{O}$, but with an extensional instance  $I'$ that is maximally contained in $I$. The consistent answers to a query posed to $\mc{O}$ are those answers shared by all the repairs of the latter.
\qa \ under this semantics is \np-hard in the size of $I$, already for DL \cite{lembo15} or \dpm \ ontologies \cite{lukasiewicz12,lukasiewicz15} with relatively low expressive power.

Repairing the inconsistent ontology by changing the extensional instance amounts, in the case of an MD ontology $\mc{O}^M$, to possibly changing the MD instance. In this regard, we might want to respect the MD structure of data, in particular, semantic constraints that apply at that level, e.g. enforcing summarizability constraints mentioned earlier in this section. Repairs and consistent answers from MD databases have been investigated in
 \cite{amw09,sinaAMW,sina}, and also in \cite{maka}, which proposes  {\em the path  schema} for MD databases as a better relational schema for dealing with the kinds of inconsistencies that appear in them.

\subsection{Quality Data Extraction as Inconsistency Handling}\label{sec:repairs}

As pointed out in Section \ref{sec:intr}, context-based quality data extraction is reminiscent of database repairing and consistent query answering
\cite{bertossi11,bertossi06}. Actually, we can reproduce from our context-based approach to data cleaning a scenario where cleaning can be seen as consistent query answering.

The initial database $D$ may not be subject to integrity constraints.\footnote{\ We have developed this case, but in principle we could have constraints on $D$, satisfied or not, and they could be mapped into the context for combination with the other elements there.} However, as we can see  in Example \ref{ex:framework}, the rule (\ref{frm:expm3}) could be seen as a rewriting of the query $\mc{Q}(t,p,v,n)\!: \  {\it Temperatures}'(t,p,v,n)$, performed
to obtain the consistent answers (or consistent contents of {\it Temperatures}' in this case) w.r.t. the contextual inclusion dependency  \ $\psi\!: \ {\it Temperatures}'(t,p,v,n) \rightarrow {\it TakenWithTherm}(t,n,{\sf b1})$.
The rewriting reflects a {\em repair semantics} based on deletions of tuples from  ${\it Temperatures}'$ when the constraint is not satisfied \cite{bertossi11}. That is, predicate $\nit{TakenWithTherm}(t,n,{\sf b1})$ acts as a filter on
predicate ${\it Temperatures}'$.

The quality version $\nit{Core}^q(D)$ of the initial instance $D$, defined in (\ref{eq:core}), can be seen then as the intersection of all repairs of $D$ w.r.t. these contextual constraints (more precisely, as the intersection
of the instances in $\nit{Qual}(D,\mc{O}^c)$).
Doing quality (or consistent) query answering directly from the intersection of all repairs is sound, but possibly incomplete. However, this has been a predominant approach to OBDA with inconsistent ontologies~\red{\cite{lembo15,lukasiewicz12,lukasiewicz15}}: the ontology is repaired by repairing the extensional instance and considering the intersection of its repairs (cf. also Section \ref{sec:inco}).

In Section \ref{sec:intr} we characterized our context-based approach to data quality mainly as one confronting incompleteness of data. However, we can also see it as addressing inconsistency w.r.t. constraints imposed at the contextual level rather than directly at the database level.

\ignore{
\comlb{How can we know if a MD ontology is inconsistent w.r.t. \egds? What kinds of inconsistencies can appear, both with \ncs \ and \egds? The authors of papers on ontology repairs start from the assumption that they know that there are inconsistencies and which they are?\\
More generally, I think the data-repair approach to inconsistent MD ontologies may be a very interesting research topic.}
\commos{Leo, I think generally the constraints have to be continuously be checked during the process of data generation and completion (e.g. chase). In some cases, e.g. ncs and separable egds, we might be able to postpone it to the end of data generation, or sometimes even we might be allowed to check them before starting the chase. In any case, I agree that detecting inconsistency and also the data-repair approaches for ontologies and specifically MD ontologies would be challenging.

Specially, in the data cleaning, I think there is interest in working with richer onotlogies for capturing data quality concerns of any form that immediately leads to repairing the ontologies for cleaning. Basically, data quality specification using ontologies and also cleaning it, which possibly leads to repairing ontologies.}
}

\subsection{\red{Categorical Value Invention and Closed Predicates}}\label{sec:clo}

We assumed in Section \ref{sec:omd} that \tgds \ do not have existential quantifiers on variables for categorical attributes. This has two important consequences. First, the \omd \ programs become weakly-sticky (cf. Proposition~\ref{prop:weaklysticky}); second, we can apply the CWA to categories and categorical attributes (actually, without existential quantifications on categorical attributes, making the CWA or the OWA does not matter for CQA).
Relaxing this condition has two immediate effects on the \md \ ontology: (a) We cannot make the CWA on dimension categories (and categorical attributes) anymore (without violating the \ncs \ in (\ref{frm:referential})); and (b) The set of \tgds \ of an OMD ontology may not be weakly-sticky anymore. The following example shows both issues.

\ignore{
\cwa \ would not hold for the underlying \md \ database since: According to the semantics of \dpm, new values could appear in the positions of categorical values. (b) The dimensional rules would not necessarily be weakly-sticky (Proposition~\ref{prop:weaklysticky} would be falsified). These are shown in the following example.}

\begin{table}[h]
\hspace*{8mm}\begin{minipage}[t]{0.45\linewidth}\centering
\setcounter{rownum}{0}
\setlength{\tabcolsep}{0.3em}
\setlength{\arrayrulewidth}{0.75pt}
\renewcommand*\arraystretch{1.1}
\caption{\it \small DischargePatients}
\vspace{-2mm}
\label{tab:discharge}
\vspace{0mm}
{\footnotesize \begin{tabular}{c|c|c|c|}
\cline{2-4}
 & \textbf{Inst.} & \textbf{Day} & \textbf{Patient}\\
\cline{2-4}
{\tiny \addtocounter{rownum}{1}\arabic{rownum}} & $H_1$ & Sep/9 & Tom Waits \\
\cline{2-4}
{\tiny \addtocounter{rownum}{1}\arabic{rownum}} & $H_1$ & Sep/6 & Lou Reed \\
\cline{2-4}
{\tiny \addtocounter{rownum}{1}\arabic{rownum}} & $H_2$ & Oct/5 & Elvis Costello \\
\cline{2-4}
{\tiny \addtocounter{rownum}{1}\arabic{rownum}} & $H_1$ & Dec/16 & Elvis Costello \\
\cline{2-4}
\end{tabular} }
\end{minipage}
\begin{minipage}[t]{0.45\linewidth}\centering
\setlength{\tabcolsep}{0.3em}
\setcounter{rownum}{0}
\setlength{\arrayrulewidth}{0.75pt}
\renewcommand*\arraystretch{1.1}
\caption{\it \small PatientUnit}
\vspace{-3mm}
\label{tab:punit}
\vspace{1mm}
{\footnotesize \begin{tabular}{c|c|c|c|}
\cline{2-4}
 & \textbf{Unit} & \textbf{Day} & \textbf{Patient}\\
\cline{2-4}
{\tiny \addtocounter{rownum}{1}\arabic{rownum}} & Standard & Sep/5 & Tom Waits \\
\cline{2-4}
{\tiny \addtocounter{rownum}{1}\arabic{rownum}} & Standard & Sep/9 & Tom Waits \\
\cline{2-4}
{\tiny \addtocounter{rownum}{1}\arabic{rownum}} & Intensive & Sep/6 & Lou Reed \\
\cline{2-4}
\end{tabular} }
\end{minipage}
\end{table}

\begin{example} \label{ex:udown} Consider categorical relations {\it DischargePatients} (Table~\ref{tab:discharge}) and {\it PatientUnit} (Table~\ref{tab:punit}), containing data on patients leaving an institution and on locations of patients, resp. Since, a patient was in a unit when discharged, we can use {\it DischargePatient} to generate data for {\it PatientUnit}, at the {\it Unit} level, down from the {\it Institution} level, through the \tgd \ (with a conjunction in the head that can be eliminated),\footnote{\ E.g. with  ${\it DischargePatients}(i,d;p) ~\rightarrow~ \exists u\;{\it TempPatient(i,u,d;p)}$, \
${\it TempPatient(i,u,d;p)} ~\rightarrow$ ~ ${\it UnitInstitution(u,i)}$, \ and \
${\it TempPatient(i,u,d;p)} ~\rightarrow~ {\it PatientUnit(u,d;p)}$.}

\vspace{-4mm}
\begin{align}
\hspace{-3mm}{\it DischargePatients}(i,d;p) ~\rightarrow~ \exists u\;({\it UnitInstitution(u,i)} \wedge {\it PatientUnit(u,d;p)}),\label{frm:nondeter}
\end{align}
\vspace{-4mm}

\noindent which invents values downwards, in the categorical position (for units) $\nit{PatientUnit}[1]$ and in the child-parent predicate \nit{UnitInstitution} in its head. \ignore{The $\exists$-variable $u$ in (\ref{frm:nondeter})  appears in the ``downward" attribute of the child-parent relation {\it UnitInstitution}, which} This may  invent new category members, which could be in conflict with the a CWA applied to \ignore{. According to the semantics of \dpm, this leads to {\em open world assumption} (OWA)~\cite{reiter} for} category predicates, child-parent predicates, and indirectly via the \ncs, to categorical attributes. 

Let's now add the following  \tgds, respectively, saying that every patient eventually leaves the hospital, and defining the patients' relationships of being on a day in the same unit.

\vspace{-4mm}
\begin{align}
\hspace{-3mm}{\it PatientUnit(u,d;p)},{\it UnitInstitution(u,i)} ~&\rightarrow~ \exists d'{\it DischargePatients}(i,d';p),\label{frm:upnondeter}\\
\hspace{-3mm}{\it PatientUnit(u,d;p)},{\it PatientUnit(u,d;p')} ~&\rightarrow~ {\it SameDay}(d;p,p'),\label{frm:nonws}
\end{align}

The set of rules~(\ref{frm:nondeter})-(\ref{frm:nonws}) is not weakly-sticky.\footnote{According to their dependency graph (cf. Section \ref{sec:wa}), ${\it PatientUnit}[1]$ has infinite rank. Rule~(\ref{frm:nonws}) breaks weak-stickiness, because $u$ is a repeated marked variable that appears only in ${\it PatientUnit}[1]$.}\boxtheorem\end{example}

If we accept value invention in OMD ontologies, then their weak-stickiness cannot be guaranteed, and has to be analyzed for each particular ontology. However, the issue raised by the example in relation to the invention of category members still persists.

Sometimes an existential quantifier is used  to  refer  to an unspecified element in a specified set or domain, as a disjunction over its elements. This interpretation of quantifiers is possible
if we have a metalevel CWA assumption or a domain closure axioms \cite{reiter} over (some) predicates, none of which is part of  \dpm.

Recent work in OBDA addresses this problem, allowing the combination of open and closed predicates in
\dpm \ ontologies, but previously tractable CQA may become intractable \cite{ahmet1}. Similar extensions and results hold for light-weight DLs \cite{seylan,franconi,lutz13,lutz15}.

\ignore{
\comlb{We should give the more prominent references for Simkus. I think Lutz has work on closed predicates in DL.}
\commos{Sure, Here I added these references from Lutz (Seylan is also the co-author in both Lutz's and Franconi's papers): \cite{seylan,lutz13,lutz15}. About Simkus, the better reference is~\cite{ahmet2} and I commented the other one.} }

In our case, if we accept value invention for category members, the natural candidates to be declared as closed in the new setting are the
unary category predicates and the child-parent predicates: we do not want to create new category members or new children for a given parent, nor, under upward data propagation,   a new parent for a given child since parents are unique, as captured by the ``upward" \egds \ (\ref{frm:key}), which
 will force the invented nulls to take the given parent values.
More problematic becomes downward data propagation with existential quantifiers over  categorical positions. Even under a closure assumption on child-parent predicates, we may end up creating new children (we  stand for existing do not have any ``downward" \egds).

If we accept  \tgds \ such as (\ref{frm:nondeter}) and  we consider  category predicates and child-parent predicates as closed, then we start departing from the usual \dpm \ semantics, and some of the results we have  for weakly-sticky programs (with \owa \ semantics) have to be reconsidered.

Having existential variables over categorical predicates may lead to new forms of inconsistency, involving category values. Adopting a repair semantic based on changes on the extensional data leads to repairs of a MD database, which should be treated as such and not as an ordinary relational database (cf. Section \ref{sec:inco}).

 As an alternative to existential categorical variables  as choices from  given (possibly closed) sets of values, we could think of using {\em disjunctive} \dpm, with   disjunctive \tgds \ \cite{alviano12,bourhis}, in particular for downward navigation. However, CQA under disjunctive sticky-\dpm \ may be undecidable in some cases  \cite{reasW,morakThesis}. Furthermore, disjunctive rule heads may become data dependant.

\red{\subsection{Related Work} \label{sec:related}}



\ignore{We admit that our work is also related to some other areas of research, such as multidimensional data models.} As a logical extension  of a multidimensional data model, our model is similar in spirit to the {\em data warehouse conceptual data model}~\cite{franconi99} that extends the  {\em entity-relationship} data model with dimensions by means of the expressive description logic (DL) $\mc{ALCFI}$~\cite{horrocks99}. They concentrate on the model and reasoning about the satisfiability of the MD schema and implied statements, but not OBDA. In \cite{malaki}, preliminary work motivated by data quality on specifying MD ontologies in light-weight DLs is reported, without going much into quality aspects or query answering.
\vspace{0.25cm}


\ignore{Our approach to data quality is declarative.} The existing declarative approaches to data quality~\cite{bertossi13} mostly use classic \ics, such as \fds \ and \ideps, and denial constraints (i.e. the \ncs \ of \dpm).  Newer classes of dependencies have been introduced to capture data quality  conditions and directly support data cleaning processes~\citep{fan08}. Examples are {\em conditional dependencies} (conditional \fds \ and \ideps), and {\em matching dependencies}~\citep{fan09-pvldb,fan11}. We claim that more expressive contexts are necessary in real life databases to express stronger conditions and semantics of (quality) data.

Models of context \cite{bolchini-survey} have been proposed and investigated in the  data management and knowledge representation literature.  Concentrating mostly on the former, in the following we briefly  review, in an itemized manner, some of those models. After describing them, we make comparisons with our model of context and its use.

\noindent A. \ {\em Multi Context Systems} (MCS)~\citep{giunchiglia94} and {\em Local Models Semantics} (LMS)~\citep{ghidini01,ghidini} are related logic-based frameworks for formalizing contexts and reasoning with them. MCS provides a  proof-theoretic framework with a hierarchy of \fo \ languages, each of them for knowledge representation and reasoning within a specific context. LMS is a model-theoretic framework based on the principles of {\em locality}, i.e. reasoning uses only part (the context) of what is potentially available; and {\em compatibility} of the kinds of reasoning performed in different contexts.


\noindent B. \ In~\citep{motschnig95,motschnig}, a general framework is proposed based on the concept of {\em viewing} for decomposing {\em information bases} into possibly overlapping fragments, called {\em contexts}, in order to be able to better manage and customize information. Viewing refers to giving partial information on conceptual entities by observing them from different viewpoints or situations.

\noindent C. \ In~\citep{analyti,theodorakis}, a model of contexts in information bases is proposed. A context is defined as a set of objects,  each of them with possibly several names. Operations, such as create, copy, merge, and browse, are defined for manipulating and using contexts. Contextual objects can be structured through traditional abstraction mechanisms, i.e. classification, generalization, and attribution. A theory of contextualized information bases is introduced. It includes a set of validity constraints for contexts, a model theory, as well as a set of sound and complete inference rules.

\noindent D. \ In~\citep{ghidini98},  ideas from~\citep{ghidini01}, specially LMS, are applied  to information integration and federated database management, where each database may have its own local semantics.  {\em LMS for federated databases} is used, as an extension of LMS. A notion of logical consequence between formulae (queries) in different databases is defined, and becomes the theoretical basis for the implementation of algorithms for query answering and their optimization.

\noindent E. \ Context-aware data tailoring~\cite{bolchini-cdt} proposes {\em context dimension trees} (CDTs) for modeling multidimensional aspects of context. It allows sub-dimensions with values of finer granularity. A user's context is modeled as a  ``chunk configuration", consisting of a  set of dimension labels and their assigned values, and is used to specify the relevant portion of a target database for the user. This user's {\em view} is computed by combining the sub-views linked to dimension values~\cite{bolchini-view,bolchini-is}. 

\noindent F. \ In~\citep{martinenghi-vldb} dimensions, as in multidimensional databases, are used for modeling contexts. A context-aware data model is proposed in which the notion of context is implicit and indirectly captured by {\em contextual attributes}, i.e. relational attributes that take as values members of dimension categories. In particular, in a contextual relation the context of a tuple is captured by its values in dimensions, while the categories of these members specify the granularity level of the context. They present a query language that extends the relational algebra, by introducing new operators for manipulating the granularity of contextual attributes.


\noindent G. \ In~\citep{pitoura11,stefanidis11} contexts are used in preference database systems to support context-aware queries whose results depend on the context at the time of their submission. Data cubes are used to store the dependencies between context-dependent preferences, database relations, and OLAP techniques for processing context-aware queries. This allows for the manipulation of the captured context data at various levels of abstraction. \ignore{To improve query performance, they use an auxiliary data structure, called context tree, that stores results of past context-aware queries indexed by the context of their execution.}

\noindent H. \ The {\em context relational model} (CR) model~\citep{rousoss} extends the relational model with contexts, which are treated as first-class citizens, at the level of database models and query languages. A relation in this model has different schemas and extensions in different contexts. \ignore{An attribute may not exist under some contexts or that the attribute may have different values under different contexts.} A set of basic operations is introduced that extend relational algebra so as to take context into account.



\vspace{2mm}
In the following we compare our  \omd \ contexts with the context models in~ D.-H., which have been used in data management and are relatively close to ours.

Our model of context is relational in that  data are represented as relations only. However, the relational context models described above are not completely relational: they use an extension of relations with new data entities. In~H., a {\em collection of relations} represents a contextual relation. Accordingly, creating, manipulating and querying those contextual relations requires additional tools and care. In E.-G., no relational representation of dimensions is given. The formalizations of context in F.-G. use a \md \ data model for modeling dimensions, and those in E. propose CDTs and chunk configurations, which are not represented by relational terms.

 With respect to languages for querying context, E.-H. use extensions of relational algebra, from which shortcomings are inherited, in particular, the lack of recursion queries and the inability  to capture incomplete data. Both features are  supported the \omd \ model. The work under~D. studies and formalizes the problem of querying federated databases using the notion of logical consequence. \ \omd \ extends the work in~F. and its query language (cf. \cite[chap. 4]{milaniThesis}).

Concerning the applications of these context models, the context model in~G. can be used in particular for context-aware preference databases. Context-aware data tailoring~E. is  a methodology for managing small databases (possibly obtained from larger sources) aimed at being hosted by portable devices. The work in ~D. focuses on using LMS for federated databases. It is not clear how these models can be adapted for other purposes. The work on context-aware databases in~F. is fairly general and can be applied in many applications in data management.  Our \md \ context model is not restricted to the problem of data quality specification and extraction, and can have wide applicability. However, it is an open problem to find ways to provide and include in \omd \ the specific applications and tools that those other models provide.

\ignore{+++
\newpage
\appendix

\comlb{I shortened and modified below. Please, check. I haven't made up my mind yet about including these topics.}

\commos{I do my best to properly answer the comments, may be the result is helpful to be included in the paper.}

\section*{APPENDIX}
\setcounter{section}{0}

\subsection*{Repairing On-the-Fly}

We have suggested above two ways of dealing with possible inconsistencies. One is the detection of separability, and the other applying a repair semantics. In the first case, the constraints are checked on the result of the chase
(at least implicitly) (cf. Section~\ref{sec:kappas}). In the second case, depending on the result of the chase and the discovery of inconsistencies, the extensional data has to be repaired. As an alternative to offer
inconsistency-tolerant \qa \ (cf. Section~\ref{sec:inco}), we could  try integrating constraint checking with data generation, restricting the latter process. This can be achieved by compiling the constraints into the \tgds,  to restrict  data generation that may lead to inconsistency.

\begin{example} \label{ex:cont} Consider the categorical relations {\it PatientUnit} and {\it PatientWard}, and the  dimensional rule $${\it PatientUnit}(u,d;p),{\it WardUnit}(w,u)~\rightarrow~ {\it PatientWard}(w,d;p),$$ and a dimensional \nc \ saying that ``there is no patient in the wards of the hospital during {\em September}": \
  ${\it PatientWard}(w,d;p),{\it DayMonth}(d,{\sf september})~\rightarrow~ \bot$. It can be complied into the  \tgd \ above, obtaining:

\vspace{-4mm}
\begin{align}
  \!\!\!\!\!{\it PatientUnit}(u,d;p),\!\!{\it WardUnit}(w,u),\!\lnot {\it DayMonth}(d,{\sf september})\! \rightarrow\! {\it PatientWard}(w,d;p).\!\!\!\!\label{frm:incons-compiled}
\end{align}
\vspace{-4mm}

 \red{The negation in the body of the modified \tgd \ (\ref{frm:incons-compiled}) is {\em stratified} \cite{cali10} since it is applied on a dimensional predicate with complete data in the extensional database.}\boxtheorem\end{example}

This form of stratified negation in \dpm \ programs has been investigated in the literature ~\cite{alviano15,cali10}.

\comlb{If we do this transformation, is there any reasonable assumption under which we stratified ontology will fall into a well-behaved (stratified) class identified in a paper? After all, you are offering tractability at the beginning of this section. Obviously stratification alone is not enough to guarantee tractability in \dpm.}

\commos{You are right, we need to guarantee stratified (or other good form of) negation. In the example, we have stratification because negation is applied on a closed predicate from dimensions. This might not be always the case and we might need to add conditions to lead to a good form of negation, like stratified.}

This approach for dealing with constraints is deserves a full formalization and further exploration, in particular, a way of generating ontology repairs on-the-fly. The properties of the generated repairs should be investigated and compared with those obtained  with the repair semantics in the literature (cf. Section \ref{sec:inco}).

\subsection*{Navigational vs. Static Constraints}\label{sec:discICs}

\comlb{What's the advantage of this separation of constraints. What problem does it solve?}

\commos{The only point of this separation is to study navigation in isolation. I believe that constraints (denial constraints and egds) can be totally separated from the navigation (join on parent-child in the body of rules), although in dimensional egds and ncs we allow them both in a single formula. The same with the value invention (through $\exists$-variables). I think they can be separated from navigation and studied in isolation.}

In the \omd \ model, dimensional constraints are \egds \ and \ncs \ of the form~(\ref{frm:dimensional-egd}) and (\ref{frm:dimensional-nc}), resp., with body atoms of the child-parent predicates for dimensional navigation as explained in Section~\ref{sec:omd}. By static constraints we refer to \egds \ and \ncs \ without these child-parent atoms in their bodies (for example, key constraints and \fds). Here, we briefly study the connection between the two types of constraints.

Dimensional constraints can be transformed into static constraints and dimensional rules of the form (\ref{frm:dimensional-rule}), as it is shown by the next example.

\begin{example} \label{ex:ndim} (ex.~\ref{ex:ont} cont.) Consider the dimensional \egd \ (\ref{frm:expegd}):

\vspace{-4mm}
\begin{align*}
  {\it Therm(w,t;n)},{\it Therm(w',t';n')},\!{\it WardUnit(w,u)},\!{\it WardUnit(w',u)} \rightarrow t=t'.
\end{align*}
\vspace{-4mm}

\noindent We can split it into a dimensional rule of the form~(\ref{frm:dimensional-rule}) and a static \egd \ as follows:

\vspace{-4mm}
\begin{align*}
{\it Therm(w,t;n)},{\it WardUnit(w,u)} ~\rightarrow~& {\it ThermTemp(u,t;n)}.\\
{\it ThermTemp(u,t;n)},{\it ThermTemp(u,t';n')} ~\rightarrow~& t=t'.
\end{align*}
\vspace{-4mm}

\noindent Similarly, the dimensional constraint (\ref{frm:expnc}),

\vspace{-4mm}
\begin{align*}
\nit{WorkSchedules}({\sf intensive},d;n,s),\nit{DayMonth}(d,{\sf jan})  ~\rightarrow~ \bot,
\end{align*}
\vspace{-4mm}

\noindent can be expressed by the following dimensional rule and static \nc:

\vspace{-4mm}
\begin{align*}
[\nit{WorkSchedules}(u,d;n,s),\!{\it WardUnit}(w,u),\hspace{1.5cm}&\\
{\it DayMonth}(d,m)]~\rightarrow&~ {\it SchedulesTemp}(u,m).\\
{\it SchedulesTemp}({\sf intensive},{\sf jan}) ~\rightarrow&~ \bot.
\end{align*}

\vspace*{-5mm}
\boxtheorem
\end{example}

Dimensional constraints of the forms (\ref{frm:dimensional-egd}) and (\ref{frm:dimensional-nc}) prevent some unnecessary steps during the chase by avoiding data propagation for the additional predicates (e.g. {\it ThermTemp} and {\it SchedulesTemp}).

\red{Dimensional constraints can represent static constraints since static \ncs \ and \egds \ are special cases of the general forms~(\ref{frm:dimensional-egd}) and (\ref{frm:dimensional-nc}), resp., without child-parent atoms in their bodies. Notice that the only \egds \ in the model are those obtained from the transformation we just illustrated in the example. However, these new \egds \ are still expected to be separable from the dimensional rules for good computational properties. For example, they could satisfy the condition in Proposition~\ref{prop:sep}.}
+++}

\ignore{+++
\subsection{Summarizability in multidimensional ontologies}\label{sec:semantic}

Like in the relational data model, semantic constraints can be applied on the \hm \ model. {\em Strictness} and {\em homogeneity} are two important constraints on dimensions that ensure the {\em summarizability property}, a desirable property that guarantees the correct computation of cube views~\cite{hurtado-pods}. A dimension is {\em strict}, i.e. each member in a category has at most one parent in each higher category. It satisfies {\em homogeneity (a.k.a. covering)} if each member in a category has at least one parent in a parent category.

\begin{figure}[ht]
\begin{center}
\includegraphics[width=11cm]{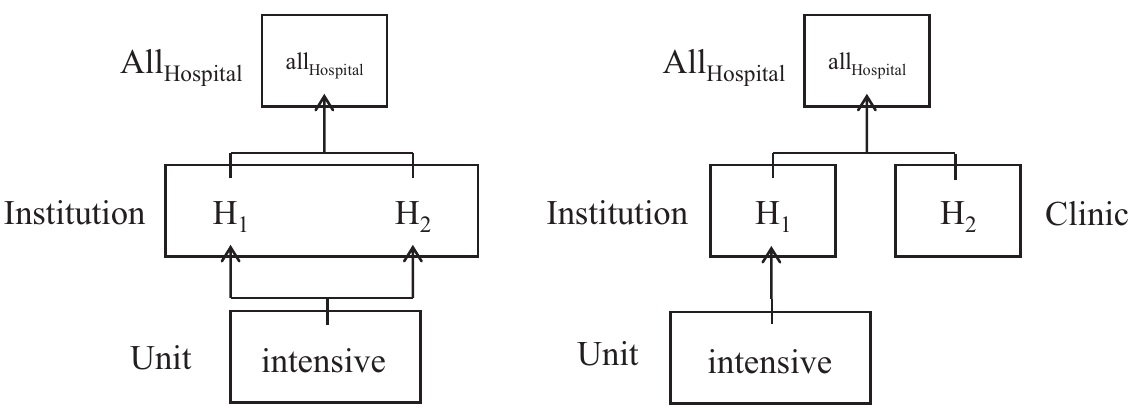}
\caption{Homogenously and strictness of dimensions}\label{fig:dim-constraint}
\end{center}
\end{figure}

\begin{example}\label{ex:sum} The {\sf Hospital} dimension (Figure~\ref{fig:hm-dimension}) satisfies both strictness and homogeneity. In Figure~\ref{fig:dim-constraint}, the dimension on the left-side is not strict, because the {\it Intensive} child member in the {\it Unit} category has two parent members, $H_1$ and $H_2$ in the {\it Institution} category. Assuming that {\it Clinic} and {\it Institution} are both parent categories of the {\it Unit} category, the dimension on the right-side is not homogeneous, because the child member of {\it Intensive} does not have a parent member in the {\it Clinic} category.\boxtheorem
\end{example}

We can represent homogeneity and strictness in the \md \ ontology through \tgds, \egds, and \nc s. To do that, we first introduce a binary predicate $T_H$ defined by rules (\ref{frm:aux}) and (\ref{frm:trans}). There is one rule of this form for every child-parent predicate $P \in \mc{L}$.

\vspace{-4mm}
\begin{align}
P(e_1,e_2) ~\rightarrow~  T_H(e_1,e_2).\label{frm:aux}
\end{align}
\vspace{-4mm}

\noindent The transitivity of $T_H$ is imposed by the following rule:

\vspace{-4mm}
\begin{align}
T_H(e_1,e_2), T_H(e_2,e_3) ~\rightarrow~  T_H(e_1,e_3).\label{frm:trans}
\end{align}
\vspace{-4mm}

The strictness constraint on dimension $H$ can be captured by \egds \ of the form~(\ref{frm:strict}). There is an \egd \ for every {\em intermediate category} $K \in \mc{K}$.\footnote{\ An intermediate category is a category that is not a base or top category.}

\vspace{-4mm}
\begin{align}
T_H(e_1,e_2),T_H(e_1,e_3),K(e_2),K(e_3) ~\rightarrow~  e_2=e_3.\label{frm:strict}
\end{align}
\vspace{-4mm}

\red{Under strictness and homogeneity, each member of a child category has a unique member in the parent category.} Homogeneity of $H$ can be represented by \tgds \ of the form (\ref{frm:hom}). There is a \tgd \ for every child-parent predicate $P\in \mc{L}$, between the child category $K$ and the parent category $K'$.

\vspace{-4mm}
\begin{align}
K(e) ~\rightarrow~ \exists e'\;P(e,e').\label{frm:hom}
\end{align}
\vspace{-4mm}

\noindent Notice that (\ref{frm:hom}) can also be expressed by the following \tgd \ and \nc:

\vspace{-4mm}
\begin{align}
P(e,e') ~\rightarrow&~ C_K(e).\label{frm:hom-nc-pre}\\
K(e),\lnot C_K(e) ~\rightarrow&~ \bot.\label{frm:hom-nc}
\end{align}
\vspace{-4mm}

\noindent Here, $C_K$ is an \red{auxiliary, defined predicate} that collects the child members \red{from} the child-parent predicate $P$, and the negation in (\ref{frm:hom-nc}) is stratified. In fact, this representations is preferred to (\ref{frm:hom}) since it expresses homogeneity as \red{a \nc} \ rather than \red{as} a data generation rule.

\begin{example} (ex.~\ref{ex:sum} cont.) \label{ex:sumrules} In the {\sf Hospital} dimension, the following rules capture strictness:

\vspace{-4mm}
\begin{align*}
\nit{T}_\nit{Hospital}(e_1,e_2),\nit{T}_\nit{Hospital}(e_1,e_3),\nit{Unit}(e_2),\nit{Unit}(e_3) ~\rightarrow~  e_2=e_3.\\
\nit{T}_\nit{Hospital}(e_1,e_2),\nit{T}_\nit{Hospital}(e_1,e_3),\nit{Institution}(e_2),\nit{Institution}(e_3) ~\rightarrow~  e_2=e_3.
\end{align*}
\vspace{-4mm}

\noindent The homogeneity constraint is imposed by the following rule, among others:

\vspace{-4mm}
\begin{align*}
{\it Ward}(w) ~\rightarrow~ \exists u\;\nit{WardUnit}(w,u),{\it Unit}(u).
\end{align*}
\vspace{-4mm}

\noindent This can be expressed by the following \tgd \ and \nc:

\vspace{-4mm}
\begin{align*}
\nit{WardUnit}(w,u) ~\rightarrow&~ C_\nit{Ward}(w).\\
{\it Ward}(w), \lnot C_\nit{Ward}(w) ~\rightarrow&~ \bot.
\end{align*}

\vspace*{-6mm}\boxtheorem
\end{example}

\subsection{Reconstruction of the context-aware databases} \label{sec:omd-ca}

Our \md \ ontologies builds upon {\em context-aware databases}~\cite{martinenghi-qa,martinenghi-er,martinenghi-vldb}. In this section, we briefly review them.

In {\em the context-aware data model}~\cite{martinenghi-qa}, the notion of context is implicit and indirectly captured by relational attributes that take as values members of dimension categories.\footnote{\ Dimensions are defined as in the \hm \ model.} In particular, in a relation in this model, the context of a tuple is captured by its values in dimensions, while the categories of these members specify the granularity level of the context.

\begin{example} \label{ex:ca1} Consider relation $\nit{Schedules}(\nit{Nurse},\nit{Shift},\nit{Unit},\nit{Day})$, with the tuples $({\sf cathy},{\sf night},{\sf terminal},{\sf sep/5})$ and $({\sf helen},{\sf morning},{\sf standard},{\sf sep/6})$ in its extension. The values of \nit{Unit} and \nit{Day} attributes are members from \nit{Unit} and \nit{Day} categories in the {\sf Hospital} and {\sf Temporal} dimensions, resp. So, $({\sf terminal},{\sf sep/5})$ and $({\sf standard}, {\sf sep/6})$ define the context of these tuples, with the granularity level specified by \nit{Unit} and \nit{Day} categories.\boxtheorem \end{example}

The context-aware data model has a query language that extends the relational algebra, by introducing new operators for manipulating the granularity of {\em contextual attributes} (i.e. attributes with values as members of dimensions). These operators add new contextual attributes and their values to a relation. The new attributes are associated with higher or lower categories of the original contextual attributes, and they make it possible to specify contexts with coarser or finer granularities. The language inherits the standard operators from the relational algebra, i.e. projection, selection and join operators.

In the following, we review the context-aware data model in detail, using our running Example~\ref{ex:ca1}.

Let $\mc{H}$ be a set of dimensions. $R^c=(C_1\!:\!l_1,...,C_m\!:\!l_m)$ is a {\em context schema}, where each $C_i$ is an {\em attribute name} and each $l_i$ is a level or category of some dimensions in $\mc{H}$. A {\em context} $\vectt{c}$ over $R^c$ is a function that maps each attribute $C_i$ to a member of $l_i$. Notice that multiple attributes can share an attribute name: they represent the same attribute name at different granularity levels. For example, $C:l$ and $C:l'$ represent $C$ in levels $l$ and $l'$, resp.

\begin{example} \label{ex:ca2} (ex.~\ref{ex:ca1} cont.) $\nit{Schedules}^c=(\nit{Loc}\!\!:\!\!\nit{Unit},\nit{Date}\!\!:\!\!\nit{Day})$ is a context schema, where $\nit{Loc}\!\!:\!\!\nit{Unit}$ and $\nit{Date}\!\!:\!\!\nit{Day}$ are attributes associated with \nit{Unit} and \nit{Day} attributes in {\sf Hospital} and {\sf Temporal} dimensions, resp.\footnote{\ \nit{Loc} is short for \nit{Location}.} Two possible contexts over $\nit{Schedules}^c$ are $({\sf terminal},{\sf sep/5})$ and $({\sf standard},{\sf sep/6})$. \boxtheorem \end{example}

As in the relational data model, $R^r=(A_1\!:\!V_1,...,A_k\!:\!V_k)$ is a {\em relation schema} (which is different from a context schema), where each $A_i$ is a distinct attribute and each $V_i$ is a set of values called the {\em domain of $A_i$}. A {\em tuple} $\vectt{t}$ over a relation schema $R^r$ is a function that associates with each $A_i$ occurring in $R^r$ a value taken from $V_i$. A {\em relation over a relation schema $R^r$} is a finite set of tuples over $R^r$.

$R(R^r\;\!|\!|\;R^c)$ is a {\em contextual relation (c-relation) schema}, where $R^r$ is a {\em relation schema}, and $R^c$ is a {\em context schema}. A {\em c-relation (instance)} over $R$ is a set of tuples $\vectt{t}=(\vectt{r}\;\!|\!|\;\vectt{c})$, where $\vectt{r}$ is a tuple over $R^r$, and $\vectt{c}$ is a context over $R^c$.

\begin{example} \label{ex:ca3} (ex.~\ref{ex:ca2} cont.) $\nit{Schedules}(\nit{Nurse}\!\!:\!\!\nit{String},\nit{Shift}\!\!:\!\!\nit{String}\;|\!|\;\nit{Loc}\!\!:\!\!\nit{Unit},$ \ $\nit{Date}\!\!:\!\!\nit{Day})$ is a c-relation schema, where $(\nit{Nurse}\!\!:\!\!\nit{String},\nit{Shift}\!\!:\!\!\nit{String})$ is a relation schema and $(\nit{Loc}\!\!:\!\!\nit{Unit},\nit{Date}\!\!:\!\!\nit{Day})$ is a context schema, separated by ``$\;|\!|\;$''. A possible extension of \nit{Schedules} contains $({\sf cathy},{\sf night}\;|\!|\;{\sf terminal},{\sf sep/5})$ and $({\sf helen},{\sf morning}\;|\!|\;$ \ ${\sf standard},{\sf sep/6})$, with $({\sf terminal},{\sf sep/5})$ and $({\sf standard},{\sf sep/6})$ as their contexts, resp.\boxtheorem \end{example}

{\em Context-relational algebra (CRA)} is the query language in the context-aware data model that extends relational algebra by introducing two new operators, {\em upward extension} and {\em downward extension}, explained below.

Let $R$ be a c-relation with schema $R(R^r\;\!|\!|\;R^c)$ and contextual attribute $C$ in $R^c$ associated to the level $l$, such that $l$ rolls up to a level $l'$ (cf. Section~\ref{sec:hm} for roll-up relationships). The {\em upward extension of $R$ from the attribute $C:l$ to $l'$}, denoted by $\upext{C:l}{C:l'}{R}$, is the c-relation of schema $R(R^r\;\!|\!|\;R^c \cup \{C:l'\})$, defined as follows,

\vspace{-4mm}
\begin{align*}
\upext{C:l}{C:l'}{R}=\{\vectt{t}'\;|\;\exists \;\vectt{t} \in R, \vectt{t}'[R^c]=\vectt{t}[R^c], \vectt{t}'[R^r]=\vectt{t}[R^r], \vectt{t}'[C:l']=L_{l}^{l'}(\vectt{t}[C:l])\},
\end{align*}
\vspace{-4mm}

\noindent where $\vectt{t}[C:l]$ is the value of attribute $C:l$ in $\vectt{t}$, $\vectt{t}[R]$ are the values of attributes of $R$ in $\vectt{t}$, and $L_{l}^{l'}$ is the roll-up relation between levels $l$ and $l'$ (cf. Section~\ref{sec:hm}). Intuitively, $\upext{C:l}{C:l'}{R}$ has the same schema of $R$ with additional contextual attribute $C:l'$ that represents $C$ in level $l'$. Members of the new attribute are specified by roll up (using $L_{l}^{l'}$) from members of $C:l$ to level $l'$.

\begin{example} \label{ex:ca4} (ex.~\ref{ex:ca3} cont.)  $\upext{\nit{Loc}:\nit{Unit}}{\nit{Loc}:\nit{Inst}}{\nit{Schedules}}$ is the upward extension of \nit{Schedules} from $\nit{Loc}\!:\!\nit{Unit}$ to the level \nit{Institution} (\nit{Inst} in short), with schema $(\nit{Nurse}\!\!:\!\!\nit{String},\nit{Shift}\!\!:\!\!\nit{String}\;|\!|\;\nit{Loc}\!\!:\!\!\nit{Unit},\nit{Date}\!\!:\!\!\nit{Day},\nit{Loc}\!\!:\!\!\nit{Inst})$, where $\nit{Loc}\!\!:\!\!\nit{Inst}$ is the additional contextual attribute. There are two tuples $({\sf cathy},{\sf night}\;|\!|\;{\sf terminal},{\sf sep/5}, {\sf H}_2)$ and $({\sf helen},{\sf morning}\;|\!|\;{\sf standard},{\sf sep/6}, {\sf H}_1)$ in the extension of $\upext{\nit{Loc}:\nit{Unit}}{\nit{Loc}:\nit{Inst}}{\nit{Schedules}}$, where {\sf terminal} and {\sf standard} roll up to ${\sf H}_2$ and ${\sf H}_1$, resp.\boxtheorem
\end{example}

Now let $l''$ be a level such that $l$ {\em drills down} to $l''$, i.e. $l''$ rolls up to $l$. The {\em downward extension of $R$ from the attribute $C:l$ to $l''$}, denoted by $\downext{C:l''}{C:l}{R}$, is the c-relation with schema $R(R^r\;\!|\!|\;R^c \cup \{C:l''\})$, defined as follows:

\vspace{-4mm}
\begin{align*}
\upext{C:l}{C:l''}{R}=\{\vectt{t}''\;|\;\exists \;\vectt{t} \in R, \vectt{t}''[R^c]=\vectt{t}[R^c], \vectt{t}''[R^r]=\vectt{t}[R^r], \vectt{t}[C:l]=L_{l''}^{l}(\vectt{t}''[C:l''])\}.
\end{align*}
\vspace{-4mm}

\noindent Here, members of the new attribute $C:l''$ are specified by drill down from members of $C$ in level $l$ to level $l''$.

\begin{example} \label{ex:ca5} (ex.~\ref{ex:ca3} cont.) $\downext{\nit{Loc}:\nit{Ward}}{\nit{Loc}:\nit{Unit}}{\nit{Schedules}}$ is the downward extension of \nit{Schedules} from $\nit{Loc}:\nit{Unit}$ to the level \nit{Ward}, with schema $(\nit{Nurse}\!\!:\!\!\nit{String},\nit{Shift}\!\!:\!\!\nit{String}\;|\!|\;\nit{Loc}\!\!:\!\!\nit{Unit},\nit{Date}\!\!:\!\!\nit{Day},\nit{Loc}\!\!:\!\!\nit{Ward})$, where $\nit{Loc}\!\!:\!\!\nit{Ward}$ is the additional contextual attribute. There are {\em three} tuples in the extension of $\downext{\nit{Loc}:\nit{Ward}}{\nit{Loc}:\nit{Unit}}{\nit{Schedules}}$: $({\sf cathy},{\sf night}\;|\!|\;{\sf terminal},{\sf sep/5}, {\sf W}_4)$, $({\sf helen},{\sf morning}\;|\!|\;{\sf standard},{\sf sep/6},{\sf W}_1)$, $({\sf helen},{\sf morn}$ \- ${\sf ing}\;|\!|\;{\sf standard},{\sf sep/6},{\sf W}_2)$. This is because {\sf terminal} drills down to ${\sf W}_4$ and {\sf standard} drills down to {\em two} members ${\sf W}_1$ and ${\sf W}_2$.\boxtheorem
\end{example}

The main rationale behind the upward and downward extensions is the need to relax a query with respect to the level of detail of the relations. For example, in the \nit{Schedule} c-relation, one might want to find schedules of a nurse in an institution, even though the schedules might be stored with a lower level (e.g., unit). Both downward and upward extensions meet needs that arise naturally in several application domains~\cite{martinenghi-er}.

The combination of the standard operators of the relation algebra and the new upward and downward extensions makes new operators, e.g. {\em upward selection} and {\em downward selection}, that are explained in detail in~\cite{martinenghi-vldb}. Context-aware databases have applications beyond context modeling, and they are referred by the more general term of {\em taxonomy-based databases}~\cite{martinenghi-er,martinenghi-vldb}.

Our extension of the \hm \ model in Section~\ref{sec:extension} has similarities with c-relations and the context-aware data model that we investigate in detail in Section~\ref{sec:omd-ca}.

The {\em context-aware data model}~\cite{martinenghi-qa,martinenghi-er,martinenghi-vldb} can be fully reconstructed in terms of the \omd \ data model. The standard operators of the relational algebra, i.e. selection, projection, and natural join, are supported by \da \ and inherited by the \md \ ontology. The upward and downward extensions are also expressible by means of dimensional rules of the form~(\ref{frm:dimensional-rule}). The \md \ model can additionally express recursion and contain incomplete data, both not expressible in the context-aware databases.

\begin{example} (ex. \ref{ex:ca3} cont.) \label{ex:extension} The \nit{Schedules} c-relation in the context-aware data model can be represented as a categorical relation with the categorical and non-categorical attributes that correspond to the contextual attributes and relation attributes of the c-relation, resp. In particular, \nit{Schedule}(\nit{Loc},\nit{Date};\nit{Nurse},\nit{Shift}) is a categorical relation that represents the c-relation, $\nit{Schedules}(\nit{Nurse}\!\!:\!\!\nit{String},\nit{Shift}\!\!:\!\!\nit{String}\;|\!|\;\nit{Loc}\!\!:\!\!\nit{Unit},\nit{Date}\!\!:\!\!\nit{Day})$. In the categorical relation, {\it Loc} and {\it Date} are categorical attributes taking values from {\it Unit} and {\it Day} categories in the {\sf Hospital} and {\sf Temporal} dimensions; and {\it Nurse} and {\it Shift} are non-categorical attributes.

The result of upward extension, $\upext{\nit{Loc}:\nit{Unit}}{\nit{Loc}:\nit{Inst}}{\nit{Schedules}}$, is a categorical predicate, ${\it Schedules}'$, which is defined by a dimensional rule of the form~(\ref{frm:dimensional-rule}):

\vspace{-4mm}
\begin{align*}
{\it Schedules}(u,d;n,s), \nit{UnitInstitution}(u,i) ~\rightarrow~ {\it Schedules}'(u,d,i;n,s).
\end{align*}
\vspace{-4mm}

\noindent Similarly, the result of downward extension, $\downext{\nit{Loc}:\nit{Ward}}{\nit{Loc}:\nit{Unit}}{\nit{Schedules}}$, is a categorical predicate, $\nit{Schedules}''$, which is defined by a rule:

\vspace{-4mm}
\begin{align*}
\nit{Schedules}(u,d;n,s), \nit{WardUnit}(w,u) ~\rightarrow~ \nit{Schedules}''(u,d,w;n,s).
\end{align*}

\vspace*{-6mm}\boxtheorem
\end{example}

The context-aware data model and its query language inherits the limitations of relational algebra, including the following (that are necessary in many applications of the \omd \ data model~\cite{milani14,milani15ruleml}): (1) It can not capture recursive queries on the hierarchical data, (2) It is unable to represent incomplete data.
+++}

\ignore{+++
\newpage
\section{APPENDIX: Datalog$^\pm$ \ \ (long version)} \label{sec:extension}

\ignore{+++
\comlb{This section has to be shortened a lot! I'll leave this for later. I will not touch it until new notice. I extracted some stuff from here to put earlier.}

A \dpm \ program $\prg=\rules \cup \constraints \cup D$ is, in general, formulated by a set of rules $\rules$ of the form~(\ref{frm:tgd}), a (possibly empty) set of constraints $\constraints$ as in (\ref{frm:egd}) and (\ref{frm:nc}), and a database $D$ that provides extensional data for the programs.\footnote{\ For simplicity of notation, when a program $\prg$ has only rules (without constraints, i.e. $\constraints=\emptyset$), we use $\prg$ to refer to the program (i.e. set of rules plus extensional data) and also its set of rules.} The semantics of \tgds, \egds, and \ncs \ in a \dpm \ program is notably different from their semantics in relational databases. With \dpm, we make the {\em open world assumption (\owa)}, which allows incomplete data for all program predicates, and \tgds \ are used to complete the data through data generation, and \egds \ and \ncs \ to restrict this process.

A \dplus \ program is a program without program constraints, i.e. $\constraints = \emptyset$. The set of models of $\prg$, denoted by $\nit{Mod}(\prg)$, contains all instances $I$, such that $I \supseteq D$ and $I \models \rules \cup \constraints$. Given a  \cq \ $\mc{Q}$, the set of answers to $\mc{Q}$ from $\prg$ is defined by $\nit{ans}(\mc{Q}, \prg):= \bigcap_{I \in \nit{Mod}(\prg)} \mc{Q}(I)$, a {\em certain answer} semantics.

\comlb{What about the chase, usual program classes, separability, etc.?}

\comlb{Here, in the following, I add sections about chase and program constraints. They are shorter versions of what we had in the thesis. I change the part about \egds \ just a little bit, I was not sure which parts of it can be shortened, while conveying the whole idea of integrating \egds.}

\subsection{The Chase Procedure} \label{sec:chase}

The \emph{chase} procedure~\cite{aho,beeri} is a fundamental algorithm used for various database problems, including implication of database dependencies, query containment, \cq \ answering under dependencies, and data exchange~\cite{beeri,cali03,fagin,johnson,maier}. The idea is that, given a set of dependencies over a database schema and an instance as input, the chase enforces the dependencies by adding new tuples into the instance, so that the result satisfies the dependencies.

Here, we review the \tgd-based chase procedure that is used with \dplus \ programs, i.e. programs without constraints. In Section~\ref{sec:constraints}, we discuss adding program constraints.

The chase procedure on a \dplus \ program $\Pi$, i.e. a \dpm \ program with set of rules $\rules$ and database $D$ (without program constraints, $\constraints=\emptyset$), starts from the extensional database $D$, and iteratively applies the \tgds \ in $\rules$ through some \tgd-based chase steps. A \tgd-chase step applies a pair of rule/assignment $(\sigma,\theta)$, called {\em applicable pair}, on the current instance $I$, by mapping the body of $\sigma$ into $I$ using $\theta$ and adding the atom $A$ obtained from the head of $\sigma$ to $I$, which results to a new instance $I'$. The atom $A$ is obtained by applying $\theta'$, i.e. an extension of $\theta$ with mapping for $\exists$-variables to fresh nulls.



The chase step explained above is called {\em oblivious}~\cite{cali13}, as it applies a rule when its body can be mapped to an instance, ignoring whether the rule is satisfied. In a sequence of chase steps, each applicable rule/assignment pair is applied only once. The sequence terminates if every applicable pair has been applied. The instances in a sequence are monotonically increasing, but not necessarily strictly increasing, because a chase step can generate an atom that is already in the current instance. Depending on the program and its extensional database, the instances in a chase sequence may be properly extended indefinitely. Different orders of chase steps may result in different sequences. The chase procedure uses the notion of the level of atoms to define a {\em ``canonical"} sequence of chase steps~\cite{cali13}.

\begin{example} \label{ex:chase} Consider a program $\Pi$ with extensional database $D=\{R(a,b)\}$ and set of rules:

\vspace{-6mm}
\begin{align*}
\sigma&:&\hspace{-2cm}R(x,y) \! ~&\rightarrow~ \! \exists z \ R(y,z).\\
\sigma'&:&\hspace{-2cm}R(x,y),R(y,z) \! ~&\rightarrow~ \! S(x,y,z).
\end{align*}
\vspace{-3mm}

With the instance $I_0:=D$, the pair $(\sigma, \theta_1)$, with $\theta_1\!: \ x\mapsto a, y\mapsto b$, is applicable: $\theta_1(\nit{body}(\sigma))=\{R(a,b)\} \subseteq I_0$. The chase inserts a new tuple $R(b,\zeta_1)$ into $I_0$ ($\zeta_1$ is a fresh null, i.e. not in $I_0$), resulting in instance $I_1$. 

Now, $(\sigma',\theta_2)$, with $\theta_2\!: \ x\mapsto a, y\mapsto b, z\mapsto \zeta_1$, is applicable, because $\theta_2(\nit{body}(\sigma'))=\{R(a,b),R(b,\zeta_1)\} \subseteq I_1$. The chase adds $S(a,b,\zeta_1)$ into $I_1$, resulting in $I_2$.\boxtheorem
\end{example}


The result of the chase procedure is an instance called ``{\em the chase}", denoted by ${\it chase}(\Pi)$ or ${\it chase}(D,\rules)$. If the chase does not terminate, the chase is an infinite instance: ${\it chase}(\Pi):=\bigcup_{i=0}^\infty(I_i)$, with $I_0:=D$, and, $I_i$ is the result of the i-th chase step for $i > 0$. If the chase stops after $m$ steps, ${\it chase}(\Pi):=\bigcup_{i=0}^m(I_i)$. 

\begin{example} (ex.~\ref{ex:chase} cont.) The chase continues, without stopping, creating an infinite instance: ${\it chase}(\Pi) = \{R(a,b), R(b,\zeta_1), S(a,b,\zeta_1), R(\zeta_1,\zeta_2), R(\zeta_2,\zeta_3), S(b,\zeta_1,\zeta_2), \ldots\}$.\boxtheorem
\end{example}

Given a program $\Pi$, its chase (instance) is a \emph{universal model}~\cite{fagin}, i.e. a representative of all models in $\nit{Mod}(\Pi)$, in the sense that, for every model in $\nit{Mod}(\Pi)$, there is a homomorphism that maps the universal model to that model. For this reason, as it is shown in~\cite[Proposition 2.6]{fagin}, the (certain) answers to a \cq \ $\mc{Q}$ under $\Pi$, i.e. those in $\nit{ans}(\mc{Q}, \Pi)$, can be computed by evaluating $\mc{Q}$ over the chase instance (and discarding the answers containing nulls).

There are various chase procedures~\cite{cali13,deutsch,fagin,marnette} that compute universal models. They differ in: (a) The definition of applicable rule/assignment pairs in chase steps. For example, in a {\em restricted} chase step~\cite{cali13} a \tgd \ is applicable only if it is not satisfied. \ (b) The order of their chase step applications. For example, {\em core chase}~\cite{deutsch} applies all applicable pairs simultaneously. 

+++}

\subsection{Programs Classes and \dpm} \label{sec:pclasses}

\cq \ answering over \dplus \ programs with arbitrary sets of \tgds \ is in general undecidable~\cite{beeri-icalp}, and it becomes decidable for those programs with a terminating chase. However, it is in general undecidable if the chase terminates, even for a fixed instance~\cite{beeri-icalp,deutsch}. Several sufficient conditions, syntactic~\cite{deutsch,fagin,krotzsch,marnette}, or data-dependent~\cite{meier}, that guarantee chase termination have been identified. {\em Weak-acyclicity}~\cite{fagin} and {\em joint-acyclicity}~\cite{krotzsch} are two kinds of syntactic conditions that use a static analysis of a dependency graph for the predicate positions in the program.

A non-terminating chase does not imply that \cq \ answers are uncomputable. Several program classes are identified for which the chase may be infinite, but \qa \ is still decidable. That is the case for {\em linear}, {\em guarded}, {\em sticky}, {\em weakly-sticky \dpm}~\cite{cali09,cali10vldb,cali11sigrec,cali12is}, {\em shy \de}~\cite{leone11}, and {\em finite expansion sets (fes)}, {\em finite unification sets (fus)}, {\em bounded-treewidth sets (bts)}~\cite{baget09,baget11,baget11ai}. Each program class defines conditions on the program rules that lead to good computational properties for \qa. In the following, we focus on sticky and weakly-sticky \dpm \ programs because of their relevance to \md \ ontologies.

\ignore{++++
\subsubsection{Weakly-acyclic programs} \label{sec:wa}

{\em Weakly-acyclic} programs are defined using dependency graphs. The {\em dependency graph (DG)} of a program $\prg$ with schema $\mc{R}$ (cf. Figure~\ref{fig:dg}) is a directed graph whose vertices are the positions of $\mc{R}$. The edges are defined as follows: for every $\sigma \in \prg$, and every universally quantified variable ($\forall$-variable)\footnote{\ Every variable that is not existentially quantified is implicitly universally quantified.} $x$ in ${\it head}(\sigma)$ in position $p$ in ${\it body}(\sigma)$ (among possibly other positions where $x$ appears in ${\it body}(\sigma)$): \ (a) for each occurrence of $x$ in position $p'$ in ${\it head}(\sigma)$, create an edge from $p$ to $p'$, (b) for each $\exists$-variable $z$ in position $p''$ in ${\it head}(\sigma)$, create a {\it special (dashed) edge} from $p$ to $p''$.

The {\it rank of a position} $p$ in the graph, denoted by $\nit{rank}(p)$, is the maximum number of special edges over all (finite or infinite) paths ending at $p$. $\finiteRank(\prg)$ denotes the set of finite-rank positions in $\prg$. A program is {\em Weakly-acyclic (\WA)} if all of the positions have finite-rank~\cite{fagin}.

\begin{example} \label{ex:dg} Let $\Pi$ be a program with rules:\end{example}

\begin{minipage}[t]{0.45\textwidth}
{\centering
\vspace{-1.75cm}
\begin{align*}
  U(x) &~\rightarrow~ \exists y\;R(x,y),\\
  R(x,y) &~\rightarrow~ P(y,x).
\end{align*}
}
\end{minipage}
\hfill
\begin{minipage}[t]{0.45\textwidth}
\resizebox{!}{2.0cm}{\begin{picture}(0,0)%
\includegraphics{./fig/dg}%
\end{picture}%
%
%
\setlength{\unitlength}{3947sp}%
\begingroup\makeatletter\ifx\SetFigFont\undefined%
\gdef\SetFigFont#1#2#3#4#5{%
  \reset@font\fontsize{#1}{#2pt}%
  \fontfamily{#3}\fontseries{#4}\fontshape{#5}%
  \selectfont}%
\fi\endgroup%
\begin{picture}(2220,1194)(631,-1597)
\put(2836,-578){\makebox(0,0)[lb]{\smash{{\SetFigFont{12}{14.4}{\rmdefault}{\mddefault}{\updefault}{\color[rgb]{0,0,0}$P[1]$}%
}}}}
\put(2836,-1523){\makebox(0,0)[lb]{\smash{{\SetFigFont{12}{14.4}{\rmdefault}{\mddefault}{\updefault}{\color[rgb]{0,0,0}$P[2]$}%
}}}}
\put(1655,-578){\makebox(0,0)[lb]{\smash{{\SetFigFont{12}{14.4}{\rmdefault}{\mddefault}{\updefault}{\color[rgb]{0,0,0}$R[1]$}%
}}}}
\put(1655,-1523){\makebox(0,0)[lb]{\smash{{\SetFigFont{12}{14.4}{\rmdefault}{\mddefault}{\updefault}{\color[rgb]{0,0,0}$R[2]$}%
}}}}
\put(646,-1021){\makebox(0,0)[lb]{\smash{{\SetFigFont{12}{14.4}{\rmdefault}{\mddefault}{\updefault}{\color[rgb]{0,0,0}$U[1]$}%
}}}}
\end{picture}%
}
\captionof{figure}{Dependency graph}\label{fig:dg}
\vspace{3mm}
\end{minipage}

According to the \dg \ of $\prg$, shown in Figure~\ref{fig:dg}, the ranks of $U[1]$, $R[1]$, and $P[1]$ are $0$, and the ranks of $R[2]$ and $P[2]$ are $1$. $\Pi$ is \WA \ since all positions have finite-rank.\boxtheorem

The problem of \bcq \ answering over a \WA \ program is \ptime-complete in data complexity~\cite{fagin}. This is because the chase for these programs stops in polynomial time w.r.t. the size of the data~\cite{fagin}. The same problem is 2\exptime-complete in combined complexity, i.e. w.r.t. the size of both program rules and the data~\cite{kolaitis}.

\subsubsection{Sticky programs} \label{sec:sticky}

The class of sticky programs is characterized through a body variable {\em marking procedure} whose input is the set of rules in the program $\prg$ (the extensional data do not participate in it). The procedure has two steps:

\begin{itemize}
  \item[(a)] {\em Preliminary step}: For each $\sigma \in \prg$ and variable $x$ in ${\it body}(\sigma)$, if there is an atom $A$ in ${\it head}(\sigma)$ where $x$ does not appear, mark each occurrence of $x$ in ${\it body}(\sigma)$.
  \item[(b)] {\em Propagation step}: For each $\sigma \in \prg$, if a marked variable in ${\it body}(\sigma)$ appears at position $p$, then for every $\sigma' \in \prg$ (including $\sigma$), mark each occurrence of the variables in ${\it body}(\sigma')$ that appear in ${\it head}(\sigma')$ in the same position $p$.
\end{itemize}

$\prg$ is {\em sticky} when, after applying the marking procedure, there is no rule with a marked variable appearing more than once in its body. Notice that a variable never appears both marked and unmarked in a  same body.

\vspace{-2mm}
\begin{example} \label{example:sticky}
The original set of three rules is shown on the left-hand side below. The second rule already shows marked variables (with a hat) after the preliminary step. The set of rules on the right-hand side are the result of whole marking procedure.\end{example}\vspace{-7mm}

\vspace{2mm}
\[ \arraycolsep=0pt
\begin{array}{rl c rl}
R(x,y),P(x,z) ~\rightarrow~& S(x,y,z). &\hspace*{1.0cm}& R(\hat{x},y),P(\hat{x},\hat{z}) ~\rightarrow~& S(x,y,z).\\
S(\hat{x},y,\hat{z}) ~\rightarrow~& U(y).&\hspace*{1.0cm}&S(\hat{x},y,\hat{z}) ~\rightarrow~& U(y).\\
U(x) ~\rightarrow~& \exists y\;R(y,x).&\hspace*{1.0cm}&U(x) ~\rightarrow~& \exists y\;R(y,x).
\end{array}
\]

Variables  $x$ and $z$ in the first rule-body end up marked  after the propagation step: they appear in the same rule's head, in positions where marked variables appear in the second rule ($S[1]$ and $S[3]$). Accordingly, the set of rules is {\em not} sticky: $x$ in the first rule's body is marked and occurs twice in it (in $R[1]$ and $P[1]$). \boxtheorem

Sticky \dpm \ is properly included in \SCh. That is programs with sch-property may not be syntactically sticky.

\begin{example} \label{ex:sch-non-sticky} Let $\prg$ be a \dplus \ program with extensional data $D=\{R(a,b)\}$ and the \tgd \ rule, $R(x,y),R(y,z)~\rightarrow~R(x,z)$. $\prg$ is {\em not} sticky as $y$ is marked and appears twice in the body of the rule. The chase of $\prg$ does not apply the rule. So, the program trivially has the {\em sch-property}.\boxtheorem\end{example}

With sticky programs, \qa \ can be done in \ptime \ in data complexity and \exptime-complete in combined complexity~\cite{cali12}. In fact, \cq \ answering over sticky programs is {\em first-order rewritable}~\cite{cali10vldb,gottlob11}. \fo \ rewritability is a desirable property as it is well known that the evaluation of \fo \ queries is in the highly tractable class \acz \ (in data complexity)~\cite{vardi}.




\subsubsection{Weakly-sticky programs} \label{sec:ws}

Weakly-sticky programs form a syntactic class that extends those of \WA \ and sticky programs.  Its characterization does not depend on the extensional data, and uses the notions of finite-rank and marked variable introduced in Section~\ref{sec:wa} and, resp., Section~\ref{sec:sticky}: A set of rules $\prg$ is {\em weakly-sticky} (WS) if, for every rule in it and every repeated variable in its body, the variable is either non-marked or appears in some positions in $\finiteRank(\prg)$.

\vspace{-2mm}
\begin{example} Consider $\prg$ with the set of rules:

\vspace{-4mm}
\begin{align*}
R(x,y) ~\rightarrow&~ \exists z\; R(y,z).\\
R(x,y),U(y),R(y,z) ~\rightarrow&~ R(x,z).
\end{align*}
\vspace{-4mm}

According to the graph of $\prg$, $\finiteRank(\prg)=\{U[1]\}$, and $\infiniteRank(\prg)=\{R[1],R[2]\}$. After applying the marking procedure, every body variable in $\prg$ is marked. $\prg$ is \WS \ since the only repeated marked variable is $y$, in the second rule, and it appears in $U[1] \in \finiteRank(\prg)$.

Now, let $\prg'$ be the program with the first rule of $\prg$ and the second rule as follows:

\vspace{-4mm}
\begin{align*}
R(x,y),R(y,z) ~\rightarrow&~ R(x,z).
\end{align*}
\vspace{-4mm}

\noindent Now, $\finiteRank(\prg')=\emptyset$ and $\infiniteRank(\prg')=\{R[1],R[2]\}$. After applying the marking procedure, every body variable in $\prg'$ is marked. $\prg'$ is {\em not} \WS \ since $y$ in the second rule is repeated, marked and appears in $R[1]$ and $R[2]$, both in $\infiniteRank(\prg)$.\boxtheorem\end{example}

Intuitively, \WS \ generalizes the syntactic stickiness condition by prohibiting repeated marked variables appearing only in infinite-rank positions. The \WS \ condition guarantees tractability of \cq \ answering, because a \cq \ can be answered on an initial fragment of the chase whose size is polynomial in the size of the extensional database (it also depends on the query). In fact, the data complexity and combined complexity of \cq \ answering over \WS \ programs are  \ptime-complete and 2\exptime-complete, respectively~\cite{cali12is}.

+++}

\vspace{1cm}
Table~\ref{tab:complexity} is a summary of complexity of \bcq \ answering under programs that have been reviewed in this section.

\begin{table}[h]
  \centering
\setlength{\tabcolsep}{0.3em}
\setlength{\arrayrulewidth}{0.75pt}
\renewcommand*\arraystretch{1.35}
\begin{tabular}{p{1.5cm} p{3cm} p{4cm}}
\hline
& {\bf Data complexity} & {\bf Combined complexity}\\
\hline
\WA    & \ptime-complete \hfill & 2\exptime-complete \hfill \\
{\em sticky}    & in \acz & \exptime-complete \\
\WS    & \ptime-complete & 2\exptime-complete\\
\hline
\end{tabular}
\caption{Complexity of \bcq \ answering under programs in Section~\ref{sec:pclasses}}\label{tab:complexity}
\end{table}
+++}

\ignore{+++
\subsection{Program Constraints} \label{sec:constraints}

So far in this section we have considered programs without constraints, i.e. only programs with extensional databases and rules of the form~(\ref{frm:tgd}). In this section, we extend these programs with \ncs \ and \egds \ of the forms (\ref{frm:nc}) and (\ref{frm:egd}), resp. These \ics \ are called {\em program constraints} in the context of \dpm \ programs.

\subsubsection{Negative constraints} \label{sec:ncs}

We recall the syntax and the semantics of \ncs \ introduced in Section~\ref{sec:relational}. A \nc \ is of the form~(\ref{frm:nc}), $\eta:P_1(\vectt{x}_1), \ldots, P_n(\vectt{x}_n) \rightarrow \bot$; and $\eta$ holds in an instance $I$ if there is no assignment $\theta$ that maps $P_1(\vectt{x}_1), \ldots, P_n(\vectt{x}_n)$ into $I$. This can be checked by evaluating a \bcq \ associated to $\eta$, $\mc{Q}_\eta: \exists x_1\;...\;\exists x_n\;(P_1(\vectt{x}_1)\wedge \ldots\wedge P_n(\vectt{x}_n))$ in $I$. The \nc \ $\eta$ holds in $I$ if and only if $\mc{Q}_\eta$ is false in $I$.





For a program $\prg$, adding \ncs \ may change \cq \ answers under the program, by eliminating some models from $\nit{Mod}(\prg)$~\cite{cali09}. According to~\cite[Theorem~11]{cali09}, \cq \ answering under a program $\prg$ with rules $\rules$ and \ncs \ $\constraints$ can be reduced to \cq \ answering under a program, $\prg'$, with only the rules $\rules$ (and without the \nc s). This can be done by:

\begin{itemize}
  \item [(a)] Checking if the \ncs \ in $\constraints$ are satisfied by $\prg'$. More precisely, for $\eta \in \constraints$, we evaluate the \bcq \ $\mc{Q}_\eta$ over $\prg'$. If at least one of such queries answers positively, $\prg$ is inconsistent, and thus \qa \ is trivial since every query is entailed.
  \item [(b)] If the \ncs \ are not satisfied by $\prg'$, for every \bcq \ $\mc{Q}$, $\prg \models \mc{Q}$ if and only if $\prg' \models \mc{Q}$, i.e. we can answer queries over $\prg'$, ignoring the \nc s.
\end{itemize}

Notice that in (a), if $\prg'$ (the program without \egds) is \WS, for which the chase might not terminate, we can answer $\mc{Q}_\eta$ on the limited portion of the chase of $\prg'$, as explained in Section~\ref{sec:ws}.

\begin{example} Consider a program $\prg$ with the database $D=\{U(a)\}$, the \tgd, $\sigma:U(x) ~\rightarrow~ \exists y\; R(x,y)$, and the \nc, $\eta: U(x)~\rightarrow~\bot$, and let $\prg^\prime$ be $\prg$ without the $\eta$. For $\prg$ as in (1), we evaluate $\mc{Q}_{\eta}$ under $\prg'$. The answer is true, which means $\eta$ does not hold, and $\prg$ is inconsistent. Every \cq \ is trivially true under $\prg$.

Now let the program $\prg''$ be $\prg'$ with the \nc, $\eta': R(x,x)~\rightarrow~\bot$. For $\prg''$, we first evaluate $\mc{Q}_{\eta'}$ under $\prg'$ and since it is false, $\eta'$ is satisfied by $\prg''$. So, we ignore the constraint: $\prg''\not\models \mc{Q}$ because $\prg' \not\models \mc{Q}$, with $\mc{Q}:\exists x\; R(x,x)$.\boxtheorem \end{example}


We can see that answering \bcq s under \dpm \ programs with \ncs \ has the same data complexity of answering \bcq s on \dplus \ programs with \tgds \ alone.


\subsubsection{Equality-generating dependencies} \label{sec:egds}

Let us retake \egds \ of the form~(\ref{frm:egd}) in Section~\ref{sec:relational}, with their semantics defined as \fo \ sentences. An \egd \ $\epsilon$ holds in an instance $I$ if and only if any assignment that maps $\nit{body}(\epsilon)$ to $I$, maps the head variables of $\epsilon$ to the same terms. So as with \nc s, adding \egds \ to a \dplus \ program $\prg$ may eliminate certain models from $\nit{Mod}(\prg)$, which in turn may change \cq \ answers. However, imposing \egds \ on a \dplus \ program is different from imposing \nc s. This is specially due to possible interactions between the \egds \ and \tgds \ during the chase procedure, as we show now.

\begin{example} \label{ex:non-separable} Consider a program $\prg$ with $D=\{R(a,b)\}$ and the following rules:

\vspace{-4mm}
\begin{align}
R(x,y) ~&\rightarrow~ \exists z\; \exists w\;S(y,z,w).\label{frm:sep1}\\
S(x,y,y) ~&\rightarrow~ P(x,y).\label{frm:sep2}
\end{align}

\noindent $\nit{chase}(\prg)=\{R(a,b),S(b,\zeta_1,\zeta_2)\}$, with $\zeta_1$ and $\zeta_2$ fresh nulls. Rule~(\ref{frm:sep2}) is not applied since $\zeta_1$ and $\zeta_2$ are not equal, as required by the body. The answer to a \bcq \ $\mc{Q}:\exists x\;\exists y\;P(x,y)$ is {\em false} under $\prg$ as $\nit{chase}(\prg) \not \models \mc{Q}$. Now consider $\prg'$ that is obtained by adding the following \egd \ to $\prg$:

\vspace{-0.7cm}
\begin{align}
S(x,y,z) ~&\rightarrow~ y=z.\label{frm:sep3}
\end{align}


\noindent The chase of $\prg'$ first applies rule~(\ref{frm:sep1}) and results in $I_1=\{R(a,b),S(b,\zeta_1,\zeta_2)\}$. Now, there is no more tgd/assignment applicable pair. But, if we apply the \egd~(\ref{frm:sep3}), it equates $\zeta_1$ and $\zeta_2$, and results in $I_2=\{R(a,b),S(b,\zeta_1,\zeta_1)\}$. (This kind of egd-chase step explained in Section~\ref{sec:egds}.) Now, rule~(\ref{frm:sep2}) and $\theta': x\mapsto b, y\mapsto \zeta_1$ are applicable and they add $P(b,\zeta_1)$ to $I_2$, generating $I_3=\{R(a,b),S(b,\zeta_1,\zeta_1),P(b,\zeta_1)\}$.

The procedure terminates since no more \tgds \ or \egds \ can be applied. The chase result, $\nit{chase}(\prg')$, is $I_3$. $\mc{Q}$ holds under $\prg'$: $\nit{chase}(\prg') \models \mc{Q}$.\boxtheorem\end{example}

Example~\ref{ex:non-separable} shows that adding an \egd \ to a program may change query answers. Also, the chase of a program may apply an \egd \ between tgd-chase steps. Actually, there might be interactions between \tgds \ and an \egd, i.e. the application of a tgd activates the \egd, which in turn might make some \tgds \ applicable. This confirms that, unlike \nc s, checking \egds \ can not be postponed until all \tgds \ have been applied: they have to be applied during the chase procedure, through {\em egd-chase steps}.

Let $\prg$ be a program with database $D$, \tgds \ $\rules$, and \egds \ $\constraints$. The \egd \ $\epsilon:P_1(\vectt{x}_1), \ldots, P_n(\vectt{x}_n) \rightarrow x=x'$ in $\constraints$ and assignment $\theta$ are {\em applicable} on an instance $I$ if $\theta(\nit{body}(\epsilon)) \in I$ and $\theta(x)\neq\theta(x')$. In this case, the {\em effect} of the application of the pair, $(\epsilon,\theta)$, an {\em egd-chase step}, is as follows:

\begin{itemize}
  \item[(a)] If $\theta(x)$ and $\theta(x')$ are two distinct constants,\footnote{\ This includes the constants that the \tgds \ may introduce.} then the result is a {\em hard constraint violation}, which causes the {\em failure of the chase}, and the halting of its computation. We say the program is {\em inconsistent}.\footnote{\ Notice that here we defined the failure of the chase with \egds.}

  \item[(b)] Otherwise (i.e. at least one of them is a null), the result is the replacement of all occurrences of $\theta(x')$ in $I$ by $\theta(x)$, where $\theta(x)$ precedes $\theta(x')$ in the lexicographical order.\footnote{\ We assume a lexicographical order between constants in $\mc{C}$ and also between nulls in $\mc{N}$, in which constants precede all null values.}
\end{itemize}

The (combined) chase procedure of a program, with \tgds \ and \egds, iteratively applies both \tgd \ and \egd \ chase steps, as follows: (a) Apply applicable pairs of \egd/assignment exhaustively, as long as they exist, and according to a pre-established order (such as tgd-chase steps). (b) Apply a tgd-chase step. In other words, a sequence of steps in the (combined) chase procedure is formed by a sequence of tgd-chase steps, while before each tgd-chase step every possible egd-chase steps are applied.

The (combined) chase {\em terminates} if it either fails (always due to a failed \egd-step) or there are no more applicable pairs of \egd/assignment or \tgd/assignment. The (combined) chase failure results in an inconsistent program that answers every \bcq \ positively. If the (combined) chase does not fail, the result is a possibly infinite universal model, that satisfies both the \tgds \ and \egds~\cite{cali13}.

The interaction of \tgds \ and \egds \ during the chase procedure (as shown in Example~\ref{ex:non-separable}) may lead to undecidability of \qa~\cite{beeri-icalp}. In fact, this is true even in simple cases, such as combinations of {\em functional dependencies (\fds)}  and {\em inclusion dependencies} (\ideps)~\cite{chandra}, or {\em key constraints} and \ideps~\cite{cali12}. A {\em separability condition} on the combination of \egds \ and \tgds \ guarantees a harmless interaction, i.e. \cq \ answering becomes decidable~\cite{cali12is,cali12}.

\begin{definition}[Separability]~\cite{cali12is}  Let $\prg$ be a program with a database $D$, a set of \tgds \ $\rules$, and a set of \egds \ $\constraints$, and let $\prg'$ be the program with $D$ and $\rules$ (without the \egds). $\rules$ and $\constraints$ are {\it separable} if either (a) the chase of $\prg$ fails, or (b) for any \bcq \ $\mc{Q}$, $\prg \models \mc{Q}$ if and only if $\prg' \models \mc{Q}$.\boxtheorem \end{definition}

In Example~\ref{ex:non-separable}, the \tgds \ and the \egd \ are not separable as the chase does not fail, and the \egd \ changes \cq \ answers (in that case, $\prg \not\models \mc{Q}$ and $\prg' \models \mc{Q}$).

Separability is a semantic condition, relative to the chase, and depends on a program's extensional data. It guarantees that, as for programs with \tgds \ and \nc s, \cq \ answering under a program $\prg$ with \tgds \ $\rules$ and \egds \ $\constraints$  can be reduced to \cq \ answering under a program, $\prg'$, with only the \tgds \ in $\rules$ (and without the \egds)~\cite{cali12is}. More precisely, if separability holds,

\begin{itemize}
  \item[(a)] Combined chase failure can be decided by posing the \bcq s obtained from the \egds \ directly to the program without the \egds~\cite[Theorem~1]{cali12amw}. More specifically, for the \egds \ in $\constraints$, $\epsilon:P_1(\vectt{x}_1), \ldots, P_n(\vectt{x}_n)\rightarrow x=x'$, the obtained \bcq s are $\mc{Q}_\epsilon:\exists \vectt{x}_1,...,\exists \vectt{x}_n\;(P_1(\vectt{x}_1)\wedge \ldots\wedge P_n(\vectt{x}_n)\wedge x\neq x')$. The chase fails iff the answer is positive at least for one of them.
  \item[(b)] If it does not fail, \cq \ answering can be done with the \tgds \ alone~\cite{cali12is,cali12amw}.
\end{itemize}

Notice that in (a), if $\prg'$ is \WS, for which the chase might not terminate, we can answer $\mc{Q}_\epsilon$ on the limited portion of the chase of $\prg'$, explained in Section~\ref{sec:ws}.

\begin{example} (ex.~\ref{ex:non-separable} cont.) Let $\prg''$ be $\prg$ with an additional \egd:

\vspace{-4mm}
\begin{align}\epsilon'\!:\;\;R(x,y)~\rightarrow~x=y.\end{align}
\vspace{-4mm}

\noindent The \tgds \ and $\epsilon'$ are separable. Intuitively, this is because $R$ in the body of $\epsilon'$ does not appear in the head of the \tgds, and as a result, $\epsilon'$ can only equate values from $\nit{Adom}(D)$ during the (combined) chase of $\prg''$. Therefore, the application of $\epsilon$ either causes failure, or it does not change the chase result or \cq \ answers. In fact, this observation leads to a sufficient syntactic condition for separability (cf. Condition (a) Definition~\ref{df:non-conf}).

Since the \tgds \ and $\epsilon'$ are separable, we can decide if the chase fails by posing the \bcq \ $\mc{Q}_{\epsilon'}:\exists x\;\exists y\;(R(x,y)\wedge x\neq y)$ to $\prg$ (the program without the egd). The answer is positive, which means the (combined) chases fails, and the program is inconsistent.\boxtheorem\end{example}

The problem of deciding if a set of \tgds \ and \egds \ is separable is undecidable~\cite{cali10vldb}. For {\em functional dependencies (FDs)}, as opposed to general \egds, a syntactic sufficient condition for separability is a {\em non-conflicting} combination of \tgds \ and \fds~\cite{cali10vldb,cali12amw}.

\begin{definition}[Non-conflicting FDs]~\cite{cali10vldb} \label{df:non-conf} Let $\prg$ be a program with a set of \tgds, $\rules$, and a set $\constraints$ of \fds. $\rules$ and $\constraints$ are {\it non-conflicting} if, for every pair formed by a \tgd \ $\sigma \in \rules$ and an \fd \ $\epsilon$ of the form $R:\vectt{A} \rightarrow \vectt{B}$ in $\constraints$, at least one of the following holds: (a) $\nit{head}(\sigma)$ is not an $R$-atom, (b) $U_\sigma \not\supseteq \vectt{A}$, or (c) $U_\sigma = \vectt{A}$ and each $\exists$-variable in $\sigma$ occurs just once in the head of $\sigma$. Here, $U_\sigma$ is the set of positions of $\forall$-variables in the head of $\sigma$.\boxtheorem\end{definition}

\begin{example} \label{ex:non-conflicting} Consider $\mc{R}$, a schema with a ternary predicate $S$ and a unary predicate $V$, a \tgd \ $\sigma: V(x) ~\rightarrow~ \exists y\;\exists z\;S(x,y,z)$, and the \fd \ $\epsilon: \{S[1],S[2]\}\rightarrow \{S[3]\}$. The \fd \ $\epsilon$ can be written as an \egd: $S(x,y,z),S(x,y,z') ~\rightarrow~ z=z'$. Here, $\sigma$ and $\epsilon$ are non-conflicting, because (b) holds: $U_\sigma \not\supseteq A$, with $U_\sigma=\{S[1]\}$ and $A=\{S[1],S[2]\}$.

Now, consider the \tgd \ $\sigma': V(x) ~\rightarrow~ \exists y\;S(x,y,y)$, and the \fd \ $\epsilon':\{S[1]\}\rightarrow \{S[2],S[3]\}$. They are not non-conflicting, because none of (a)-(c) holds: For (a), $S$ appears in the head of $\sigma$ and the body of $\epsilon$. For (b) and (c), $A=U_{\sigma'}=\{S[1]\}$, but $y$ appears twice in the head of $\sigma'$.
\boxtheorem\end{example}

Conditions (a) and (b) in Definition~\ref{df:non-conf} imply separability, by ensuring that the application of a \tgd \ can not make an \egd \ applicable. In particular for (a), the atoms introduced by a \tgd \ never appear in the body of an \egd. For (b), these atoms do not make the \egd \ applicable since they introduce fresh nulls in the positions in $A$. With respect to (c), the atoms can make the \egd \ applicable, but applying the \egd \ does not change \cq \ answers (as shown in Example~\ref{ex:c}), which still guarantees separability. Notice that the non-conflicting condition is decidable.

\begin{example} \label{ex:c}Let $\prg$ be a program with $D=\{P(a,b),V(a)\}$, \fd \ $\epsilon: \{P[1]\}\rightarrow \{P[2]\}$, and \tgd \ $\sigma: V(x) \rightarrow \exists y\;P(x,y)$. According to (c), $\epsilon$ and $\sigma$ are non-conflicting: $A=U_{\sigma}=\{P[1]\}$ and $y$ appears once in the head of $\sigma$.

The chase of $\prg$ applies $(\sigma,\theta)$, with $\theta: x \mapsto a$, and results in $I_1=\{P(a,b),V(a),$ \ $P(a,\zeta_1)\}$. Now, $\epsilon$ is applied, which converts $\zeta_1$ into $b$, and results in $I_2=D$. This \egd \ application does not change \cq \ answers since for every \cq \ $\mc{Q}$, it holds $\mc{Q}(I_1)=\mc{Q}(I_2)$.\boxtheorem\end{example}

+++}


\vspace{4mm}
\begin{thebibliography}{1}
\bibliographystyle{plainnat}

\bibitem[Abiteboul et al.(1995)]{abiteboul}
Abiteboul, S., Hull, R. and Vianu, V.
\newblock{\em Foundations of Databases}.
\newblock{Addison-Wesley}, 1995.



\bibitem[Ahmetaj et al.(2016a)]{ahmet1}
Ahmetaj, S., Ortiz, M. and \v{S}imkus, M.
\newblock{Polynomial Datalog Rewritings for Ontology Mediated Queries with Closed Predicates.}
In \newblock{\em Proc. of the Alberto Mendelzon International Workshop on Foundations of Data Management (AMW)}, CEUR-WS Proc. Vol. 1644, 2016.

\ignore{\bibitem[Aho et al.(1979)]{aho}
Aho, A. V., Beeri, C. and Ullman, J. D.
\newblock{The Theory of Joins in Relational Databases}.
\newblock{\em ACM Transactions on Database Systems (TODS)}, 1979, 4(3): 297-314.}

\bibitem[Alviano et al.(2012)]{alviano12}
Alviano, M., Faber, W., Leone, N. and Manna, M.
\newblock{Disjunctive Datalog with Existential Quantifiers: Semantics, Decidability, and Complexity Issues}.
\newblock{\em Theory and Practice of Logic Programming (TPLP)}, 2012, 12(4-5): 701-718.

\bibitem[Alviano et al.(2012)]{alviano12-datalog}
Alviano, M., Leone, N., Manna, M., Terracina, G. and Veltri, P.
\newblock{Magic-Sets for Datalog with Existential Quantifiers}.
In \newblock{\em Proc. of the Int. Conference on Datalog in Academia and Industry 2.0}, 2012, Springer LNCS 7494, pp. 31-43.


\bibitem[Analyti et al.(2007)]{analyti}
Anality, A., Theodorakis, M., Spyratos, N. and Constantopoulos, P.
\newblock{Contextualization as an Independent Abstraction Mechanism for Conceptual Modeling}.
\newblock{\em Information Systems}, 2007, 32(1): 24-60.

\ignore{
\bibitem[Arenas et al.(1999)]{arenasPods99}
Arenas, M., Bertossi, L. and Chomicki, J.
\newblock{Consistent Query Answers in Inconsistent Databases}.
In \newblock{\em Proc. of the ACM SIGMOD-SIGACT Symposium on Principles of Database Systems (PODS)}, 1999, pp. 68-79. }


\bibitem[Ariyan \& Bertossi(2011)]{sinaAMW}
Ariyan, S. and Bertossi, L. \newblock{Structural Repairs of Multidimensional Databases}. In \newblock{\em Proc. of the Alberto Mendelzon International \WS \ of Foundations of Data Management (AMW)}, 2011. CEUR-WS, Vol-749.

\bibitem[Ariyan \& Bertossi(2013)]{sina}
Ariyan, S. and Bertossi, L. \newblock{A Multidimensional Data Model with Subcategories for Flexibly Capturing Summarizability}. In \newblock{\em Proc. of the International Conference on Scientific and Statistical Database Management (SSDBM)}, 2013.







\bibitem[Baget et al.(2011b)]{baget11ai}
Baget, J. F., Lecl\'{e}re, M., Mugnier, M.L. and Salvat, E.
\newblock{On Rules with Existential Variables: Walking the Decidability Line}.
\newblock{\em Artificial Intelligence}, 2011, 175(9-10): 1620-1654.

\bibitem[Baget et al.(2015)]{baget15}
Baget, J. F., Bienvenu, M., Mugnier, M.L. and Rocher, S.
\newblock{Combining Existential Rules and Transitivity: Next Steps}.
In \newblock{\em Proc. of the International Joint Conference on Artificial Intelligence (IJCAI)}, 2015, pp. 2720-2726.

\bibitem[Bahmani et al.(2012)]{kr12}
Bahmani, Z., Bertossi, L., Kolahi, S. and Lakshmanan, L. \newblock{Declarative Entity Resolution via Matching Dependencies and Answer Set Programs}. In \newblock{\em Proc. of the International Conference on Principles of Knowledge Represenattion and Reasoning (KR)}, 2012, AAAI Press, pp. 380-390.

\ignore{
\bibitem[Bahmani \& Bertossi(2017)]{flairs17}
\red{Bahmani, Z.} and Bertossi, L. \newblock{Enforcing Relational Matching Dependencies with Datalog for Entity Resolution}. To appear in \newblock{\em Proc. the International Florida Artificial Intelligence Research Society Conference (FLAIRS), 2017.}   }

\bibitem[Barcelo(2009)]{barcelo}
Barcelo, P. \newblock{Logical Foundations of Relational Data Exchange}. {\em ACM SIGMOD Record}, 2009, 38(1):49-58.

\bibitem[Batini \& Scannapieco(2006)]{batini}
Batini, C. and Scannapieco, M.
\newblock{\em Data Quality: Concepts, Methodologies and Techniques}. Second edition,
\newblock{Springer}, 2016.

\bibitem[Beeri \& Vardi(1981)]{beeri-icalp} 
Beeri, C. and Vardi, M. Y.
\newblock{The Implication Problem for Data Dependencies}.
In \newblock{\em Proc. of the Colloquium on Automata, Languages and Programming (ICALP)}, 1981, Springer LNCS 115, pp. 73-85.



\bibitem[Bertossi(2006)]{bertossi06}
Bertossi, L. \newblock{Consistent Query Answering in Databases}. \newblock {\em ACM Sigmod Record}, June 2006, 35(2):68-76.

\bibitem[Bertossi et al.(2009)]{amw09}
Bertossi, L., Bravo, L. and Caniupan, M. \newblock{Consistent Query Answering in Data Warehouses.}
In \newblock{\em Proc. of the Alberto Mendelzon International Workshop on Foundations of Data Management (AMW)},  2009. CEUR-WS, Vol-450.

\bibitem[Bertossi et al.(2011a)]{bertossi-brite}
Bertossi, L., Rizzolo, F. and Lei, J.
\newblock{Data Quality is Context Dependent}.
In \newblock{\em Proc. of the Workshop on Enabling Real-Time Business Intelligence (BIRTE) Collocated with the International Conference on Very Large Data Bases (VLDB)}, Springer LNBIP 84, 2011, pp. 52-67.


\bibitem[Bertossi(2011b)]{bertossi11}
Bertossi, L.
\newblock{\em Database Repairing and Consistent Query Answering}.
\newblock{Morgan \& Claypool}, 2011.

\bibitem[Bertossi \& Bravo(2013)]{bertossi13}
Bertossi, L. and Bravo, L.
\newblock{Generic and Declarative Approaches to Data Quality Management}.
In \newblock{\em Handbook of Data Quality -  Research and Practice}, 2013, Springer, pp. 181-211, DOI: 10.1007/978-3-642-36257-6\_9.

\bibitem[Bertossi et al.(2016)]{bertossi16}
Bertossi, L. and Rizzolo, F.
\newblock{Contexts and Data Quality Assessment}. Corr Arxiv Paper cs.DB/1608.04142, 2016.

\bibitem[Bienvenu at al.(2014)]{bienvenu14}
Bienvenu M., Bourgaux, C. and Goasdou\`{e}, F.
\newblock{Querying Inconsistent Description Logic Knowledge Bases under Preferred Repair Semantics}.
In \newblock{\em Proc. of the National Conference on Artificial Intelligence (AAAI)}, 2014, pp. 996-1002.

\bibitem[Bienvenu at al.(2016)]{bienvenu16}
Bienvenu M., Bourgaux, C. and Goasdou\`{e}, F.
\newblock{Explaining Inconsistency-tolerant Query Answering over Description Logic Knowledge Bases}.
In \newblock{\em Proc. of the National Conference on Artificial Intelligence (AAAI)}, 2016, pp. 900-906.







\ignore{
\bibitem[Bolchini et al.(2007a)]{bolchini}
Bolchini, C., Schreiber, F. and Tanca, L.
\newblock{A Methodology for a Very Small Data Base Design}.
\newblock{\em Information Systems}, 2007, 32(1): 61-82.}


\bibitem[Bolchini et al.(2007a)]{bolchini-survey}
Bolchini, C., Curino, C. A., Quintarelli, E., Schreiber, F. A. and Tanca, L.
\newblock{A Data-Oriented Survey of Context Models}.
\newblock{\em ACM SIGMOD Record}, 2007, 36(4): 19-26.

\bibitem[Bolchini et al.(2007b)]{bolchini-view}
Bolchini, C., Quintarelli, E., Rossato, R. and Tanca, L.
\newblock{Using Context for the Extraction of Relational Views}.
In \newblock{\em Proc. of the International and Interdisciplinary Conference on Modeling and Using Context}, 2007, pp. 108-121.


\bibitem[Bolchini et al.(2009)]{bolchini-cdt}
Bolchini, C., Curino, C. A., Quintarelli, E., Schreiber, F. A. and Tanca, L.
\newblock{Context Information for Knowledge Reshaping}.
\newblock{\em International Journal of Web Engineering and Technology}, 2009, 5(1): 88-103.

\bibitem[Bolchini et al.(2013)]{bolchini-is}
Bolchini, C., Quintarelli, E. and Tanca, L.
\newblock{CARVE: Context-Aware Automatic View Definition over Relational Databases}.
\newblock{\em Information Systems}, 2013, 38(1): 45-67.



\ignore{\bibitem[Bourhis et al.(2013)]{bourhis}
Bourhis, P., Morak, M. and Pieris, A.
\newblock{The Impact of Disjunction on Query Answering Under Guarded-Based Existential Rules}.
In \newblock{\em Proc. of the International Joint Conference on Artificial Intelligence (IJCAI)}, 2013, pp. 796-802.}

\bibitem[Bourhis et al.(2015)]{bourhis}
Bourhis, P., Manna, M., Morak, M. and Pieris, A. \newblock
Guarded-Based Disjunctive Tuple-Generating Dependencies. {\em ACM Trans. Database Syst.}, 2016, 41(4).




\bibitem[Cal\`{i} et al.(2003)]{cali03} 
Cal\`{i}, A., Lembo, D. and Rosati, R.
\newblock{On the Decidability and Complexity of Query Answering over Inconsistent and Incomplete Databases}.
In \newblock{\em Proc. of the ACM SIGMOD-SIGACT Symposium on Principles of Database Systems (PODS)}, 2003, pp. 260-271.


\bibitem[Cal\`{i} et al.(2009)]{cali09} 
Cal\`{i}, A., Gottlob, G. and Lukasiewicz, T.
\newblock{Datalog$^\pm$: A Unified Approach to Ontologies and Integrity Constraints}.
In \newblock{\em Proc. of the International Conference on Database Theory (ICDT)}, 2009, pp. 14-30.

\ignore{
\bibitem[Cal\`{i} et al.(2010a)]{cali10vldb}
Cal\`{i}, A., Gottlob, G. and Pieris, A.
\newblock{Advanced Processing for Ontological Queries}.
In \newblock{\em Proc. VLDB Endowment (PVLDB)}, 2010, 3(1-2): 554-565.}

\ignore{
\bibitem[Cal\`{i} et al.(2010b)]{cali10} 
Cal\`{i}, A., Gottlob, G., Lukasiewicz, T., Marnette, B. and Pieris, A.
\newblock{Datalog$^\pm$: A Family of Logical Knowledge Representation and Query Languages for New Applications}.
In \newblock{\em Proc. of the Annual IEEE Symposium on Logic in Computer Science (LICS)}, 2010, pp. 228-242. }



\ignore{
\bibitem[Cal\`{i} et al.(2012a)]{cali12is}
Cal\`{i}, A., Gottlob, G. and Pieris, A.
\newblock{Ontological Query Answering under Expressive Entity-Relationship Schemata}.
\newblock{\em Information Systems}, 2012, 37(4): 320-335.}

\ignore{\bibitem[Cal\`{i} et al.(2012b)]{cali12jws} 
Cal\`{i}, A., Gottlob, G. and Lukasiewicz, T.
\newblock{A General Datalog-Based Framework for Tractable Query Answering over Ontologies}.
\newblock{\em J. of Web Semantics}, 2012, 14:57-83. }

\bibitem[Cal\`{i} et al.(2012c)]{cali12} 
Cal\`{i}, A., Gottlob, G. and Pieris, A.
\newblock{Towards More Expressive Ontology Languages: The Query Answering Problem}.
\newblock{\em Artificial Intelligence}, 2012, 193:87-128.

\bibitem[Cal\`{i} et al.(2012d)]{cali12amw} 
Cal\`{i}, A., Console, M. and Frosini, R.
\newblock{On Separability of Ontological Constraints}.
In \newblock{\em Proc. of the Alberto Mendelzon International Workshop on Foundations of Data Management (AMW)}, 2012, CEUR-WS Proc. Vol. 866, pp. 48-61.

\bibitem[Cal\`{i} et al.(2013)]{cali13} 
Cal\`{i}, A., Gottlob, G. and Kifer, M.
\newblock{Taming the Infinite Chase: Query Answering under Expressive Relational Constraints}.
\newblock{\em Journal of Artificial Intelligence Research (JAIR)}, 2013, 48(1): 115-174.



\bibitem[Caniupan and Bertossi(2010)]{monica}
Caniupan-Marileo, M. and Bertossi, L. \newblock{The Consistency Extractor System: Answer Set Programs for
Consistent Query Answering in Databases}. {\em Data \& Knowledge Engineering}, 2010, 69(6):545-572.

\bibitem[Ceri et al.(1990)]{ceri}
Ceri, S., Gottlob, G. and Tanca, L.
\newblock{\em Logic Programming and Databases}.
\newblock{Springer}, 1990.

\bibitem[Chandra \& Vardi(1985)]{chandra} 
Chandra, A.K. and Vardi, M.Y.
\newblock{The Implication Problem for Functional and Inclusion Dependencies}.
\newblock{\em SIAM Journal of Computing}, 1985, 14(3): 671-677.

\ignore{
\bibitem[Chiang \& Miller(2011)]{chiang11}
Chiang, F. and Miller, R.
\newblock{A Unified Model for Data and Constraint Repair}.
In \newblock{\em Proc. of the International Conference on Data Engineering (ICDE)}, 2011, pp. 446-457. }



\ignore{
\bibitem{das}
Ganti,~V. and Das Sarma,~A. {\em Data Cleaning. A Practical Perspective}. \newblock{Morgan \& Claypool}, 2013.}




\bibitem[Eckerson(2002)]{eckerson}
Eckerson, W.
\newblock{Data Quality and the Bottom Line: Achieving Business Success Through a Commitment to High Quality Data}.
\newblock{\em Report of the Data Warehousing Institute}, 2002.

\bibitem[Enderton(2001)]{enderton} Enderton, H. B. \newblock{{\em A Mathematical Introduction to Logic}}. 2nd Edition, Academic Press, 2001.

\bibitem[Fagin et al.(2005)]{fagin}
Fagin, R., Kolaitis, P. G., Miller, R. J. and Popa, L.
\newblock{Data Exchange: Semantics and Query Answering}.
\newblock{\em Theoretical Computer Science (TCS)}, 2005, 336(1): 89-124.

\bibitem[Fan(2008)]{fan08}
Fan, W.
\newblock{Dependencies Revisited for Improving Data Quality}.
In \newblock{\em Proc. of the ACM SIGMOD-SIGACT Symposium on Principles of Database Systems (PODS)}, 2008, pp. 159-170.

\ignore{
\bibitem[Fan(2009)]{fan09}
Fan, W.
\newblock{Constraint-Driven Database Repair}.
In \newblock{\em Encyclopedia of Database Systems, Springer US}, 2009, pp. 458-463.}

\bibitem[Fan \& Geerts(2012)]{geerts}
Fan, W. and Geerts, F. \newblock{\em Foundations of Data Quality Management}. \newblock{Morgan \& Claypool}, 2012.

\bibitem[Fan(2015)]{fan}
Fan, W. \newblock{Data Quality: From Theory to Practice}. {\em SIGMOD Record}, 2015, 44(3):7-18.

\bibitem[Fan et al.(2009)]{fan09-pvldb}
Fan, W., Jia, X., Li, J. and Ma, S.
\newblock{Reasoning about Record Matching Rules}.
In \newblock{\em Proc. VLDB Endowment (PVLDB)}, 2009, 2(1): 407-418.

\bibitem[Fan et al.(2011)]{fan11}
Fan, W., Gao, H., Ji, X., Li, J. and Ma, S.
\newblock{Dynamic Constraints for Record Matching}.
\newblock{\em The International Journal on Very Large Data Bases (VLDBJ)}, 2009, 20(4): 495-520.

\bibitem[Franconi \& Sattler(1999)]{franconi99}
Franconi, E. and Sattler, I.
\newblock{A DataWarehouse Conceptual Data Model for Multidimensional Aggregation}.
In \newblock{\em Proc. of the International Workshop on Design and Management of Data Warehouses (DMDW)}, 1999, Article No. 13.

\bibitem[Franconi et al.(2011)]{franconi}
Franconi, E., Garcia, Y. and Seylan, I.
\newblock{Query Answering with DBoxes is Hard}.
\newblock{\em Electronic Notes in Theoretical Computer Science (ENTCS)}, 2011, 278(1): 71-84.










\bibitem[Ghidini \& Serafini(1998)]{ghidini98}
Ghidini, C. and Serafini, L.
\newblock{Model Theoretic Semantics for Information Integration}.
In \newblock{\em Proc. of the International Conference on Artificial Intelligence, Methodology, Systems, and Applications (AIMSA)}, 1998, Springer LNAI Vol. 1480, pp. 267-280.

\bibitem[Ghidini \& Giunchiglia(2001)]{ghidini01}
Ghidini, C. and Giunchiglia, F.
\newblock{Local Models Semantics, or Contextual Reasoning = Locality + Compatibility}.
\newblock{\em Artificial Intelligence}, 2001, 127(1): 221-259.

\bibitem{ghidini}
Ghidini,~C. and Serafini,~L. \newblock Multi-Context Logics - A General Introduction. In {\em Context in Computing},
Br\'ezillon,~P. and Gonzalez,~A.~J. (eds.), Springer, 2014, pp. 381-399.

\bibitem[Giunchiglia \& Serafini(1994)]{giunchiglia94}
Giunchiglia, F. and Serafini, L.
\newblock{Multilanguage Hierarchical Logics, or: How We Can Do without Modal Logics}.
\newblock{\em Artificial Intelligence}, 1994, 65(1): 29-70.



\bibitem[Gottlob et al.(2011)]{gottlob11}
Gottlob, G., Orsi, G. and Pieris, A.
\newblock{Ontological Queries: Rewriting and Optimization}.
In \newblock{\em Proc. of the International Conference on Data Engineering (ICDE)}, 2011, pp. 2-13.

\bibitem[Gottlob et al.(2015)]{reasW}
Gottlob, G., Morak, M. and Pieris, A. \newblock
Recent Advances in Datalog$^\pm$. {\em Reasoning Web 2015}, Springer LNCS 9203, 2015, pp. 193-217.




\bibitem[Herzog et al.(2009)]{herzog}
Herzog, T., Scheuren, F. and Winkler, W.
\newblock{\em Data Quality and Record Linkage Techniques}.
\newblock{Springer}, 2009.

\bibitem[Horrocks \& Sattler(1999)]{horrocks99}
Horrocks, I. and Sattler, S.
\newblock{A Description Logic with Transitive and Inverse Roles and Role Hierarchies}.
\newblock{\em ACM Transactions on Database Systems (TODS)}, 1999, 9(3): 385-410.


\bibitem[Hurtado \& Mendelzon(2002)]{hurtado-pods}
Hurtado, C. and Mendelzon, A.
\newblock{OLAP Dimension Constraints}.
In \newblock{\em Proc. of the ACM SIGMOD-SIGACT Symposium on Principles of Database Systems (PODS)}, 2002, pp. 169-179.

\bibitem[Hurtado et al.(2005)]{hurtado-acm}
Hurtado, C., Gutierrez, C. and Mendelzon, A.
\newblock{Capturing Summarizability with Integrity Constraints in OLAP}.
\newblock{\em ACM Transactions on Database Systems (TODS)}, 2005, 30(3): 854-886.

\ignore{
\bibitem{ilyas}
Ilyas,~I. and Chu,~X.
Trends in Cleaning Relational Data: Consistency and Deduplication. {\em Foundations and Trends in Databases}, 2015, 5(4):281-393. }

\bibitem[Imielinski \& Lipski(1984)]{imielinski}
Imielinski, T. and Lipski, W.
\newblock{Incomplete Information in Relational Databases}.
\newblock{\em Journal of the ACM}, 1984, 31(4): 761-791.

\bibitem[Jensen et al.(2010)]{jensen}
Jensen, Ch. S.,
Bach Pedersen, T. and
Thomsen, Ch. \newblock{
{\em Multidimensional Databases and
Data Warehousing}}. \newblock{Morgan \& Claypool}, 2010.

\bibitem[Jiang et al.(2008)]{lei}
Jiang, L., Borgida, A. and Mylopoulos, J.
\newblock{Towards a Compositional Semantic Account of Data Quality Attributes}.
In \newblock{\em Proc. International Conference on Conceptual Modeling (ER)}, 2008, pp. 55-68.

\bibitem[Johnson \& Klug(1984)]{johnson} 
Johnson, D. S. and Klug, A.
\newblock{Testing Containment of Conjunctive Queries under Functional and Inclusion Dependencies}.
In \newblock{\em Proc. of the ACM SIGMOD-SIGACT Symposium on Principles of Database Systems (PODS)}, 1984, pp. 164-169.

\bibitem[Juran \& Godfrey(1999)]{juran}
Juran, J.M. and A.M. Godfrey.
\newblock{\em Juran's Quality Handbook, Fifth Edition}.
\newblock{McGraw-Hill}, 1999.

\ignore{
\bibitem[Kolahi \& Lakshmanan(2009)]{kolahi}
Kolahi, S. and Lakshmanan, L.
\newblock{On Approximating Optimum Repairs for Functional Dependency Violations}.
In \newblock{\em Proc. of the International Conference on Database Theory (ICDT)}, 2009, pp. 53-62. }

\bibitem[Kolaitis et al.(2006)]{kolaitis}
Kolaitis,~P.~G., Tan,~W.~C. and Panttaja,~J.
\newblock{The Complexity of Data Exchange}.
In \newblock{\em Proc. of the ACM SIGMOD-SIGACT Symposium on Principles of Database Systems (PODS)}, 2006, pp. 30-39.



\bibitem[Lembo et al.(2010)]{lembo10}
Lembo, D., Lenzerini M., Rosati, R., Ruzzi, M. and Savo, D. F.
\newblock{Inconsistency-Tolerant Semantics for Description Logics}.
In \newblock{\em Proc. of the International Conference on Web Reasoning and Rule Systems (RR)}, 2010, pp. 103-117.

\bibitem[Lembo et al.(2015)]{lembo15}
Lembo, D., Lenzerini M., Rosati, R., Ruzzi, M. and Savo, D. F.
\newblock{Inconsistency-tolerant Query Answering in Ontology-Based Data Access}.
\newblock{\em Journal of Web Semantics}, 2015, 3:3-29.


\bibitem[Lenzerini(2002)]{lenzerini}
Lenzerini, M.
\newblock{Data Integration: A Theoretical Perspective}.
In \newblock{\em Proc. of the ACM SIGMOD-SIGACT Symposium on Principles of Database Systems (PODS)}, 2002, pp. 233-246.



\bibitem[Libkin(2014)]{libkin}
Libkin, L. \newblock{Incomplete Data: What Went Wrong, and How to Fix It}. \newblock{\em Proc. of the ACM SIGMOD-SIGACT Symposium on Principles of Database Systems (PODS)}, pp. 1-13.

\bibitem[Lukasiewicz et al.(2012)]{lukasiewicz12}
Lukasiewicz, T., Martinez, M., Pieris, A. and Simari, G.
\newblock{Inconsistency Handling in Datalog+/- Ontologies}.
In \newblock{\em Proc. of the European Conference on Artificial Intelligence (ECAI)}, 2012, pp. 558-563.

\bibitem[Lukasiewicz et al.(2015)]{lukasiewicz15}
Lukasiewicz, T., Martinez, M., Pieris, A. and Simari, G.
\newblock{From Classical to Consistent Query Answering under Existential Rules}.
In \newblock{\em Proc. of the National Conference on Artificial Intelligence (AAAI)}, 2015, pp. 1546-1552.




\bibitem[Lutz et al.(2013)]{lutz13}
Lutz, C., Seylan, I. and Wolter, F.
\newblock{Ontology-Based Data Access with Closed Predicates is Inherently Intractable (Sometimes)}.
In \newblock{\em Proc. of the International Joint Conference on Artificial Intelligence (IJCAI)}, 2013, pp. 1024-1030.

\bibitem[Lutz et al.(2015)]{lutz15}
Lutz, C., Seylan, I. and Wolter, F.
\newblock{Ontology-Mediated Queries with Closed Predicates}.
In \newblock{\em Proc. of the International Joint Conference on Artificial Intelligence (IJCAI)}, 2015, pp. 3120-3126.

\bibitem[Malaki et al.(2012)]{malaki}
Malaki, A., Bertossi, L. and Rizzolo, F. \newblock{Multidimensional Contexts for Data Quality Assessment}.
In \newblock{\em Proc. of the Alberto Mendelzon International Workshop on Foundations of Data Management (AMW)}, 2012. CEUR-WS, Vol-866.

\bibitem[Maier et al.(1979)]{maier}
Maier, D., Mendelzon, A. and Sagiv, Y.
\newblock{Testing Implications of Data Dependencies}.
\newblock{\em ACM Transactions on Database Systems (TODS)}, 1979,  4(4): 455-469.




\ignore{
\bibitem[Martinenghi \& Torlone(2009)]{martinenghi-qa}
Martinenghi, D. and Torlone, R.
\newblock{Querying Context-Aware Databases}.
In \newblock{\em Proc. of the International Conference on Flexible Query Answering Systems (FQAS)}, 2009, pp. 76-87. }


\bibitem[Martinenghi \& Torlone(2014)]{martinenghi-vldb}
Martinenghi, D. and Torlone, R.
\newblock{Taxonomy-Based Relaxation of Query Answering in Relational Databases}.
\newblock{\em The International Journal on Very Large Data Bases (VLDBJ)}, 2014, 23(5): 747-769.




\bibitem[Mitchell(1983)]{mitchell} 
Mitchell, J.
\newblock{The Implication Problem for Functional and Inclusion Dependencies}.
\newblock{\em Information and Control}, 1983, 56(1): 154-173.

\ignore{
\bibitem[Milani et al.(2014)]{milani14}
Milani, M., Bertossi, L. and Ariyan, S.
\newblock{Extending Contexts with Ontologies for Multidimensional Data Quality Assessment}.
In \newblock{\em Proc. of the International Workshop on Data Engineering meets the Semantic Web (DESWeb) collocated with the International Conference on Data Engineering (ICDE)}, 2014, pp. 242-247, DOI:10.1109/ICDEW.2014.6818333.
}

\ignore{
\bibitem[Milani \& Bertossi(2015a)]{milani15amw}
Milani, M. and Bertossi, L.
\newblock{Tractable Query Answering and Optimization for Extensions of Weakly-Sticky Datalog$\pm$}.
In \newblock{\em Proc. of the Alberto Mendelzon International Workshop on Foundations of Data Management (AMW)}, CEUR-WS Proc. Vol. 1378, 2015, pp. 101-105.
}

\bibitem[Milani \& Bertossi(2015b)]{milani15ruleml}
Milani, M. and Bertossi, L.
\newblock{Ontology-Based Multidimensional Contexts with Applications to Quality Data Specification and Extraction}.
In \newblock{\em Proc. of the International Symposium on Rules and Rule Markup Languages for the Semantic Web (RuleML)}, Springer LNCS 9202, 2015, pp. 277-293.

\ignore{
\bibitem[Milani et al.(2016a)]{milani16amw}
Milani, M., Bertossi, L. and Cal\`{i}, A.
\newblock{Query Answering on Expressive Datalog$^\pm$ Ontologies}.
In \newblock{\em Proc. of the Alberto Mendelzon International Workshop on Foundations of Data Management (AMW)}, CEUR-WS Proc. Vol. 1644, 2016. }

\bibitem[Milani \& Bertossi(2016b)]{milani16rr}
Milani, M. and Bertossi, L.
\newblock{Extending Weakly-Sticky Datalog$^\pm$: Query-Answering Tractability and Optimizations}.
In \newblock{\em Proc. of the International Conference on Web Reasoning and Rule Systems (RR)}, Springer LNCS 9898, 2016, pp. 128-143.

\bibitem[Milani et al.(2016c)]{milani16rr-cali}
Milani, M., Bertossi, L. and Cal\`{i}, A.
\newblock{A Hybrid Approach to Query Answering under Expressive Datalog$^\pm$}.
In \newblock{\em Proc. of the International Conference on Web Reasoning and Rule Systems (RR)}, Springer LNCS 9898, 2016, pp. 144-158.

\bibitem[Milani(2017)]{milaniThesis}
Milani, M. \newblock{{\em Multidimensional Ontologies for Contextual Quality Data Specification and Extraction}}. \newblock{PhD in Computer Science Thesis, Carleton University, 2017}.
\url{http://people.scs.carleton.ca/~bertossi/papers/mostafaFinal.pdf}

\bibitem[Morak(2014)]{morakThesis}
Morak, M. \newblock{{\em The Impact of Disjunction on Reasoning under Existential Rules}}. \newblock{PhD in Computer Science Thesis, University of Oxford, 2015}.


\bibitem[Motschnig(1995)]{motschnig95}
Motschnig-Pitrik, R.
\newblock{An Integrating View on the Viewing Abstraction: Contexts and Perspectives in Software Development, AI, and Databases}.
\newblock{\em Systems Integration}, 1995, 5(1): 23-60.


\bibitem[Motschnig(2000)]{motschnig}
Motschnig-Pitrik, R.
\newblock{A Generic Framework for the Modeling of Contexts and its Applications}.
\newblock{\em Data {$\&$} Knowledge Engineering}, 2000, 32(2): 145-180.



\bibitem[Pitoura et al.(2011)]{pitoura11}
Pitoura, E., Stefanidis, K. and Vassiliadis, P.
\newblock{Contextual Database Preferences}.
\newblock{\em IEE Data Engineering Bulletin}, 2011, 34(2): 19-26.

\bibitem[Poggi et al.(2008)]{poggi}
Poggi, A., Lembo, D., Calvanese, D., De Giacomo, G., Lenzerini, M. and Rosati, R.
\newblock{Linking Data to Ontologies}.
\newblock{\em Data Semantics}, 2008, 10(1): 133-173.


\bibitem[Rabin(1965)]{rabin} Rabin, M. O. \newblock{A Simple Method for Undecidability Proofs and Some Applications}. \newblock{In {\em Logic, Methodology and Philosophy of Science, Proceedings of the 1964 International Congress}}, Bar-Hillel, Y. (ed.). Studies in Logic and the Foundations of Mathematics. North-Holland Publishing Company, Amsterdam 1965, pp. 38-68.



\bibitem[Redman(1998)]{redman}
Redman, T. \newblock
\newblock{The Impact of Poor Data Quality on the Typical Enterprise}.
\newblock{\em Communications of the ACM}, 1998, 41(2): 79-82.

\bibitem[Reiter(1984)]{reiter}
Reiter, R.
\newblock{Towards a Logical Reconstruction of Relational Database Theory}.
In \newblock{\em On Conceptual Modelling}, Springer, 1984, pp. 191-233.

\bibitem[Rosati(2011)]{rosati}
Rosati, R.
\newblock{On the Complexity of Dealing with Inconsistency in Description Logic Ontologies}.
In \newblock{\em Proc. of the International Joint Conference on Artificial Intelligence (IJCAI)}, 2011, pp. 1057-1062.

\bibitem[Rousoss et al.(2005)]{rousoss}
Rousoss, Y., Stavrakas, Y. and Pavlaki, V.
\newblock{Towards a Context-Aware Relational Model}.
In \newblock{\em Proc. International Workshop on Context Representation and Reasoning}, 2005, CEUR-WS, Vol-136, pp. 5-17.


\bibitem[Seylan et al.(2009)]{seylan}
Seylan, I., Franconi, E. and De Bruijn, J.
\newblock{Effective Query Rewriting with Ontologies over DBoxes}.
In \newblock{\em Proc. of the International Joint Conference on Artificial Intelligence (IJCAI)}, 2009, pp. 923-925.


\bibitem[Stefanidis et al.(2011)]{stefanidis11}
Stefanidis, K., Pitoura, E. and Vassiliadis, P.
\newblock{Managin Contextual Preferences}.
\newblock{\em Information Systems}, 2011, 36(8): 1158-1180.

\ignore{\bibitem[Stefanidis et al.(2005)]{stefanidis05}
Stefanidis, K., Pitoura, E. and Vassiliadis, P.
\newblock{A Context-Aware Preference Database System}.
\newblock{\em Pervasive Computing and Communications}, 2005, 3(4): 439-460.

\bibitem[Stefanidis et al.(2007)]{stefanidis}
Stefanidis, K., Pitoura, E. and Vassiliadis, P.
\newblock{Adding Context to Preferences}.
In \newblock{\em Proc. of the International Conference on Data Engineering (ICDE)}, 2007, pp. 846-855.}

\bibitem[Theodorakis et al.(2002)]{theodorakis}
Theodorakis, M., Anality, A., Constantopoulos, P. and Spyratos, N.
\newblock{A Theory of Contexts in Information Bases}.
\newblock{\em Information Systems}, 2002, 27(3): 151-191.


\bibitem[Wang \& Strong(1996)]{wang}
Wang R. Y.  and Strong D. M.
\newblock{Beyond Accuracy: What Data Quality Means to Data Consumers}.
\newblock{\em Journal of Management Information Systems}, 1996, 12(4): 5-33.

\ignore{
\bibitem[Volkovs et al.(2014)]{chiang14}
Volkovs, M., Chiang, F., Szlichta, J., and Miller, R.
\newblock{Continuous Data Cleaning}.
In \newblock{\em Proc. of the International Conference on Data Engineering (ICDE)}, 2014, pp. 244-255. }


\bibitem[Yaghmaie et al.(2012)]{maka}
Yaghmaie, M., Bertossi, L. and Ariyan, S. \newblock{Repair-Oriented Relational Schemas for Multidimensional Databases}.
In \newblock{\em Proc. of the International Conference on Extending Database Technology (EDBT)}, 2012.

\end{thebibliography}
\end{document}